\documentclass[11pt,letterpaper]{article} \usepackage[utf8]{inputenc}
\usepackage[margin=1in]{geometry} \usepackage{amsmath} \usepackage{amssymb}
\usepackage{amsthm} \usepackage{mathtools} \usepackage{microtype}
\usepackage{graphicx} \usepackage{tikz} \usepackage{booktabs} \usepackage{array}
\usepackage{float} \usepackage{caption} \usepackage{subcaption}
\usepackage{xcolor} \usepackage{algorithm} \usepackage{algpseudocode}
\usepackage{aliascnt} \usepackage[numbers,sort&compress]{natbib}
\usepackage{hyperref} \usepackage[nameinlink,capitalise,noabbrev]{cleveref}
\makeatletter
\def\theHALG@line{\thealgorithm.\arabic{ALG@line}}
\makeatother
\hypersetup{hidelinks,pdftitle={NP-Hardness and Additive Inapproximability of Deterministic Communication Complexity}}
\usetikzlibrary{arrows.meta,positioning,calc,matrix,fit,backgrounds}
\theoremstyle{plain} \newtheorem{theorem}{Theorem}[section]
\newaliascnt{lemma}{theorem} \newtheorem{lemma}[lemma]{Lemma}
\aliascntresetthe{lemma} \newaliascnt{proposition}{theorem}
\newtheorem{proposition}[proposition]{Proposition}
\aliascntresetthe{proposition} \newaliascnt{corollary}{theorem}
\newtheorem{corollary}[corollary]{Corollary} \aliascntresetthe{corollary}
\newaliascnt{claim}{theorem} \newtheorem{claim}[claim]{Claim}
\aliascntresetthe{claim} \newaliascnt{observation}{theorem}
\newtheorem{observation}[observation]{Observation}
\aliascntresetthe{observation} \newaliascnt{fact}{theorem}
 \aliascntresetthe{fact}
\newaliascnt{conjecture}{theorem}
 \aliascntresetthe{conjecture}
\theoremstyle{definition} \newaliascnt{definition}{theorem}
\newtheorem{definition}[definition]{Definition} \aliascntresetthe{definition}
\newaliascnt{assumption}{theorem}
 \aliascntresetthe{assumption}
\newaliascnt{example}{theorem} 
\aliascntresetthe{example} \theoremstyle{remark} \newaliascnt{remark}{theorem}
 \aliascntresetthe{remark}

\usetikzlibrary{arrows.meta,positioning,calc,matrix,fit,decorations.pathreplacing}
\definecolor{aliceblue}{RGB}{32,96,151} \definecolor{boborange}{RGB}{183,96,27}
\definecolor{answergreen}{RGB}{34,111,81} \definecolor{linkblue}{RGB}{32,71,116}
\crefname{lemma}{Lemma}{Lemmas}
\crefname{proposition}{Proposition}{Propositions}
\crefname{corollary}{Corollary}{Corollaries}
\crefname{definition}{Definition}{Definitions}
\crefname{example}{Example}{Examples} \crefname{remark}{Remark}{Remarks}

\newcommand{\D}{\operatorname{D}}
\newcommand{\Int}[2]{\mathsf{I}_{#1}(#2)}
\newcommand{\RP}[3]{\mathcal{P}(#1;#2,#3)}
\newcommand{\card}[1]{\lvert#1\rvert}
\newcommand{\ceil}[1]{\left\lceil#1\right\rceil}
\newcommand{\floor}[1]{\left\lfloor#1\right\rfloor}
\newcommand{\ind}{\mathbf{1}}
\newcommand{\NP}{\mathsf{NP}}
\newcommand{\BPP}{\mathsf{BPP}}
\newcommand{\SAT}{\mathsf{SAT}}
\DeclareMathOperator{\rank}{rank} \crefname{claim}{Claim}{Claims}
\crefname{observation}{Observation}{Observations} \crefname{fact}{Fact}{Facts}
\crefname{conjecture}{Conjecture}{Conjectures}
\crefname{assumption}{Assumption}{Assumptions}

\begin{document}
\title{NP-Hardness and Additive Inapproximability of Deterministic Communication Complexity}
\author{Serge Gaspers\thanks{University of New South Wales} \and
Tao Zixu He\thanks{University of Massachusetts Amherst} \and
Simon Mackenzie\footnotemark[1]}
\date{}
\hypersetup{pageanchor=false}
\begin{titlepage}
\maketitle
\thispagestyle{empty}

\begin{abstract}
Given the full truth table of a Boolean function whose input is split
between two players, we study the complexity of determining the minimum
worst-case number of bits they must communicate to compute it. This is a
basic meta-complexity question, posed by Yao in 1979.

We prove NP-hardness of approximation within every fixed additive error
for deterministic communication complexity, under deterministic
polynomial-time reductions. The result holds both in the standard model
and in Yao's original model, where the players alternate after every bit
and only the receiver of the last bit must determine the answer. In
particular, it resolves the exact computation question under Yao's original
conventions. Earlier versions of this same paper established exact
NP-hardness for the standard model, independently and concurrently with
Hirahara, Ilango, and Loff.

In both models, a polynomial-time estimator with any fixed additive error
would yield a polynomial-time $3$-SAT algorithm, deterministic or
bounded-error randomized according to the estimator. For the standard
model, we also give quantitative implications for errors that grow with
the number $T$ of table entries. A bounded-error randomized polynomial-time
estimator with additive error $c\log T/\log\log T$, for any fixed
$0\le c<1/8$, yields a bounded-error randomized $3$-SAT algorithm on $N$
variables and $O(N)$ clauses running in time $2^{O(N^\alpha)}$ for every
fixed $8c<\alpha<1$. A deterministic polynomial-time estimator with additive
error $c_*\log\log T$, for an effective absolute constant $c_*>0$, yields a
deterministic algorithm running in time $2^{O((\log N)^2)}$ on such formulas.
\end{abstract}

\end{titlepage}
\setcounter{page}{1}
\hypersetup{pageanchor=true}

\section{Introduction}
\label{sec:intro}

Alice and Bob know a Boolean matrix $M$. Alice receives a row index and
Bob receives a column index. They must determine the entry at their
intersection, communicating as few bits as possible. Their local
computation is free, and they must answer correctly for every pair of
indices. The deterministic communication complexity $\D(M)$ is the
smallest worst-case number of bits used by any such protocol.

We study how hard it is to estimate $\D(M)$ when the algorithm receives
the whole table as input. The table is total, meaning every entry is specified, and we write $T$ for its number
of entries. An estimator runs in time measured in $T$ and outputs a number,
not a protocol. A randomized estimator must meet its error guarantee with
probability at least $2/3$ on every table, while the protocols whose cost it
estimates remain deterministic. Yao posed the problem of computing $\D(M)$
exactly from the table when he introduced communication complexity in
1979~\cite{yao1979distributed}.

We prove that computing deterministic communication complexity is
NP-hard, and that this remains true when any fixed additive error is
allowed. The exact NP-hardness result for the standard model was first
presented in earlier versions of this
work~\cite{gaspers-he-mackenzie-npcc}, independently and concurrently
with Hirahara, Ilango and Loff~\cite{hirahara2025communication}.

We also prove exact and fixed-error NP-hardness in Yao's original
model: the players alternate after every bit, and only the receiver
of the last bit needs to determine the answer. The earlier
exact-hardness results did not cover these speaking and output
conventions. Our alternating construction therefore resolves Yao's
question in the model in which it was posed.

\paragraph{The challenge of approximation.}
In the standard model, the exact-hardness reduction maps a $3$-CNF formula
to a table whose optimal cost is one bit larger if the formula is
unsatisfiable than if it is satisfiable. An estimate within additive error
$a$ distinguishes the two cases only if their costs differ by more than
$2a$. A natural way to enlarge the gap is to combine several such tables
and hope that their one-bit gaps add up. However, a protocol for the combined problem may share
communication among them, so their costs need not add, and the lower bound
must hold against every such protocol.
Combining problems also enlarges the truth table. Since the reduction
writes out every entry, the table must have size polynomial in the size of
the formula for each fixed gap. Given $k$, our reductions compute a table
and a number $B$ such that the optimal cost is $\le B$ if the formula is
satisfiable and $\ge B+k$ otherwise. We call this an \emph{additive gap}
of $k$ bits. We do not establish hardness of approximation within any
fixed multiplicative factor above one.

\paragraph{Two rules for speaking.}
In an \emph{unrestricted} protocol, the bits sent so far determine who
speaks next, and the same player may send several consecutive bits. In a
\emph{strictly alternating} protocol, the players take turns after every
bit. Where an unrestricted protocol lets one player send two consecutive
bits, an alternating protocol must let the other player speak in between,
so the two optimum costs can differ.

\paragraph{Interlacing.}
Interlacing combines $q$ copies of a matrix $M$ through a pointer. In the $q$-fold interlace $\Int{q}{M}$, Alice
receives a row $x$ of $M$ together with an index $i\in\{1,\ldots,q\}$, and
Bob receives a list $(y_1,\ldots,y_q)$ of $q$ columns of $M$, possibly
with repetitions. The required output is $M(x,y_i)$.
Alice's index acts as a one-step pointer into Bob's input, as in pointer
chasing~\cite{nisan1993rounds}. \Cref{fig:lookup} shows the case $q=2$
for a $2\times2$ matrix.

There is a natural protocol: Alice sends $i$ using $\log q$ bits, and the
players then run an optimal protocol for $M$ on $x$ and $y_i$, for a total
of $\D(M)+\log q$ bits. But is communicating the index an unavoidable extra
cost? Perhaps Bob could speak first, or the players could combine
information about the index and the answer, saving bits.

The interlacing framework rules out such savings when $M$ is robustly
hard, that is, when every large subtable of $M$ is nearly as hard as $M$
itself. Then any protocol that does not start by revealing Alice's index
leaves a subproblem that is still too hard for the bits that remain, so
the natural protocol is optimal~\cite{mackenzie2025refuting,gaspers-he-mackenzie-npcc}.

The interlacing method for lower bounds was developed by Mackenzie and
Saffidine~\cite{mackenzie2025refuting} to refute the exact direct-sum
conjecture, and underlies our original NP-hardness
reduction~\cite{gaspers-he-mackenzie-npcc}.
If $M$ has $n$ columns, Bob has $n^q$ possible lists. To keep its table
polynomial, that reduction already replaced them by a polynomial-size
family of lists.
Repeating the construction needs more: the table produced by one round must
again have lower bounds on its large subtables, so that the next round can
use it as the base matrix $M$. An intermediate version obtained
additive inapproximability under ETH using subexponential-time
constructions~\cite{gaspers2026interlacingeth}.

\paragraph{Polynomial-size gap amplification.}

Our deterministic construction keeps explicit input sets of polynomial
size at every amplification step, so every fixed gap is constructed in
deterministic polynomial time. Our randomized construction instead keeps
all tuples during the amplification. The resulting table is too large to
print, but each entry can be computed from short descriptions of its row
and column. At the end, the construction samples rows and columns of this
table. The table has low rank, so short protocols can be correct on only
few sets of columns, and a union bound over these sets shows that the
sample preserves the gap with high probability (\cref{thm:compression}). In both constructions, the
lower bounds for unrestricted protocols come from interlacing. Because the
two constructions control the table size differently, they give different
tradeoffs between the approximation error and the running time of the
$3$-SAT algorithm that an estimator would imply.

Strict alternation needs a separate construction. The short protocols in our
unrestricted construction send multi-bit messages. Making them alternate
adds turns, and these extra bits can exceed the gap. We therefore
construct a different table, prove its lower bound using the fixed order of
speaking, and again keep only a small explicit set of Bob's inputs.

\paragraph{Output and speaking conventions.}
By default, the conversation alone must determine the answer, which we call
\emph{public output}. We write $D_A^{\mathrm{alt}}(M)$ for the minimum
cost with public output and strict alternation, with Alice starting.
Yao's original convention instead allows the receiver of the last bit
to use their own input as well as the conversation to determine the
answer. We denote its optimum by $Y_{\mathrm{free}}(M)$, where the
protocol may choose its starting player independently of the inputs.
In both conventions, the bits exchanged must determine when the
conversation has ended.

\paragraph{From approximation to satisfiability algorithms.}
For errors growing with the table size, we prove that an estimator for
communication complexity would yield a $3$-SAT algorithm with an
explicitly bounded running time.
If one assumes that such a $3$-SAT algorithm does not exist, each
implication becomes an inapproximability result. The fixed-error
NP-hardness results need no such assumption.

\subsection{Results}
\label{sec:results}

The following table summarizes the fixed-error hardness results and two
examples of our general algorithmic implication. The full tradeoff is
stated in \cref{thm:time-tradeoff}.

\begin{center}
\small
\begin{tabular}{@{}p{.18\linewidth}p{.27\linewidth}p{.25\linewidth}p{.19\linewidth}@{}}
\toprule
Protocol model & Additive error & Implied $3$-SAT algorithm & Algorithm type\\
\midrule
Unrestricted & Every fixed error & Polynomial time & Same as estimator\\
\addlinespace
Strict alternation & Every fixed error & Polynomial time & Same as estimator\\
\addlinespace
Unrestricted & $c_*\log\log T$ & $2^{O((\log N)^2)}$ & Same as estimator\\
\addlinespace
Unrestricted & $c\log T/\log\log T$, $c<1/8$ & $2^{O(N^\alpha)}$, every $8c<\alpha<1$ & Randomized\\
\bottomrule
\end{tabular}
\end{center}

The last two rows concern formulas with $N$ variables and at most $c_sN$
clauses, for any fixed density bound $c_s$. Each row describes the algorithm
that a polynomial-time estimator would imply, and applies to both deterministic and randomized estimators. All randomized algorithms here have bounded error. 

In each reduction below,
$\Phi$ is a $3$-CNF formula. The YES (satisfiable) case is $\Phi\in\SAT$,
and the NO case is $\Phi\notin\SAT$. The reduction computes the threshold
$B$ from the formula and the requested gap, without deciding
satisfiability.

\begin{theorem}[Fixed gaps for unrestricted protocols]
\label{intro:deterministic}
For every fixed integer $k\ge1$, a deterministic polynomial-time algorithm
maps a $3$-CNF formula $\Phi$ to an explicit total Boolean table $M$ and an
integer $B$ such that
\[
\Phi\in\SAT\ \Longrightarrow\ \D(M)\le B,\qquad
\Phi\notin\SAT\ \Longrightarrow\ \D(M)\ge B+k.
\]
The polynomial may depend on $k$. The table can be taken square with
power-of-two side length.
\end{theorem}

A different construction gives the same fixed-gap hardness under
strict alternation.

\begin{theorem}[Fixed gaps under strict alternation]
\label{yao:main}
For every fixed integer $k\ge1$, a deterministic polynomial-time algorithm
maps a $3$-CNF formula $\Phi$ to an explicit total Boolean table $M$ and an
integer $B$ such that
\[
\Phi\in\SAT\ \Longrightarrow\ D_A^{\mathrm{alt}}(M)\le B,\qquad
\Phi\notin\SAT\ \Longrightarrow\ D_A^{\mathrm{alt}}(M)\ge B+k.
\]
The polynomial bounds on output size and running time may depend on $k$.
\end{theorem}

For additive error $a$, choose $k>2a$ and compare the estimate with
$B+k/2$. Thus, for every fixed error, estimating
either cost in the preceding two theorems in deterministic polynomial time
implies $\mathsf P=\NP$, and bounded-error randomized estimation implies
$\NP\subseteq\BPP$. Changing the starting player increases the optimal cost by at most one
bit, and so does requiring the receiver of the last bit to announce the
answer (\cref{yao:comparison}). By increasing the fixed
gap, we therefore also obtain hardness under Yao's output convention.

\begin{corollary}[Yao's output convention]\label{yao:bpp}
For every fixed nonnegative additive error, a deterministic polynomial-time
estimator of $Y_{\mathrm{free}}(M)$ on all explicit total Boolean tables
implies $\mathsf P=\NP$, and a bounded-error randomized one implies
$\NP\subseteq\BPP$. The same conclusions hold for estimating the public-output
alternating cost with Alice first, Bob first, or a starter chosen
independently of the inputs.
\end{corollary}

Taking error zero in \cref{yao:bpp} answers Yao's original question by
proving NP-hardness under his speaking and output rules. We do not establish
a growing-error bound for the alternating model.

For unrestricted protocols, choosing the gap as a function of the formula
size gives an explicit tradeoff between approximation error and the
running time of the resulting $3$-SAT algorithm
(\cref{thm:time-tradeoff}). Here $N$ counts variables in the formula and
$T$ counts entries in the communication table. The following two
consequences illustrate the randomized and deterministic regimes. The
constant $1/8$ uses the sharper size bound for the randomized
construction.

\begin{corollary}[Randomized estimation and $3$-SAT running time]
\label{thm:quantitative}
Fix $0\le c<1/8$. Suppose a bounded-error randomized polynomial-time
estimator approximates $\D(M)$ within additive error
$c\log T/\log\log T$ on every sufficiently large explicit total Boolean
table with $T$ entries. Then, for every fixed $8c<\alpha<1$ and every
fixed $c_s$, sparse $3$-SAT on $N$ variables and at most $c_sN$ clauses has
a bounded-error randomized algorithm with worst-case running time
$2^{O(N^\alpha)}$. It suffices for the estimator to work on square tables
whose side length is a power of two.
\end{corollary}

For the smaller error scale below, the deterministic construction turns a
deterministic estimator into a quasipolynomial-time algorithm for sparse
$3$-SAT.

\begin{corollary}[Deterministic estimation and $3$-SAT running time]
\label{det:time}
There is an effective absolute constant $c_*>0$ with the following property.
If a deterministic polynomial-time estimator approximates $\D(M)$ within
additive error $c_*\log\log T$ on every sufficiently large explicit total
Boolean table, then, for every fixed $c_s$, $3$-SAT on $N$ variables and
at most $c_sN$ clauses has a deterministic algorithm running in time
$2^{O((\log N)^2)}$. It suffices for the estimator to work on square tables
whose side length is a power of two.
\end{corollary}

\Cref{sec:time-tradeoff} gives the general tradeoff and explains how to choose
the gap for other running-time bounds.

The proofs of \cref{intro:deterministic,det:time} appear in
\cref{sec:deterministic-endpoints}. We prove
\cref{thm:time-tradeoff,thm:quantitative} in \cref{sec:time-tradeoff}, and
\cref{yao:main,yao:bpp} together at the end of
\cref{sec:alternating}. \Cref{sec:overview} explains the ideas behind these proofs.

\subsection{Formal verification and proof dependencies}
\label{sec:formal-verification}

A companion Lean development provides machine-checked proofs of our
constructions and their consequences for communication complexity and
$3$-SAT running time~\cite{interlacing2026lean}. The formal statements
retain explicit hypotheses for the cited source reductions.
The public proof inspector~\cite{interlacing2026inspector} links manuscript
statements to the corresponding Lean declarations and exposes their
definitions and hypotheses. The repository~\cite{interlacing2026lean}
provides the source, build instructions, and verification records.

The earlier direct-sum construction of Mackenzie and
Saffidine~\cite{mackenzie2025refuting} has separately been formalized in
Lean by Soltani Moakhar, Gholami, Springer, JafariRaviz, and
Hajiaghayi~\cite{moakhar2026beyond}, with statements and proofs available in
a public inspector~\cite{beyondlibrary2026inspector}.

OpenAI GPT models used through ChatGPT Pro and Codex, and Anthropic's
Claude models, assisted with proof development, correctness checking, writing,
and the subsequent Lean formalization. The
\hyperref[sec:ai-methodology]{AI methodology statement} describes their roles.

\clearpage
\section{Technical overview}
\label{sec:overview}

Our reduction encodes a graph so that failure of $3$-coloring forces one
extra bit of communication. A proper $3$-coloring gives a protocol of an
explicit cost $C-1$ (\cref{src:upper}). Conversely, interlacing forces every
protocol of that cost to assign Alice's inputs to bins, in effect choosing a
coloring, before Bob reveals which edge is tested. In the NO case, some tested
edge is monochromatic, which forces an additional bit. The difficulty is to
prove this lower bound against every protocol, and then to repeat the
construction without losing the lower bound or polynomial size.

We explain three steps: the one-bit gap, the lower bounds under input
deletion that make the gap repeatable, and the explicit short input lists
that make every fixed number of rounds a polynomial-time reduction.
Strict alternation requires a different encoding, described last.

\subsection{Constructing the source matrix}

Our source reduction maps a formula to a graph of maximum degree
$\Delta=O(1)$ that is $3$-colorable if the formula is satisfiable (the
YES case), and otherwise leaves at least a constant fraction $\gamma>0$ of
its edges monochromatic under every assignment of three colors to its
vertices (the NO case, \cref{src:gap-colouring}). This NO condition survives
small deletions of vertices. The graph has no isolated vertices, so
$|E|\ge|V|/2$, and deleting fewer than $\gamma|V|/(4\Delta)$
vertices destroys fewer than $\gamma|E|/2$ edges, so at least
$\gamma|E|/2$ monochromatic edges remain.

The construction takes a base communication matrix $R$ of cost $d=\D(R)$
and adds rows encoding the graph, described below, to interlaced copies of
$R$. In the first round, $R$ is a simple matrix of known cost (\cref{cl:seed}). We
allow a general base matrix so that the construction can be repeated, with
the output of one round as the base matrix of the next. We also need lower bounds
that survive deletion of some inputs. We measure the retained rows and columns
using fixed positive rational weights, summing to one on each side. These act
as fractional counts, which let us control how much each input family contributes,
independently of its number of labels. Later, adjusting
weights also lets us shorten tuple lists while keeping the weights induced on
small sets of coordinates (\cref{sec:small-lists}).
The weights do not change communication cost, since every protocol must still compute
every entry correctly. The deterministic construction never samples from them,
and its final output is an ordinary Boolean table with the weights discarded.

A proper $3$-coloring places the vertices in
three groups so that the endpoints of every edge lie in different groups. We call
these groups \emph{bins}. The interlacing construction uses a power-of-two
number $s\ge4$ of bins. We treat each graph vertex as an \emph{item} to be
placed in a bin, and add $s-3$ auxiliary items that occupy the extra bins,
so that the vertices cannot use them as extra colors. At an edge, its two
endpoints and the auxiliary items are \emph{active}, and all other items
are \emph{inactive}. We require that each bin contain at most one active
item at every edge. The auxiliary items are active at every edge, so no two
of them share a bin. Since every vertex of our source graph
lies on an edge, no vertex can share a bin with an auxiliary item, and the
vertices share the remaining three bins, exactly as in $3$-coloring.

Each item $i$ gives Alice one possible input $(i,u)$ for every row $u$
of $R$. A coloring partitions these inputs by the
bin of their item. Naming the bin need not identify Alice's item or her row of $R$.

The full construction uses $4s$ labeled groups, each containing a copy of
all items, so that the proof can discard damaged groups. For each edge in each group, a binary vector over all items marks the
active items of that group with 1 and every other item with 0. We index these
vectors by $\alpha$. In the NO case, every assignment of items to bins has a
constant fraction of indices at which one bin contains two active items and
a third item that is inactive there, even after deleting a small fraction of
the items (\cref{src:item-witness}). We say there is a \emph{collision}
at such an index. The inactive third item can come from another group. There
are $4s$ groups but only $s$ bins, and at most $s$ groups can have a bin
containing items of no other group. Hence for most groups the bin holding an
active pair also holds an item of another group, which is inactive at that
index because each index activates items of one group only.
The proof extracts an assignment from any short protocol.

Alice's inputs are of two kinds. An \emph{item row} consists
of an item $i$ and a row $u$ of the base matrix $R$. The other inputs,
called \emph{template rows}, are grouped into $s$ bins and are obtained
from $R$ by nested interlaces, described below. Their purpose is to force
every protocol of cost at most $C-1$ to begin with $\log s$ consecutive bits
from Alice (\cref{cl:separation}). They name a bin before Bob reveals any
index $\alpha$.

Bob's columns have two families. The first supplies a separate index
and list of base columns for each bin, together with a list used by the
item rows, and enforces the initial $\log s$ bits. The second uses the same
index $\alpha$ and the same list of base columns in every bin. At this
index, the template rows use $s-2$ positions of the list, active items
use distinct additional positions, and inactive items use one further
common position. Thus two active items and one inactive item in the
same bin expose $s+1$ positions.

\paragraph{Defining the table without a coloring.}
A template row $(p,z)$ consists of its bin label $p$ and a tuple $z$ that
specifies, for each index $\alpha$, a pair $z_\alpha=(r,u)$ of a position
$r\in\{1,\ldots,s-2\}$ and a base row $u$. Bob's first-family column
supplies $(\alpha_p,\gamma_p)$ for each bin, where $\gamma_p$ is a list
of $2s-1$ base columns. The entry is $R(u,\gamma_{p,r})$, with
$(r,u)=z_{\alpha_p}$. Thus each template row uses only its own bin's
data. Letting the data of different bins vary independently gives the
outer interlace, with one block per bin, that forces Alice's initial messages.

If a template row carried just one pair $(r,u)$ instead of the tuple $z$,
its entry after the bin was chosen would not depend on Bob's
index $\alpha_p$. The template subproblem would no longer require that
index to be communicated. With the tuple, Bob's index selects which pair
of Alice's input matters, and then Alice's position $r$ selects which entry
of Bob's list matters. The
players are left with one instance of $R$. These successive selections
are the nested interlaces, with transposition exchanging the players'
roles between levels.

Once Alice's first $\log s$ bits are forced, they split the item rows into
the same $s$ parts as the template rows, and this yields a bin assignment,
as explained in the next subsection.

An item row $(i,u)$ has no bin label, because the reduction cannot use a
coloring to assign one. Bob therefore also holds a list
$\xi=(\xi_1,\ldots,\xi_{s+1})$ of base columns shared by all bins. For each
index $\alpha$, the at most $s-1$ active items receive distinct positions
$\pi_\alpha(i)\in\{1,\ldots,s\}$ in a fixed public order.
If all Bob's indices agree on $\alpha$ and item $i$ is active there,
the entry is $R(u,\xi_{\pi_\alpha(i)})$, and otherwise it is
$R(u,\xi_{s+1})$. These rules use no coloring. In the coloring protocol,
Bob checks privately whether his indices agree, and if Alice's item is
active, her bin and the index identify it.

The second family sets every bin's index to the same $\alpha$ and every
list to the same $\gamma$, with
$\xi=(\gamma_{s-1},\ldots,\gamma_{2s-1})$.
The template and item rows now read positions of one common list.
A single protocol must handle both families, and before Bob speaks,
Alice's messages depend only on her row. Consequently, the second family
detects collisions in the partition forced by the first.

The second family has separate labels and a fixed positive share of
the column weight. In the full-product construction, its columns
already occur in the first family but may have very small total
weight there. The separate weight prevents a small deletion from
removing all collision tests. \Cref{def:source-matrix} gives the full
entry rules and weights.

\paragraph{A coloring gives a protocol.}
In the YES case, fix the bin assignment given by a proper $3$-coloring.
Alice names her bin $p$, using $\log s$ bits, and Bob sends that bin's index $\alpha_p$,
using $\log q_2$ bits, where the number $q_2$ of indices is a power of two.
Alice then sends one of
\[
\underbrace{s-2}_{\text{template positions}}
\quad+\quad
\underbrace{1}_{\text{active item}}
\quad+\quad
\underbrace{1}_{\text{inactive item}}
=s
\]
possible messages. If Alice's item is active, her bin and the index
determine it, since a proper coloring puts at most one active item in each
bin. Bob can therefore identify the selected base column from his input,
and the players finish by running a protocol for $R$.

The two Alice messages each cost $\log s$ bits, giving total cost
$C-1:=d+2\log s+\log q_2$. For an arbitrary graph, Alice can instead name
her active item's position explicitly. This uses at most $2s-1$ message
values and costs one additional bit. Thus the table always has a
cost-$C$ protocol, and a coloring saves one bit
(\cref{src:upper}).

\subsection{Why the reduction rules out every short protocol}

We prove that the NO-case lower bound survives deletion of a small fraction
of the row and column weight, so that it can be reused in later rounds.
Suppose a protocol of cost $C-1$, with any pattern of interaction, computes
the remaining matrix. From such a protocol we
extract a single bin assignment. Its first $\log s$ bits are Alice's,
sent before Bob can reveal any index. They split Alice's rows into $s$
parts, and we assign each item to a part that receives at least a
$1/(2s)$ fraction of the weight of its rows. For a NO graph, this fixed assignment must contain
collisions.

\paragraph{A short protocol fixes a bin assignment.}
Group the template rows by their bin $p$. In the first column family,
Bob holds a separate index and list of base columns for every bin,
and in the full matrix every combination occurs. The entry of a template row in bin $p$ depends only
on the data for that bin. On template rows the matrix is therefore an
$s$-fold interlace whose blocks are the bins.

Within one bin, the template rows already require
$d+\log q_2+\log s$ bits, by two applications of the interlacing lower
bound (\cref{src:profiles}). Of the $C-1$ available bits, only $\log s$
remain to tell the bins apart.
Suppose Bob sends a bit while $\ell$ bins are still possible. His heavier
child keeps every Alice row and at least half the column weight, so the interlacing
bound for $\ell$ bins still demands $\log \ell$ extra bits,
one more than remain. Similarly, if one of Alice's bits kept half the
weight of more than $\ell/2$ bins on one side, that side would need more bits
than remain. The
separation theorem (\cref{cl:separation}) therefore forces the first
$\log s$ bits to be Alice's and to split the bins evenly. Grouping her rows
by these $\log s$ bits gives $s$ parts, one for each bin. Each part contains almost all the template-row weight of one bin
and very little of the weight of the others.

All items of one type have rows of equal total weight, so a small deletion
makes few items lose more than half of the weight of their rows $(i,u)$. We discard these and assign each remaining item to a part containing at least a
$1/(2s)$ fraction of the original weight of its rows. Such a part exists by
averaging over the $s$ parts. The rows of one item need not stay in one part,
since different base rows $u$ may follow different paths. However, the
assignment is fixed before Bob has spoken, so it cannot depend on
the edge tested by his index. For a NO graph, \cref{src:item-witness} guarantees many indices
at which one assigned bin contains two active items and an inactive item.

\paragraph{A collision gives an extra position.}
Restrict Bob to the second column family at a collision index $\alpha$.
All bins now use the same index and the same list of base columns.
Within the collision's bin, the template rows and the three items read
\[
\underbrace{s-2}_{\text{template positions}}
\quad+\quad
\underbrace{2}_{\text{distinct active positions}}
\quad+\quad
\underbrace{1}_{\text{inactive position}}
=s+1
\]
distinct positions. The inactive item is essential: two active items
alone would give only $(s-2)+2=s$ positions. As $s$ is a power of two,
naming one of $s+1$ positions costs one more bit than one of $s$.

\paragraph{Why collisions force more communication.}
Counting positions alone does not prove a lower bound against an
interactive protocol. Bob's replies restrict which lists of base columns remain possible, and a lower
bound for the original column set need not survive this restriction.
In the second family, every bin has the same
index $\alpha$. After Alice's $\log s$ bits, Bob may send bits
that narrow down the possible values of $\alpha$ before Alice speaks again,
so that each branch of the protocol covers only some indices. If
collisions occurred at only a few indices, Bob could isolate them on a
branch where naming $\alpha$ costs far fewer than $\log q_2$ bits, leaving
room for the extra position. We therefore need collisions at a constant
fraction of all indices.

For each index, we track Bob's retained second-family inputs with that
index through his bits, and stop when Alice next speaks. Each index at which the deletion leaves most of these inputs gives one
tracked set, and a collision marks its set,
recording the extra bit. An unmarked set follows a child retaining at
least half its weight. A marked set splits into two unmarked sets if both
children keep enough weight, and otherwise keeps its mark. Counting marked
sets twice, a split replaces two by one plus one, preserving the count.

A stopping node reached after $r$ Bob bits can accommodate a count of at
most $q_2 2^{-r}$ within its remaining communication budget. The stopping
nodes are prefix-free, so Kraft's inequality bounds their total capacity
by $q_2$. Initially, however, the count exceeds $q_2$: collisions occur
at a constant fraction of the indices (\cref{src:item-witness}), while
each index carries the same weight, so the small deletion damages far fewer. This contradiction
rules out cost $C-1$ in the NO case
(\cref{lem:collision-tracking,src:near-full}). Together with the coloring
protocol, it yields the one-bit gap against every protocol, whatever
its pattern of interaction.

\subsection{Making the one-bit gap repeatable}

A one-bit gap does not automatically yield a larger gap. Using the output
as the next base matrix requires lower bounds on the restrictions that
remain after messages, and a hard matrix can have easy restrictions.
For example, restricting Bob's tuples in an interlace of $R$ to those
with all coordinates equal removes the need to learn Alice's index.

We therefore maintain a stronger invariant in the NO case. Measuring
inputs by the fixed positive rational weights, we require, for a
threshold $\rho$ fixed across rounds, that every restriction retaining
the weights below needs the stated communication:
\[
\begin{array}{c|c|c}
\text{retained row weight}&\text{retained column weight}
   &\text{communication lower bound}\\ \hline
\ge\rho&\ge3/5&d\\
\ge\rho&\ge3/10&d-1 .
\end{array}
\]
The threshold $\rho$ may depend on the source size. We call these the
\emph{deletion bounds} (\cref{def:profile}). The second column threshold
is half the first: after Bob sends one bit, the heavier branch retains
at least half the column weight, and the allowed lower bound drops by
one. The YES upper bounds do not require this invariant.

The source construction gives its extra bit only on restrictions
retaining almost all the weight on both sides (\cref{src:near-full}).
The deletion bounds needed for another round allow most of the row
weight and a constant fraction of the column weight to disappear. Two further
interlaces close this difference, each handling one side.

\paragraph{Preserving the extra bit after input deletion.}
Transpose the source matrix to $N$, so its two column families become
row families, and form the $w$-fold interlace $U=\Int{w}{N}$. Alice
holds a copy index and a row of that copy, and Bob holds one column for
each copy. Consider a restriction retaining almost all row weight but
only a small fixed fraction $\theta$ of the tuple weight. Suppose this
restriction has a protocol of cost $C+\log w-1$, one bit less than
naming a copy and running the cost-$C$ protocol for $N$.

Stop each path just before Bob's first bit or just after Alice's
$(\log w)$-th bit, whichever comes first, padding paths that end
earlier with fixed Alice bits.
The first-family rows force almost all their weight in most copies to
reach distinct stopping nodes at depth $\log w$, one assigned to each
such copy (\cref{res:destinations}). At these nodes $C-1$ bits remain.
Bob has not spoken, so his tuple set is still the initial restriction.
By the conditioning lemma (\cref{det:conditioning}), almost every
copy still retains almost all its column weight. If its assigned node
also kept almost all its rows, the source lower bound would require
$C$ further bits. So the node must miss a substantial part of its
copy's rows.

Almost all first-family weight is already at the assigned nodes, and
little weight was deleted initially. Summing over the assigned copies
therefore places substantial surviving second-family weight outside
the node assigned to its copy. Since only Alice has spoken, every
surviving row reaches exactly one stopping node. Each such displaced
row reaches either another copy's assigned node or a node with no
assigned copy.

Neither group can receive enough weight
(\cref{fig:reference-destinations}). An assigned node already contains
the hard first-family rows of its own copy, so much second-family
weight from another copy would need more than the remaining $C-1$ bits
(\cref{res:foreign}). The unassigned nodes have small total capacity
(\cref{res:destinations,res:incoming}). The two bounds sum to less than the displaced
weight, a contradiction (\cref{res:alignment}). Transposing $U$ exchanges
the two retained-weight conditions, yielding $W$ with its full lower
bound when a small fraction of rows and almost all columns remain.

\paragraph{Recovering the bit in the deletion bounds.}
For constant retained column weight, $W$ satisfies the deletion bounds
only one bit below its full cost. A final
interlace combines this guarantee with the full-cost bound just proved.
For full products, with product tuple weights, a tuple set of retained weight $v$ cannot have small
projections at too many coordinates. If its projection onto coordinate
$i$ has original weight $p_i$, containment in the product of its
projections gives
\[
v\le\prod_i p_i,
\qquad\text{hence}\qquad
\sum_i\log(1/p_i)\le\log(1/v).
\]
Among sufficiently many copies retaining enough row weight, at least
one therefore retains almost all its column weight, so the full-cost
bound for $W$ applies.

Finding one such copy would recover only the cost of $W$, and we must
also account for the bits needed to identify a copy. We do so by
induction over messages. At an Alice bit, one branch keeps enough
copies with substantial row weight. At a Bob bit, keeping the heavier
tuple set increases $\log(1/v)$ by at most one. Choosing enough copies
at the outset leaves room for these increases. The deletion bounds control
the intermediate restrictions, and the full-cost bound closes the
remaining one-bit deficit (\cref{cl:bridge}). For the explicit lists,
the conditioning argument described next replaces the full-product
inequality (\cref{det:restore}).

Consequently, the three steps above form one round $R\mapsto R^+$, which
has a public overhead $h$ and satisfies
the following guarantee. In the YES case,
\[
\D(R^+)\le\D(R)+h.
\]
In the NO case, if $\D(R)=d\ge3$ and $R$ satisfies the deletion bounds
at value $d$, then
\[
\D(R^+)=d+h+1,
\]
and $R^+$ satisfies the same deletion bounds at value $d+h+1$
(\cref{thm:amplifier}). This restoration is what makes iteration
possible. Starting from the initial matrix of known cost $d_0\ge3$
(\cref{cl:seed}), which satisfies the deletion bounds, the matrix $R_k$
after $k$ rounds has
\[
\text{YES: }\D(R_k)\le d_0+kh,
\qquad
\text{NO: }\D(R_k)=d_0+kh+k.
\]
The threshold $d_0+kh$ is computable without deciding whether the graph
is colorable.

\subsection{Keeping the deterministic construction polynomial}

Allowing every tuple of $q$ columns from a set $Y$ creates $|Y|^q$ inputs.
We replace each product by an explicit list of polynomially many tuples
with positive rational weights. For an absolute constant $D$, the list
has exactly the same joint weights as the full product on every set of
at most $D$ coordinates. This is \emph{limited independence}. We build
and enumerate the entire list deterministically, and the weights do not
represent random choices made by the reduction.

We extend tuples one coordinate at a time and remove
tuples by adjusting their weights. Associate with each tuple a zero--one
vector indicating its assignment on every set of at most
$D$ coordinates, plus a final entry one.
The required joint weights and the total weight specify the weighted sum of these vectors.
If the list is longer than the vector dimension, a linear dependence
lets us change the weights without changing that sum. Since total
weight is fixed, the change has both positive and negative components,
so we can move along it until one tuple has weight zero and remove it,
keeping all other weights nonnegative. Compressing after each new coordinate
keeps the intermediate lists short, avoiding enumeration of an
exponentially large product. With $q$ coordinates over $m$ labels, the list has at most
$(1+qm)^D$ tuples with polynomial-bit weights (\cref{det:code}). The original entry rules still apply to every listed
tuple, so the upper-bound protocols are preserved.

The lower bound is subtler, because conditioning on earlier messages can
destroy independence. For example, retaining only tuples whose first two
coordinates agree makes those coordinates dependent. Our conditioning
lemma shows that, after conditioning on any event of weight at least $v$,
only $\mathrm{poly}(1/v)$ coordinates are exceptional. On every
set of at most eight other coordinates, the conditional distribution
is close in total variation to the original product distribution
(\cref{det:conditioning}). By a moment bound under the original weights,
too many disjoint coordinate sets with substantially changed distributions
would contradict the weight of the conditioning event.

We choose the interlacing widths and deletion thresholds so that enough
coordinates remain outside the exceptional set. Projecting onto these
coordinates loses at most the total-variation error in column weight,
so the lower bounds of the base matrix still apply
(\cref{det:rareboost,det:restore}).

The widths are fixed polynomials in $n$, so each round takes time polynomial in the
source graph size $n$ and the encoding length of the current matrix,
with an exponent independent of the round. After $k$ rounds, size and
running time are at most $n^{O(A^k)}$ for an absolute constant $A>1$.
Every fixed gap therefore gives a deterministic polynomial-time
reduction (\cref{det:reduction,intro:deterministic}).

For sparse formulas, $n=N^{O(1)}$, so a suitable $k=\Theta(\log\log N)$ makes
the construction quasipolynomial in $N$, while $\log\log T=O(k)$.
A deterministic polynomial-time estimator with additive error $c_*\log\log T$, for a
small enough constant $c_*$, would therefore decide sparse $3$-SAT in quasipolynomial time
(\cref{det:time}).

\subsection{Two further consequences of keeping fewer inputs}

\paragraph{Sampling at the end tolerates larger errors.}
For the randomized result, we keep full products of tuples throughout the
amplification. The table is then too large to print, but we can compute
single entries from short descriptions of the inputs. After all $k$
rounds, we use the input weights as sampling probabilities: we draw $m$ rows
and $m$ columns independently and print the $m^2$ entries at their intersections.
Randomness is used only by the reduction; the protocols whose cost is
measured remain deterministic.

In the NO case, a protocol may be chosen after the sample is drawn, so the
sample must defeat every protocol of cost below the NO cost
simultaneously. Fix a protocol tree with Alice's messages and the leaf
outputs, but not Bob's bits. A column is \emph{compatible} with it and a
row set if one choice of Bob's bits, depending only on the column, makes
the tree correct on all those rows.
Over all such trees of
depth $t$ and all row sets, a matrix of real rank $r$ has at most
$2^{O(r\log(r+2)\,2^t)}$ distinct sets of compatible columns
(\cref{thm:rank-count}). The count is independent of the number of input labels.
For a tree of depth below the NO cost and a row set of weight $\ge3/5$,
the deletion bounds (with $\rho\le3/5$) give the compatible columns weight $<3/5$. A union
bound over these sets then shows that, with high probability,
$m=\Theta(r\log(r+2)\,2^t)$ random columns are not all contained in any
one of them. Once such a column sample is fixed, every row set compatible
with a short protocol on the sampled columns has weight $<3/5$, so the same
union bound applies to the sampled rows (\cref{thm:compression}). The
sample need only preserve the gap, since no further amplification follows. An interlace stacks one block per copy, so it multiplies the rank
by at most its width. One round thus multiplies it by at most $2^h$, and
$\rank(R_k)\le2^{d_0+kh}$ (\cref{prop:iteration}). For a sparse formula on $N$ variables the graph has
$N^{1+o(1)}$ vertices and $h=(1+o(1))\log N$,
so taking $k\approx N^\alpha/\log N$ gives
$\log T=2\log m=(4+o(1))k\log N$ and
$k=(\alpha/4+o(1))\log T/\log\log T$.
Sampling and evaluation take time $n^{O(k+1)}=2^{O(N^\alpha)}$
(\cref{lem:sample-access}).
An estimate within additive error $c\log T/\log\log T$ distinguishes the
two cases whenever $\alpha>8c$, giving \cref{thm:quantitative}.

\paragraph{Strict alternation needs a different encoding.}
The coloring protocol above uses consecutive bits from the same player.
Inserting dummy turns can cost more than the gap, so it does not give
hardness under strict alternation.

We instead start from the hypergraph-coloring reduction of Guruswami and
Lee~\cite{guruswami2018balanced}, whose balanced YES colorings give a
short alternating protocol. For a power of two $s$, each edge has
$s^2$ vertices. In the YES case there is an $s$-coloring in which every
edge contains at least $s-1$ vertices of every color. In the NO case,
every coloring with at most $2^s$ colors has a monochromatic edge
(\cref{yao:source}).

Alice holds a vertex and a tuple of $s/2$ addresses. Bob holds an edge,
a coordinate selecting one address, and a Boolean array indexed by an
address and an edge position. The selected
address and the vertex's position in Bob's edge identify the array bit
to output. The output is zero if the vertex is outside the edge
(\cref{def:alternating-table}).
Write $s=2^t$, and let $b$ bits name an edge. In the YES case,
the players exchange the color, address and edge information in
alternating turns. The coloring leaves at most $2s-1$ candidate
vertices of Alice's color in Bob's edge, giving total cost
$2b+4t+2$ (\cref{prop:alternating-upper}).

For the NO lower bound, fix a short protocol and, for each row, record
Alice's responses to \emph{every}
possible sequence of Bob's earlier bits during her first $t$ turns.
There are at most $2^{2^t-1}\le2^s$ such response rules. Assign each
vertex its most frequent rule, used by at least a $2^{-s}$ fraction of its
address tuples. The NO promise gives
a monochromatic edge, so many rows sharing one early response rule
must still identify many different array positions. Counting only the
transcript of one interaction would not suffice, because Bob's
messages may depend on his entire array.

We construct a polynomial-size explicit list of arrays in which each
pattern on any $j$ positions occurs in at most a $2^{-j}+\varepsilon$
fraction of the arrays. Following a
hypothetical short protocol while retaining large sets of rows and
arrays leaves many arrays. By the list property, the few output patterns
of the last turns on the retained rows match fewer arrays than survive
(\cref{yao:array-list,yao:small-bias-escape,yao:lift}).
This gives a NO lower bound of $2b+6t-6$, and hence a gap of $2t-8$.
Allowing either starter and receiver-only output loses at most two
bits (\cref{yao:comparison}). Choosing $t$ as a sufficiently large
constant therefore gives a deterministic polynomial-time reduction
for every fixed gap in Yao's convention.

\section{Related work}

\paragraph{Which optimization problem is hard?}
For explicitly given total tables, Kushilevitz and Weinreb proved that
approximating deterministic communication complexity is hard under
cryptographic assumptions~\cite{kushilevitz2009complexity}.
Ilango, Loff and Oliveira proved NP-hardness and strong
inapproximability for \emph{partial} matrices, where the protocol is
required to be correct only on specified input
pairs~\cite{ilango2020multioutput}. Our tables are \emph{total}: every
pair of a row and a column has a specified bit, and the protocol must
be correct on all of them.

For total matrices,
Hirahara, Ilango and Loff proved additive hardness for the number of
bits in a three-message protocol, and hardness of approximating the
number of leaves of a protocol tree for every fixed number $k\ge3$ of
messages~\cite{hirahara2021constant}. A message is a block of
consecutive bits from one player, and a leaf is a possible end of the
conversation. Neither restricting the number of messages nor counting
leaves gives the same optimization problem as minimizing bits under
strict alternation or unrestricted speaking order. The exact-hardness
results for the unrestricted
model~\cite{gaspers-he-mackenzie-npcc,hirahara2025communication} left open
both hardness with one bit of additive error and Yao's original
alternating model, as Hirahara, Ilango and Loff
noted~\cite{hirahara2025communication}. Our fixed-gap theorems resolve both.

\paragraph{Ingredients of our reductions.}
For the unrestricted reductions, we obtain hard graph-coloring instances
by combining reductions of Dinur and of Bogdanov, Obata and
Trevisan~\cite{dinur2007pcp,bogdanov2002lower}. In these graphs, each
vertex meets only a constant number of edges. The alternating reduction
uses the hypergraph-coloring promise of Guruswami and
Lee~\cite{guruswami2018balanced} described in \cref{sec:overview}.

Both reductions also replace large input sets by short explicit lists
that keep the statistics their lower-bound proofs use. For the deterministic
interlacing construction, we follow Koller and
Megiddo~\cite[Section~3]{koller1994sample}: we extend a list of tuples one
coordinate at a time and delete tuples without changing the joint
weights of the values on every set of a bounded number of coordinates. For strict alternation, we need a short list of Boolean arrays on which
no bit pattern at a few positions is much more frequent than for random
arrays, and we take it from the small-bias construction of
\citet*{alon1992smallbias}.

Bounds on the number of column sets on which short protocols can be
correct, expressed in terms of matrix rank, appear in graph algorithms~\cite{bui2010rankwidth,oum2014faster}
and in work on hard submatrices~\cite{hrubes2025hard}. Our count (\cref{thm:rank-count}) uses the
row-sum method of Oum, S{\ae}ther and
Vatshelle~\cite{oum2014faster}: the integer sum of some Boolean rows is zero
exactly in the columns where all those rows are zero. Mackenzie's hardness
condensation~\cite{mackenzie2026condensation} also counts the input sets
on which short protocols succeed, to shrink a communication problem while
keeping its cost high relative to the new input length. Our sampling
theorem (\cref{thm:compression}) has a different goal: a sampled
subtable whose cost stays above a given bound.

\paragraph{Reading guide.}
After the preliminaries (\cref{sec:models}), the formal sections follow the order of
\cref{sec:overview}, and the appendices contain the remaining estimates
and parameter checks.

\clearpage
\section{Preliminaries}
\label{sec:models}

\subsection{Communication and estimation}

An estimator receives the whole table, whereas each player knows only its own input.

\begin{definition}\label{def:table-estimation}
A table $M:X\times Y\to\{0,1\}$ is total, and its input sets $X$ and $Y$ are
finite and nonempty sets of labels. Equal rows or columns may have different labels.
Its size is $T=|X||Y|$.
A deterministic public-output protocol is a binary tree whose internal
nodes specify a speaker and a bit-valued function of that player's input
and the preceding transcript. Each leaf has a fixed output, and the
protocol must compute every entry correctly. Its cost is its maximum
depth, charging every bit, including fixed bits. Local computation is free.
Write $\D(M)$ for the least cost with unrestricted speaking order.

An additive-$a(T)$ estimator returns $\widehat D(M)$ satisfying
$|\widehat D(M)-\D(M)|\le a(T)$. Its output must be comparable in polynomial time with any binary rational of bit length polynomial in $T$. A bounded-error
randomized estimator satisfies this guarantee with probability at least
$2/3$ on every input. Polynomial time means worst-case time polynomial
in $T$. Estimators for the alternating costs below follow the same conventions.
\end{definition}

In the alternating model, the starting player and the output convention
can each affect the cost.

\begin{definition}[Alternating costs]\label{yao:model}
In strict alternation the players speak one bit at a time, in turn.
Let $D_A^{\mathrm{alt}}$ and $D_B^{\mathrm{alt}}$ be the minimum
public-output costs when Alice or Bob starts, respectively, and let
$D_{\mathrm{free}}^{\mathrm{alt}}=\min\{D_A^{\mathrm{alt}},D_B^{\mathrm{alt}}\}$.
The starter is chosen independently of the inputs. Every bit is charged,
and stopping is public: the transcript determines whether communication
has ended.

Yao's receiver-output cost $Y_{\mathrm{free}}$ permits the same choice of
starter, but the receiver of the last bit determines the answer from
that player's input and the transcript. A protocol of depth zero instead designates one
player, who determines the answer from that player's own input alone. Stopping is
still public and correctness is required on every input.
\end{definition}

Our quantitative statements use sparse $3$-SAT: the input has $N$ variables
and at most $c_sN$ clauses, where $c_s$ is fixed in advance. Running times
are worst-case bounds, including all random tapes for randomized algorithms;
bounded error means success probability at least $2/3$ on every formula.
Constants may depend on $c_s$ and on the estimator. We do not apply a
sparsification step with an uncharged exponential cost, and all stated running
times refer directly to these sparse inputs.

All communication protocols in this paper are deterministic and exact.
Only estimators and randomized reductions use random bits. All logarithms
are binary unless written $\ln$. For an integer $q\ge1$ write $[q]=\{1,\ldots,q\}$, and for $a\ge1$ write
$\operatorname{pow}_2(a)=2^{\ceil{\log a}}$ for the least power of two
at least $a$.

Every $n\times n$ table has $\D(M)\le\ceil{\log n}+1$: Alice sends her
row index and Bob sends the entry.

\subsection{Relabeling and input weights}

We keep the matrices small in two ways, by explicit short lists and by
sampling. In both, each new row or column label decodes to an original one.

Let $F:X\times Y\to\{0,1\}$, $f:X'\to X$ and $g:Y'\to Y$, and define $F'$ on $X'\times Y'$ by
$F'(x',y')=F(f(x'),g(y'))$. Alice computes $f(x')$, Bob computes $g(y')$,
and they run a protocol for $F$. Hence $F'$ has at most the
communication cost of $F$ in every model above, and at most its rank. Costs and ranks
are equal when $f$ and $g$ are onto, since then $F$ is obtained from $F'$ in the same way. Transposition preserves rank and unrestricted
cost, and exchanges $D_A^{\mathrm{alt}}$ and $D_B^{\mathrm{alt}}$.

In particular, duplicating rows and columns makes a table square with
power-of-two side length without changing its cost. We may also
require the side length to be at least a prescribed bound $b$ polynomial in
the input size. The side length is then less than $2\max\{b,|X|,|Y|\}$, so
the enlarged table can be printed in polynomial time.

For the unrestricted construction, we assign the labels in each input set
positive rational weights summing to one. For row weights $\mu_X$ and column weights $\mu_Y$, a set's
weight is the sum of its labels' weights, for example
$\mu_X(A)=\sum_{x\in A}\mu_X(x)$. Uniform weights give ordinary fractions
of labels. Unequal weights let us prescribe the total weight of each input family
and shorten tuple lists while preserving their marginals (\cref{sec:small-lists}). The weights
are auxiliary data for the construction and its lower-bound proof.
Communication cost depends only on the table. The final deterministic reduction discards the weights.
Distinct labels keep separate weights, even when their rows or columns agree.

If $X'$ has weights $\mu'$, a decoder $f:X'\to X$ induces weights on $X$ by adding the weights of
labels with the same image, and labels of $X$ outside the image get weight zero. A set $A\subseteq X$ therefore has induced
weight $\mu'(f^{-1}(A))$. In particular, a retained set $A'\subseteq X'$
has image of induced weight at least $\mu'(A')$, since $A'\subseteq f^{-1}(f(A'))$.

Under product weights on $Y^q$, a tuple $(y_1,\ldots,y_q)$ has weight
$\prod_{j=1}^q\mu_Y(y_j)$.
Some counting arguments are shorter in probability notation. There we regard a label
as drawn according to its weights and use the standard terms distribution,
marginal, and independence. Such a draw only serves the calculation and is not a random
choice by a communication protocol. An event is a set of labels.
Conditioning on a retained set $A$ of
positive weight means renormalizing: $(\mu\mid A)(B)=\mu(A\cap B)/\mu(A)$.
When we project retained tuples onto some coordinates, we specify whether we
measure the projection with the original or the conditional weights.

\paragraph{Counting surviving blocks.}
Later proofs need many blocks of a partition each to retain a given fraction of
their own weight, which we derive from the total retained weight as follows.
Suppose we partition an input set into $q$ blocks, each of weight $1/q$, such as the
blocks $\{i\}\times X$ of rows of $\Int{q}{R}$ in
\cref{def:interlace} below. If a retained subset
has total weight $\beta$, the fractions of their own weight that the blocks retain sum
to $q\beta$. We set aside at most $e$ blocks and suppose that at most $r$ of the other blocks
retain fraction $\ge t$. Bounding these $r+e$ fractions by one and
every remaining fraction by $t$ gives
\[
q\beta\le r+e+qt.
\]
Thus, if $q\beta>r+e+qt$, more than $r$ blocks not set aside retain
fraction $\ge t$. If the $q$ blocks instead partition only a part of the
input set, of total weight $\lambda$, that part retains fraction at least
$(\beta-(1-\lambda))/\lambda$ of its weight, since the rest of the input set
contributes at most $1-\lambda$ to $\beta$.

\subsection{Interlacing and restrictions}
\label{sec:interlacing}

In an interlace, a row is a row of the base matrix together with an
index $i$, a column is a tuple of base columns, and the entry is read from
the $i$th coordinate of the tuple.

\begin{definition}[Interlacing]\label{def:interlace}
For a Boolean matrix $R:X\times Y\to\{0,1\}$,
the full $q$-fold interlace $\Int{q}{R}$ has rows $(i,u)\in[q]\times X$,
columns $(y_1,\ldots,y_q)\in Y^q$, and entry
\[
\Int{q}{R}((i,u),(y_1,\ldots,y_q))=R(u,y_i).
\]
When $R$ has normalized input weights $\mu_X,\mu_Y$, give row $(i,u)$
weight $\mu_X(u)/q$ and column $(y_1,\ldots,y_q)$ weight
$\prod_{j=1}^q\mu_Y(y_j)$.
\end{definition}

\paragraph{Why column deletion matters.}
For a fixed base row $u$, the rows $(i,u)$ with different indices $i$ read
different coordinates of Bob's tuple, but these rows may coincide after
columns are deleted. For example, consider the $2\times2$ base matrix
\[
R=\begin{array}{c|cc}
     & 1 & 2\\ \hline
  a  & 0 & 1\\
  b  & 1 & 0
\end{array}.
\]
The left-hand table in \cref{fig:lookup} is its two-fold interlace.
We obtain the right-hand table by keeping only columns $(1,1)$ and $(2,2)$.
For each fixed $u\in\{a,b\}$, the rows $(1,u)$ and $(2,u)$ then coincide.

\begin{figure}[H]
\centering
\begin{tikzpicture}[>=Stealth,font=\small,
 every node/.style={align=center}]
\matrix (before) at (0,0) [matrix of math nodes,
 nodes={minimum width=9mm,minimum height=6mm,inner sep=1pt},
 column 1/.style={nodes={minimum width=7mm}},
 column 2/.style={nodes={minimum width=6mm}},
 row 1/.style={nodes={font=\footnotesize}}] {
 u&i&(1,1)&(1,2)&(2,1)&(2,2)\\
 a&1&0&0&1&1\\
 a&2&0&1&0&1\\
 b&1&1&1&0&0\\
 b&2&1&0&1&0\\
};
\matrix (after) at (8,0) [matrix of math nodes,
 nodes={minimum width=10mm,minimum height=6mm,inner sep=1pt},
 column 1/.style={nodes={minimum width=7mm}},
 column 2/.style={nodes={minimum width=6mm}},
 row 1/.style={nodes={font=\footnotesize}}] {
 u&i&(1,1)&(2,2)\\
 a&1&0&1\\
 a&2&0&1\\
 b&1&1&0\\
 b&2&1&0\\
};
\node[above=3mm of before,font=\small\bfseries] {Rows may depend on $i$};
\node[above=3mm of after,font=\small\bfseries] {Rows no longer depend on $i$};
\draw[->,semithick] ([xshift=4mm]before.east)
 -- node[above=3mm,font=\footnotesize] {keep $y_1=y_2$}
    ([xshift=-4mm]after.west);
\draw (before-1-1.south west) -- (before-1-6.south east);
\draw (after-1-1.south west) -- (after-1-4.south east);
\draw[black!35] (before-3-1.south west) -- (before-3-6.south east);
\draw[black!35] (after-3-1.south west) -- (after-3-4.south east);
\begin{scope}[on background layer]
\fill[black!10] (before-1-4.north west) rectangle (before-5-5.south east);
\fill[answergreen!12,rounded corners=2pt]
 ([xshift=.8mm,yshift=-.8mm]after-2-3.north west) rectangle
 ([xshift=-.8mm,yshift=.8mm]after-3-4.south east);
\fill[answergreen!12,rounded corners=2pt]
 ([xshift=.8mm,yshift=-.8mm]after-4-3.north west) rectangle
 ([xshift=-.8mm,yshift=.8mm]after-5-4.south east);
\end{scope}
\node[right=3mm of after,align=left,font=\footnotesize,text=answergreen!70!black] {identical rows\\for fixed $u$};
\node[below=3mm of before,font=\footnotesize] {Shaded columns are deleted.};
\end{tikzpicture}
\caption{Deleting columns with $y_1\ne y_2$ removes the dependence on
the row index $i$. For each fixed $u$, the rows $(1,u)$ and $(2,u)$ agree
on every retained column.}
\label{fig:lookup}
\end{figure}

This collapse occurs for every base matrix $R$. On the columns
$(y,\ldots,y)$, the entry of $\Int{q}{R}$ is $R(u,y)$ for every row index
$i$. Hence a lower bound for the full interlace need not survive arbitrary
column deletion. For this reason, \cref{def:profile} requires lower bounds only for
restrictions that retain a prescribed weight of rows and of columns.

\paragraph{Keeping fewer column tuples.}
Full interlacing creates $|Y|^q$ columns, too many for an explicit
polynomial-time reduction. We therefore
replace $Y^q$ by a short list with rational weights. For every assignment of
values to at most $D$ coordinates, we preserve the total weight of tuples
matching that assignment. The definition below calls these exact counting
identities $D$-wise independence. \Cref{det:code} constructs such lists
deterministically (\cref{sec:small-lists,sec:deterministic}).

\begin{definition}[Weighted lists of tuples]\label{def:exact-tuples}
Fix integers $q,D\ge1$, the tuple length and the independence order. For a finite alphabet $Y$ with prescribed
normalized rational weights $\mu$, an indexed tuple list consists of a
finite nonempty label set $\mathcal L$, positive normalized rational
weights $\nu$ on its labels, and a decoder $z:\mathcal L\to Y^q$.
For $J\subseteq[q]$, write $z(\ell)_J$ for the restriction of $z(\ell)$ to $J$.
The list is \emph{exactly $D$-wise independent} if, for every $J$ with $|J|\le D$,
the map $z_J:\ell\mapsto z(\ell)_J$ induces the product weights $\mu^J$. These induced
weights are the marginal on $J$.
No independence on larger sets is required. A set of labels may contain one label
but not another with the same decoded tuple.

For a base matrix $R:X\times Y\to\{0,1\}$ with column weights $\mu$,
the interlace on this list has a column for each label $\ell$, with entry
$R(u,z_i(\ell))$ at row $(i,u)$. Rows keep the weights of \cref{def:interlace}, and columns
get the weights $\nu$. For both kinds of interlace, \emph{block} $i$ is the row set
$\{i\}\times X$, and we measure row weight within a block by the base row
weights, without division by $q$.
\end{definition}

The full product is the list with labels $Y^q$, identity decoder and
product weights. For a label set $K\subseteq\mathcal L$ and a set $J\subseteq[q]$, write
$\pi_JK=\{z(\ell)_J:\ell\in K\}\subseteq Y^J$ for the projection of its decoded tuples.

To apply a lower bound to a set $J$ of surviving blocks, we project the retained
column tuples onto the coordinates in $J$. By the next observation, the resulting
matrix is a submatrix of the restriction.

\begin{observation}[Restrictions projected to a set of blocks]\label{obs:representatives}
Let $G$ be a restriction of an interlace of $R$, on the full product or on
a tuple list, with retained label set $K$, and let $J$ be a set of blocks.
For each $i\in J$, the rows in block $i$ read only coordinate $i$ of the
decoded tuple. For each $w\in\pi_JK$, choose
a label $\ell_w\in K$ with $z(\ell_w)_J=w$. Identifying $\ell_w$ with $w$, the restriction of $G$ to these labels and to the
retained rows of the blocks in $J$ is then the full-product interlace on $J$,
restricted to the columns $\pi_JK$ and to the same rows. This matrix is a submatrix of $G$,
so every communication lower bound for it holds for $G$. In that interlace,
its column set has weight $\mu^J(\pi_JK)$, and each
block keeps its base row weights. If the marginal on $J$ is $\mu^J$, as for
full products and for exactly $|J|$-wise independent lists, then
$\mu^J(\pi_JK)\ge\nu(K)$, since $K\subseteq z_J^{-1}(\pi_JK)$. We never use
the weights of the chosen representatives, and we assert no independence of
the retained labels.
\end{observation}

\section{A reusable gap construction}
\label{sec:unrestricted}

In this section we state the one-round amplifier \(R\mapsto R^+\) and then
iterate it. In the NO case the amplifier raises the cost by one bit more than
in the YES case, and its output again satisfies the deletion bounds
(\cref{def:profile}) that the next round needs.

\subsection{Lower bounds that survive restrictions}

The deletion bounds use two column-weight thresholds, \(3/5\) and \(3/10\).
The second is half the first because, after Bob sends one bit, the heavier
branch retains at least half the column weight. At the lower threshold we
allow the communication lower bound to drop by one bit.

\begin{definition}[Deletion bounds]
\label{def:profile}
Let \(F:X\times Y\to\{0,1\}\) have positive normalized rational input weights. For
\(0<\rho\le1\) and an integer \(d\), \(F\) satisfies the
\emph{deletion bounds} \(\RP{F}{\rho}{d}\) if, for every \(j\in\{0,1\}\) and all \(U\subseteq X\), \(V\subseteq
Y\),
\[
\mu_X(U)\ge\rho\ \text{and}\ \mu_Y(V)\ge(3/5)2^{-j} \quad\Longrightarrow\quad
\D(F[U,V])\ge d-j.
\]
Here \(F[U,V]\) is the restriction of \(F\) to \(U\times V\).
No matching upper bound is required.
\end{definition}

Whether the source graph is $3$-colorable determines whether the constructed
matrix has the smaller (YES) or the larger (NO) cost. For an unsatisfiable formula, we need the graph
to stay non-$3$-colorable after a small fraction of its vertices is deleted,
since otherwise deleting inputs could erase the lower bound. The reductions
of Dinur and of Bogdanov, Obata and Trevisan provide such
graphs~\cite{dinur2007pcp,bogdanov2002lower}.

\begin{theorem}[Bounded-degree coloring source]
\label{src:gap-colouring}
There are constants \(\Delta\ge1\) and rational \(\gamma>0\), and a
deterministic polynomial-time reduction from \(3\)-SAT to nonempty
loopless multigraphs \(G=(V,E)\) with no isolated vertices and maximum degree at
most \(\Delta\), such that:
\begin{enumerate}
\item a satisfiable formula produces a $3$-colorable graph;
\item an unsatisfiable formula produces a graph for which every $3$-coloring has at least \(\gamma|E|\) monochromatic edge occurrences.
\end{enumerate}
For sparse formulas on \(N\ge1\) variables, there is a fixed positive integer
\(p\) such that the output satisfies, for \(n=|V|\),
\[
N\le |E|,\qquad n,|E|\le N(\log(N+2))^p,\qquad n/2\le |E|\le\Delta n/2.
\]
\end{theorem}

The proof in \cref{app:graph-source} composes the two cited reductions and
checks the size bounds and graph conventions stated above.

We call graphs from satisfiable formulas YES and those from unsatisfiable
formulas NO. We build the amplifier in two ways, which satisfy the same
guarantee. In the \emph{full-product case}, used for the randomized
reduction, every interlace uses all column tuples, and
\cref{lem:full-schedule} fixes the widths and thresholds. In the \emph{explicit case},
used for the deterministic reduction, the interlaces use the small tuple lists of
\cref{det:construction}, and \cref{lem:generated-schedule} fixes the
widths and thresholds. In each case a final interlace restores the deletion
bounds at the full output cost for the next round, by \cref{cl:bridge} in the full-product case and by
\cref{det:restore} in the explicit case.

\begin{theorem}[Amplification under input deletion]
\label{thm:amplifier}
Let \(G=(V,E)\) be a YES or NO graph produced by \cref{src:gap-colouring},
and put \(n=|V|\). In either case there are a public row-weight threshold
\(\rho\), a public integer \(h\), and a deterministic, implicitly represented
construction \(R\mapsto R^+\) on base matrices \(R\) with positive normalized
rational input weights. In the full-product case, \(n\) must exceed an
effectively computable constant, \(\rho=2^{-L}\) with \(L=\ceil{\log n}\), and
\(h=L+O_{\Delta,\gamma}(\log L)\). In the explicit case, \(\rho=1/\mathsf N\)
and \(h=10\log\mathsf N\), with \(\mathsf N=s^4\) for the power of two \(s\) of
\cref{lem:generated-schedule}. In both cases:
\begin{enumerate}
\item[(i)] On either source promise,
\[
\D(R^+)\le\D(R)+h+1,
\]
and, if \(G\) is YES, the stronger bound \(\D(R^+)\le\D(R)+h\) holds.
\item[(ii)] If \(G\) is NO and the base matrix \(R\) satisfies \(\D(R)=d\ge3\)
and \(\RP{R}{\rho}{d}\), then
\[
\D(R^+)=d+h+1,\qquad \RP{R^+}{\rho}{d+h+1}.
\]
\item[(iii)] On either source promise, \(\rank(R^+)\le2^h\rank(R)\).
\end{enumerate}
\end{theorem}

The proof of \cref{thm:amplifier} is assembled at the end of \cref{sec:amplification-proof}
(p.~\pageref{proof:thm:amplifier}). It uses the interlacing estimates of
\cref{sec:interlacing-details} and, in the explicit case, the small lists of
\cref{sec:small-lists}. The rank bound~(iii) follows from \cref{lem:amplifier-rank}.

\subsection{Iteration}

One round adds $h$ bits to the YES upper bound and $h+1$ bits to the NO
lower bound. Starting from the initial matrix of \cref{cl:seed}, of known cost
$d_0$, the two bounds therefore differ by $i$ bits after $i$ rounds.

\begin{lemma}[Initial matrix]\label{cl:seed}
For $L\ge1$, let $p_0=\operatorname{pow}_2(64(L+1))$ and
$R_0(z,j)=z_j$ on $\{0,1\}^{p_0}\times[p_0]$, with uniform input weights. Then
\[
d_0:=\log p_0+1,\qquad \D(R_0)=d_0,\qquad
\rank(R_0)=p_0=2^{d_0-1},\qquad \RP{R_0}{2^{-L}}{d_0}.
\]
For $L=\ceil{\log n}$ and $n\ge4$, the initial matrix has polynomial size in $n$
and can be printed in polynomial time.
\end{lemma}
The proof is in \cref{sec:initial-template-proofs}
(p.~\pageref{proof:cl:seed}).

\begin{proposition}[Iterated amplification]\label{prop:iteration}
Fix one case of \cref{thm:amplifier}, whose threshold is $\rho$ and whose
overhead is $h$, and a source graph to which it applies. Iterate the theorem from the
matrix $R_0$ of \cref{cl:seed} with $L=\log(1/\rho)$, an integer in both
cases, using the same
parameters in every round, and let $R_i$ be the matrix after $i\ge0$ rounds.
Put $B_i=d_0+ih$, $H_i=B_i+i$, and $r_i=2^{B_i-1}$. A YES graph
gives $\D(R_i)\le B_i$. A NO graph gives
$\D(R_i)=H_i$ and $\RP{R_i}{\rho}{H_i}$. On either promise,
$\rank(R_i)\le r_i$ and $\D(R_i)\le H_i$.
\end{proposition}
\begin{proof}
\Cref{cl:seed} gives the case $i=0$, with $d_0\ge3$. By
\cref{thm:amplifier}, each round raises the YES bound by $h$, the NO cost
and the general upper bound by $h+1$, and rank by a factor at most $2^h$.
In the NO case, $H_i\ge3$ and $\RP{R_i}{\rho}{H_i}$ give the hypotheses of
(ii), which restores the deletion bounds at threshold $\rho$, so
the induction continues.
\end{proof}

\section{Proof of amplification under input deletion}
\label{sec:amplification-proof}

We now prove \cref{thm:amplifier}, for the matrices $N,U,W$ of
\cref{def:source-matrix}. By \cref{src:profiles,cl:separation}, a depth-$(C-1)$
protocol for a near-full restriction of $N$ begins with $\log s$ Alice bits, and the collision argument
then gives \cref{src:near-full}, the lower bound for $N$ after near-full
deletion. For $U$, the template bounds and \cref{det:conditioning,det:rareboost}
give the stopping nodes of \cref{res:destinations}. The arrival bounds
\cref{res:incoming,res:foreign} and \cref{src:near-full} give
\cref{res:alignment}: $W$ has full cost when little column weight is
deleted, and deletion bounds one bit below full cost. Finally, \cref{cl:bridge}
for full products and \cref{det:restore} for explicit lists combine these
two conclusions to restore the deletion bounds at full cost.

\subsection{Widths and tolerances}

We need enough coordinates to encode the graph, and enough interlaced
copies that the lower bounds survive the restrictions left by a
protocol's messages. Both cases use one recipe with two scales: $\rho$, the retained
fraction of base rows for which hardness must survive between rounds, and
$s$, the number of bins used to group inputs within a round.

\begin{definition}[Amplification parameters]\label{def:amplification-parameters}
Let $n=|V|$ be the number of vertices of the source graph. Choose powers of two $s,\rho^{-1}$ with
$n\le\rho^{-1}$, $\rho\le s^{-4}$, and $s\ge s_0$, the effective constant
of \cref{lem:common-schedule}. Set
\[
\begin{gathered}
q_1=s-2,\qquad q_2=s^2/\rho,\qquad w=s^8,\qquad Q=s^{24},\\
x=\rho^8,\qquad \eta=s^{-2},\qquad
\theta=(512Q)^{-1},\qquad \delta_U=\eta/(16w).
\end{gathered}
\]
The source construction of \cref{def:source-matrix} has $q_1$ template
positions and $q_2$ indices, and the two further interlaces have widths
$w$ and $Q$. The deletion bounds of \cref{src:profiles} for the matrices
$A$, $R_{\rm temp}$ and $N_{\rm ref}$ of the source construction use
row-weight threshold $x$, and the lower bound for the source matrix $N$
survives deleting an $\eta$ fraction of each input set's weight
(\cref{src:near-full}). For the transpose $W$ of the first further interlace
(\cref{def:source-matrix}), $\theta$ is the row-weight threshold of its
deletion bounds, and restrictions retaining row weight at least $\theta$ and
column weight at least $1-\delta_U$ keep the full cost (\cref{res:alignment}).
\end{definition}

In the full-product case, we take $L=\ceil{\log n}$, $\rho=2^{-L}$, and
$s=\operatorname{pow}_2(2^{16}(L+1))$. \Cref{lem:full-schedule} shows that
these choices meet the constraints above once $n$ exceeds an effective
constant. In the explicit case, we take the least dyadic $s\ge s_0$ with
$s^4\ge n$, after enlarging $s_0$ if necessary, and set $\rho=s^{-4}$
(\cref{lem:generated-schedule}). \Cref{sec:parameters} verifies the
numerical bounds \eqref{eq:width-margins} shared by both cases and the
additional conditions of each case. All
these choices are fixed across rounds.

\subsection{Encoding the graph}

We construct the source matrix of \cref{sec:overview} from interlaced copies
of the base matrix $R$ and rows encoding the graph. In a NO instance, every
assignment of the items defined below to bins has a \emph{collision}
(\cref{src:item-witness}), that is, a coordinate at which one
bin contains two active items and an inactive item. We show in the rest of
the proof that a collision forces the extra bit of the NO case. Because of
the template rows, defined below, every protocol cheaper than the NO-case cost
must begin with $\log s$ bits from Alice, and these bits distinguish the $s$
bins (\cref{cl:separation}). The item rows encode the graph.

\begin{definition}[Source construction]
\label{def:source-matrix}
Let \(R:X\times Y\to\{0,1\}\) have positive normalized rational input weights, and let
\(s\ge4\) be dyadic, meaning a power of two. Coordinates ranging over integer sets such as \([q_2]\) are uniform below,
and product sets carry the product of the specified factor weights.

\smallskip\noindent\textit{Active items.}
We repeat the graph in several groups so that the proof can discard groups
that lose too many items.
An item is a vertex or auxiliary label in one of \(4s\) groups:
\(\mathcal I=[4s]\times(V\sqcup[s-3])\).
Each edge occurrence in each group has a coordinate at which its two
endpoints and the \(s-3\) auxiliary items of that group are \emph{active}.
All other items are inactive there.
Pad these \(4s|E|\) coordinates with inactive coordinates to length
\(q_0=\operatorname{pow}_2(4s|E|)\), then repeat each coordinate
\(q_2/q_0\) times, where \(q_2\) is a dyadic multiple of \(q_0\).
Write \(v_i(\alpha)\in\{0,1\}\) for the indicator that item \(i\) is active
at index \(\alpha\in[q_2]\), and \(\pi_\alpha\) for the map numbering the at most
\(s-1\) active items \(1,2,\ldots\) in a fixed order on \(\mathcal I\).
Neither depends on any assignment of items to bins.

\smallskip\noindent\textit{Template interlaces.}
Recall \(q_1=s-2\), which leaves two of the \(s\) values of Alice's last
message in \cref{src:upper} for active and inactive items.
Define
\[
\Gamma=Y^{2s-1},\qquad J=[q_1]\times X,\qquad S=J^{q_2},\qquad
J_2=[q_2]\times\Gamma.
\]
The template rows use the first \(q_1\) positions of a list
\(\gamma\in\Gamma\), and we reserve its last \(s+1\) positions for the item
rows, since \(q_1+s+1=2s-1\). The two successive interlace matrices are
\[
A(\gamma,(r,u))=R(u,\gamma_r),\qquad
R_{\rm temp}(z,(\alpha,\gamma))=A(\gamma,z_\alpha).
\]
Thus $A$ has rows indexed by $\Gamma$ and columns by $J$, and $R_{\rm temp}$
has rows indexed by $S$ and columns by $J_2$. Moreover, $A^{\mathsf T}$ is a
$q_1$-fold interlace of $R$ with repeated column labels,
and $R_{\rm temp}^{\mathsf T}$ is a $q_2$-fold interlace of $A$. A template label \(z\in S\)
is a tuple of pairs \(z_\alpha=(r,u)\in J\), one for each index \(\alpha\).
Against a list \(\gamma\), the pair gives the entry \(R(u,\gamma_r)\).

\smallskip\noindent\textit{Rows and entries of \(B_{\rm raw}\).}
The matrix \(B_{\rm raw}\) has template rows \((p,z)\), indexed by a bin
\(p\in[s]\) and a template label, and item rows \((i,u)\), indexed by an item
and a base row:
\[
\mathcal K=([s]\times S)\sqcup(\mathcal I\times X).
\]
We give the template rows most of the weight, so that deleting an \(\eta\)
fraction of all row weight removes at most an \(\eta/\lambda\) fraction of theirs.
Set \(\lambda=255/256\) and \(\mu=1/256\).
Template rows have total weight \(\lambda\), and vertex rows and auxiliary rows
each have total weight \(\mu/2=1/512\). Within each type, the bin or item is
uniform and \(z\) or \(u\) has the weights of \(S\) or \(X\).
For a column \(Z=(\xi,j_1,\ldots,j_s)\), where \(\xi\in Y^{s+1}\) and
\(j_p=(\alpha_p,\gamma_p)\in J_2\), set
\[
B_{\rm raw}((p,z),Z)=R_{\rm temp}(z,j_p),
\]
\[
B_{\rm raw}((i,u),Z)=
 \begin{cases}
 R(u,\xi_{\pi_\alpha(i)}),
   &\alpha_1=\cdots=\alpha_s=\alpha,\ v_i(\alpha)=1,\\
 R(u,\xi_{s+1}),&\text{otherwise}.
 \end{cases}
\]
Thus an item row uses its assigned position when the indices
\(\alpha_1,\ldots,\alpha_s\) agree and the item is active there, and otherwise
it uses the last entry of the shared list.

\smallskip\noindent\textit{Two column families.}
The first column family \(\Omega=Y^{s+1}\times J_2^s\) has product weights.
The second column family is
\[
\mathcal D = \{(\operatorname{tail}(\gamma),j,\ldots,j):j=(\alpha,\gamma)\in
J_2\}, \qquad \operatorname{tail}(\gamma)=(\gamma_{q_1+1},\ldots,\gamma_{2s-1}),
\]
with the weights induced from \(J_2\). We give the two families
the weights
\[
\begin{array}{ccl}
\text{Family}&\text{Weight}&\text{Role}\\
\Omega&\lambda&\text{Force Alice's initial messages}\\
\mathcal D&\mu&\text{Read the three tail positions of a collision}.
\end{array}
\]
The two families have distinct labels even where their columns coincide,
so each keeps its own weight. At every bin \(p\), the label \(j_p\) has the weights prescribed on \(J_2\) in
either family. We call the columns of the two families \emph{$\Omega$-inputs} and
\emph{$\mathcal D$-inputs}. After transposition they become rows, and we
then also call them \emph{$\Omega$-rows} and \emph{$\mathcal D$-rows}.

\smallskip\noindent\textit{The transpose and two further interlaces.}
Transpose \(B_{\rm raw}\), and restrict the transpose to the rows from the
first column family:
\[
N=B_{\rm raw}^{\mathsf T},\qquad
N_{\rm ref}=N[\Omega,\mathcal K],
\]
where the weights on \(\Omega\) are normalized. The matrix \(N_{\rm ref}\) is used only in
the lower-bound analysis. Finally set
\[
U=\Int{w}{N},\qquad W=U^{\mathsf T},\qquad
R^+=(\Int{Q}{W})^{\mathsf T}.
\]
\end{definition}

As in \cref{sec:overview}, in a protocol below the NO-case cost the
$\Omega$-inputs force Alice's first \(\log s\) bits to name a bin
(\cref{cl:separation}), and this fixes an assignment of the items to
bins. Against a $\mathcal D$-input with label \((\alpha,\gamma)\), the rows
of two items active at \(\alpha\) and of one inactive item read three
distinct positions of the same list \(\gamma\). The interlace \(U\),
transposed to \(W\), extends the lower bound for \(N\) to restrictions
retaining only row weight \(\theta\). For \(W\) we obtain the deletion bounds
only one bit below full cost (\cref{res:alignment}), and the final
interlace restores that bit.

Throughout, we write \(d=\D(R)\). With the parameters of
\cref{def:amplification-parameters}, we define the costs and the threshold
used in the lower-bound analysis (in the explicit case, we drop the
subscript \(\rm g\) that \cref{sec:deterministic} attaches to its
parameters):
\begin{equation}\label{eq:source-costs}
H_1:=d+\log s,\qquad H_2:=H_1+\log q_2,\qquad C:=H_2+\log s+1,\qquad
x_{\rm cap}:=2s\rho.
\end{equation}
In the NO case, we will show \(\D(A)\ge H_1\), \(\D(R_{\rm temp})\ge H_2\)
and \(\D(N)\ge C\). The collision bound of \cref{src:profiles}
requires the \(q_1\) template positions and the three tail positions to
retain base row weight at least \(x_{\rm cap}\) within their blocks.

The following example shows the \(s+1\) positions exposed by a collision.
Fix \(s=4\) and a coordinate \(\alpha\) at which \(u,v\) are active
and \(h\), from another group, is inactive. Suppose these three items
lie in the same bin, with \(\pi_\alpha(u)=1\) and
\(\pi_\alpha(v)=2\). On a \(\mathcal D\)-input, their rows use
\(\gamma_3,\gamma_4,\gamma_7\), respectively. Together with the
bin's \(s-2=2\) template positions, they expose \(s+1=5\) positions
of the same list.

\begin{figure}[H]
\centering
\begin{tikzpicture}[
  >=Stealth,
  font=\small,
  collisioncell/.style={
    draw=black!65, line width=.45pt,
    minimum width=16mm, minimum height=8.5mm,
    inner sep=0pt, outer sep=0pt
  },
  collisionactive/.style={
    circle, draw=answergreen, fill=answergreen!12,
    line width=.7pt, minimum size=6.5mm, inner sep=0pt
  },
  collisioninactive/.style={
    rectangle, draw=boborange, fill=boborange!7, dashed,
    line width=.7pt, minimum size=6.5mm, inner sep=0pt
  }
]

\draw[rounded corners=3pt, line width=.65pt]
  (-.4,1.15) rectangle (11.6,2.55);

\node[font=\small\bfseries] at (5.6,2.92)
  {One bin at a fixed coordinate \(\alpha\)};

\node[
  draw=aliceblue,
  fill=aliceblue!7,
  rounded corners=2pt,
  line width=.65pt,
  minimum width=27mm,
  minimum height=8mm
] (templates) at (1.6,1.78) {Template rows};

\node[text=answergreen] at (4.8,2.28) {Active items};
\node[collisionactive] (u) at (4.0,1.70) {\(u\)};
\node[collisionactive] (v) at (5.6,1.70) {\(v\)};

\node[text=boborange] at (10.4,2.28) {Inactive item};
\node[collisioninactive] (h) at (10.4,1.70) {\(h\)};

\node[collisioncell, fill=aliceblue!10]
  (g1) at (0.8,0) {\(\gamma_1\)};
\node[collisioncell, fill=aliceblue!10]
  (g2) at (2.4,0) {\(\gamma_2\)};

\node[collisioncell, fill=answergreen!12]
  (g3) at (4.0,0) {\(\gamma_3\)};
\node[collisioncell, fill=answergreen!12]
  (g4) at (5.6,0) {\(\gamma_4\)};

\node[collisioncell, fill=black!4, text=black!45]
  (g5) at (7.2,0) {\(\gamma_5\)};
\node[collisioncell, fill=black!4, text=black!45]
  (g6) at (8.8,0) {\(\gamma_6\)};

\node[collisioncell, fill=boborange!12]
  (g7) at (10.4,0) {\(\gamma_7\)};

\draw[->, aliceblue, line width=.7pt]
  (templates.south) -- (g1.north);
\draw[->, aliceblue, line width=.7pt]
  (templates.south) -- (g2.north);

\draw[->, answergreen, line width=.7pt]
  (u.south) -- (g3.north);
\draw[->, answergreen, line width=.7pt]
  (v.south) -- (g4.north);

\draw[->, boborange, dashed, line width=.7pt]
  (h.south) -- (g7.north);

\node at (5.6,-.78)
  {One shared list \(\gamma\) on a \(\mathcal D\)-input};

\node[font=\normalsize] at (5.6,-1.68) {\(
  \underbrace{2}_{\text{template}}
  +\underbrace{2}_{\text{active}}
  +\underbrace{1}_{\text{inactive}}
  =\underbrace{5}_{\text{positions}}>4=s
\)};

\end{tikzpicture}
\caption{A collision exposes \(s+1\) positions, illustrated for \(s=4\).
On a \(\mathcal D\)-input with label \((\alpha,\gamma)\), the template
rows in the depicted bin use positions \(1,2\), the active items
\(u,v\) use positions \(3,4\), and the inactive item \(h\) uses
position \(7\). Activity and the positions read are determined by
\(\alpha\), independently of the bin assignment. Only the positions
used by the displayed rows are highlighted, and other items are omitted.}
\label{fig:source-collision}
\end{figure}

The next lemma shows that, for a NO graph, every assignment of the items
remaining after a few deletions still has collisions as in
\cref{fig:source-collision} at a constant fraction \(\zeta\) of the
coordinates.

\begin{lemma}
\label{src:item-witness}
A YES graph admits an assignment of \(\mathcal I\) to \([s]\) with at most one
active item in each bin at every coordinate. For a NO graph, delete fewer than
\((s/2)\delta_v n\) vertex items and fewer than \(s/2\) auxiliary items, where
\[
\delta_v=\frac{\gamma|E|}{2\Delta n},\qquad \zeta=\frac{\gamma}{24}.
\]
Every assignment of the remaining items to the \(s\) bins then has at least
\(\zeta q_2\) coordinates at which one bin contains two active items and an
inactive item.
\end{lemma}

We prove the collision guarantee in \cref{sec:collision-conversation}.
First, the YES assignment gives a short protocol. With \(d=\D(R)\) and \(C\) as in
\eqref{eq:source-costs}, its cost is at most \(C-1\).

\begin{proposition}
\label{src:upper}
If \(\D(R)\le y\), then \(\D(N)\le y+2\log s+\log q_2+1\). For a YES graph, the
final \(+1\) can be omitted. Both bounds remain valid after restricting either
input set.
\end{proposition}

\begin{proof}
For a YES graph, fix a public bin assignment $\chi$ from
\cref{src:item-witness}. We describe a protocol for $B_{\rm raw}$.
Alice holds a template row $(p,z)$ or an
item row $(i,u)$, and Bob holds $Z=(\xi,j_1,\ldots,j_s)$ with
$j_{p'}=(\alpha_{p'},\gamma_{p'})$ for each $p'\in[s]$. They exchange:
\[
\begin{array}{lll}
\text{Alice} & \text{bin }p\text{ (use }p=\chi(i)\text{ for an item row)}
  & \log s\text{ bits},\\
\text{Bob} & \alpha_p & \log q_2\text{ bits},\\
\text{Alice} & \text{one of }q_1\text{ template positions, active, inactive}
  & \log s\text{ bits}.
\end{array}
\]
The third message has $q_1+2=s$ possibilities. For a template row, Alice
writes $z_{\alpha_p}=(r,u)$ and sends code $r$, and the answer is
$R(u,(\gamma_p)_r)$, where Bob knows $(\gamma_p)_r$. For an item row, she sends
active precisely when $v_i(\alpha_p)=1$. Bob can test from his own input
whether $\alpha_1=\cdots=\alpha_s$. If so and the code is active, Bob
identifies the active item in bin $p$, unique by the choice of $\chi$, and uses $\xi_{\pi_{\alpha_p}(i)}$. Otherwise he
uses $\xi_{s+1}$. For both row types, the players
then finish by running a cost-$y$ protocol for $R$ once.

For an arbitrary graph, we use any public assignment and replace the active
code by the $s$ possible values of $\pi_{\alpha_p}(i)$. There are now
$q_1+s+1=2s-1$ message values, so the third message costs one additional bit.
These rules apply to $\Omega$-inputs and $\mathcal D$-inputs alike, since
both have the form $(\xi,j_1,\ldots,j_s)$. Exchanging the players gives the
protocol for $N=B_{\rm raw}^{\mathsf T}$, and restriction preserves its cost.
\end{proof}

\subsection{A collision survives the conversation}
\label{sec:collision-conversation}

The NO-case lower bound for $N$ uses the NO promise only
through \cref{src:item-witness}. We need that lemma's collision count for every assignment of the retained items, because Alice's
first messages fix the assignment before any index is revealed.

\begin{proof}[Proof of \cref{src:item-witness}]
\phantomsection\label{proof:src:item-witness}
For a YES graph, let \(c:V\to[3]\) be a proper $3$-coloring, which exists by
\cref{src:gap-colouring}. Assign each vertex \(v\) to bin \(c(v)\), and assign
the \(s-3\) auxiliary labels to the remaining bins. Every edge coordinate
activates that edge's two endpoints and those auxiliaries in one group, so all
its active items occupy distinct bins. Padded coordinates activate no items.

Now suppose that \(G\) is NO. By the deletion hypotheses in
\cref{src:item-witness}, fewer than \(s/2\) groups lose an auxiliary item and
fewer than \(s/2\) groups lose more than \(\delta_v n\) vertices. Hence more
than \(3s\) groups retain every auxiliary item and all but at most \(\delta_v
n\) vertices.

Fix such a group. If two auxiliaries share a bin, they are active together at every
edge coordinate. Otherwise the auxiliaries occupy \(s-3\) distinct bins,
leaving three free bins. Color the retained vertices in those free bins by
their bins, and extend the coloring arbitrarily to all other vertices. The
argument below works for every such extension.
At least \(\gamma|E|\) edge occurrences are monochromatic by the NO promise,
and at most \(\Delta\delta_v n=\gamma|E|/2\) meet a deleted vertex.
Every remaining monochromatic edge occurrence gives two active items in one bin.
If an endpoint lies in the bin of an auxiliary, it shares that bin with the auxiliary.
Otherwise both endpoints lie in free bins and, having the same color, share a bin.
In both cases this group has
at least \(\gamma|E|/2\) edge coordinates with two active items in one bin.

It remains to find an inactive item in the same bin. For each counted
coordinate, we choose one bin containing the two active items.
We discard a group if one of its chosen bins contains retained items of that
group only, and call such a bin exclusive to the group. A bin is exclusive to at most
one group, since it contains items of that group, so at most \(s\) groups are discarded.
In each of the more than \(2s\) remaining groups, every chosen
bin also contains a retained item from another group. That item is inactive
at the chosen coordinate, which activates items of one group. We have therefore found more than
\(\gamma s|E|\) coordinates with a collision.
Since \(q_0<8s|E|\), their fraction among the \(q_0\) padded coordinates exceeds
\(\gamma/8>\zeta=\gamma/24\). Repeating each coordinate \(q_2/q_0\) times
preserves this fraction.
\end{proof}

The next estimate is used to prove the template-layer deletion bounds
below for full-product interlaces. For explicit lists we instead use
\cref{det:conditioning,det:rareboost}. The proof of the deletion bounds in both cases is in
\cref{sec:initial-template-proofs}, and the collision argument uses only the resulting bounds.

\begin{lemma}\label{lem:many-product-blocks}
Suppose $\RP{M}{x}{H}$ holds for a rationally weighted matrix $M$, where
$H\ge3$. Let $u\ge16$ be dyadic with $ux\le1$. Every restriction of a
full-product interlace of $M$ retaining at least $9u/16$ blocks, each of base row weight at least
$ux$, and a column set of weight at least $2^{-u/1024}$, has cost at least
$H+\log u$.
\end{lemma}

The proof is in \cref{sec:interlace-induction}, on
p.~\pageref{proof:lem:many-product-blocks}.

For the collision argument, we need deletion bounds for $A$, $R_{\rm temp}$
and $N_{\rm ref}$, and a \emph{collision bound}, which says that the template positions
together with the three positions of a collision require at least $H_1+1$ bits.

\begin{lemma}[Template-layer deletion bounds]\label{src:profiles}\label{det:profiles}
Consider either case of \cref{thm:amplifier}: the full-product construction of
\cref{def:source-matrix} with the parameters of \cref{lem:full-schedule}, where
$\rho=2^{-L}$, or the explicit construction of \cref{det:construction} with the
parameters of \cref{lem:generated-schedule}, where $\rho=\rho_{\rm g}=1/\mathsf N$
and the subscript $\rm g$ is suppressed. Suppose $\D(R)=d\ge3$ and $\RP{R}{\rho}{d}$,
let $x=\rho^8$ be the block threshold, and define $H_1,H_2,C,x_{\rm cap}$ by
\eqref{eq:source-costs}. Then
\[
\RP{A}{x}{H_1},\qquad \RP{R_{\rm
temp}}{x}{H_2},\qquad \RP{N_{\rm ref}}{x}{C-1}.
\]
In the interlace of $R$ on $\Gamma$ (on $\Gamma_{\rm g}$ in the explicit case),
the first $q_1$ positions together with any three distinct tail positions
require at least $H_1+1$ bits whenever every position retains base row weight
at least $x_{\rm cap}$ within its block and the retained set of list labels
has weight greater than $1/2$.
\end{lemma}

The proof is in \cref{sec:initial-template-proofs}, after the initial-matrix
proof (p.~\pageref{proof:src:profiles}).

Three proofs below (\cref{lem:collision-tracking,res:destinations,cl:two-stream})
follow retained input sets along a protocol prefix and count them where the
prefix stops. A mark lets a set count twice until it splits into two sets. In
the collision application, the mark records the extra bit forced by a collision.

\begin{lemma}[Kraft count]\label{lem:kraft-count}
Let $q$ be a power of two, and let $V$ be a set of nodes of a binary protocol
tree, none an ancestor of another, all at depth at most $\log q$. For $v\in V$
at depth $r_v$ put $\kappa_v=q2^{-r_v}$. Suppose each node $v\in V$ holds $h_v$
objects, $z_v\le h_v$ of them marked, and $\sum_{v\in V}(h_v+z_v)>q$. Then some $v$
has $h_v+z_v>\kappa_v$. For this $v$, if $\kappa_v\ge2$ then
$h_v\ge\kappa_v/2+1$, and if $\kappa_v=1$ then $h_v\ge2$ or $h_v=z_v=1$.
\end{lemma}
\begin{proof}
Kraft's inequality gives $\sum_v\kappa_v\le q$, so some $v$ has
$h_v+z_v>\kappa_v$, and $z_v\le h_v$ gives $2h_v>\kappa_v$. If
$\kappa_v\ge2$, then $\kappa_v$ is even and $h_v\ge\kappa_v/2+1$. If
$\kappa_v=1$, then $h_v+z_v\ge2$, so $h_v\ge2$ or $h_v=z_v=1$.
\end{proof}

In the applications, the objects are retained input sets. We follow these sets
through the bits of one player, starting from sets of weight at least
$t$, by the following \emph{tracking rule}, which keeps every tracked set at
weight at least $t2^{-r}$ after $r$ such bits. An unmarked set follows a child retaining at least half its weight.
After $r$ bits, a marked set splits into two unmarked sets if both children retain
weight at least $t2^{-r-1}$. Otherwise it follows a child meeting that threshold,
which exists since one child retains half its weight, and keeps
its mark. Thus every mark splits at most once. Before it splits, a marked
set loses less than $\sum_{r\ge0}t2^{-r-1}=t$ of its weight. A split
replaces the contribution $2$ of a marked set to $\sum_v(h_v+z_v)$ by $1+1$. Hence
$\sum_v(h_v+z_v)$ equals the number of initial sets plus the number of
initial marks.

\begin{lemma}[Following collision witnesses]\label{lem:collision-tracking}
Let $q\ge2$ be dyadic, let $0\le b<q$ be an integer, and let $x>0$
satisfy $2qx\le1/8$.
Suppose a protocol with at most
$H+\log q$ bits left has $q-b$ indices, each with a retained set of
list inputs of weight at least $3/4$. Each set consists of inputs with that
fixed index. Mark $w_0>b$ of these sets.
Assume the following lower bounds, where all set weights are computed from
the original input weights.
\begin{enumerate}
\item[(i)] Until Alice speaks, a set of retained lists at one index with weight at least $x$, against the
unchanged set of Alice's inputs, costs at least $H>0$.
\item[(ii)] At the first Alice bit, one child retains a set of Alice's inputs on which,
for every dyadic $u\ge2$ with $ux\le1$, retained sets at any $u/2+1$ distinct indices, each
of weight at least $ux$, cost at least $H+\log u$.
\item[(iii)] A marked set retaining weight greater than $1/2$ costs at least
$H+1$ before Alice speaks.
\end{enumerate}
Then no such protocol exists.
\end{lemma}
\begin{proof}
We track the retained list sets of each index through Bob bits by the
tracking rule with $t=2qx$, and we stop at the first Alice bit. A marked set loses less
than $2qx\le1/8$ before it splits, so an unsplit mark retains weight greater
than $1/2$. By (i), a tracked set cannot end at a leaf, nor reach Bob depth
$\log q+1$: it would still have weight at least $x$ but only $H-1$ bits left.
The stopping nodes therefore satisfy the hypotheses of \cref{lem:kraft-count},
where $v$ holds $h_v$ tracked sets and $z_v$ unsplit marks, and
$\sum_v(h_v+z_v)=q-b+w_0>q$. Let $v$ be the node given by that lemma, with
$\kappa=\kappa_v$. Its sets have distinct indices, because descendants of a
split never rejoin, and each has weight at least $2\kappa x$.

If $\kappa\ge2$, then (ii) with $u=\kappa$ requires $H+\log\kappa$ bits after
Alice's bit, but only $H+\log\kappa-1$ remain. If $\kappa=1$ and $h_v\ge2$, then (ii) with $u=2$ and the
available weights $2x$ requires $H+1$ bits after Alice's bit, but only $H-1$
remain. If $h_v=z_v=1$, then by (iii) the unsplit marked set needs $H+1$ bits
at $v$, where only $H$ remain because $v$ has depth $\log q$.
\end{proof}

\begin{theorem}[Source hardness after deletion]\label{src:near-full}
Use either the full-product construction of \cref{def:source-matrix} or the
explicit-list construction of \cref{det:construction}, with a NO graph and
$\D(R)=d\ge3$, and define $H_1,H_2,C,x_{\rm cap}$ by \eqref{eq:source-costs}.
Suppose the deletion bounds and the collision bound stated in
\cref{src:profiles} hold. Assume the numerical inequalities
\eqref{eq:width-margins} of \cref{lem:common-schedule}, that the weights on $S$ give exact $8$-wise
independence with the prescribed weights on $J$ as marginal, and that the bin tuple $(j_1,\ldots,j_s)$ of the $\Omega$-inputs has
exactly $8$-wise independent weights with the prescribed $J_2$ marginal.
Then $\D(N)=C$, and every restriction retaining at least $1-\eta$ of each
input set's total weight has cost at least $C$. In either case of
\cref{thm:amplifier} with $\RP{R}{\rho}{d}$, these hypotheses follow from \cref{src:profiles},
\cref{lem:common-schedule}, and the exact independence of the constructed weights.
\end{theorem}
\begin{proof}
In outline, Alice's first $\log s$ bits assign the items
to bins. The collisions of this assignment give the marked sets of
\cref{lem:collision-tracking}, and \cref{src:profiles} gives its hypotheses
(i)--(iii). \Cref{lem:collision-tracking} then rules out depth $C-1$.

Suppose a restriction of $N$ retaining at least $1-\eta$ of each input set's
weight has a protocol of depth at most $C-1$.
We transpose it to a protocol of the same depth for the corresponding
restriction of $B_{\rm raw}$. Each template bin has row weight $\lambda/s$, so it
loses at most $\eta s/\lambda<1/1000$ of its total $S$ weight. The first column family $\Omega$
retains normalized weight at least $1-\eta/\lambda>99/100$.
Template entries do not depend on the component \(\xi\) of a column. We therefore project the retained
$\Omega$-inputs onto their bin tuple $(j_1,\ldots,j_s)$ (in the explicit
case, onto their $\mathcal J_{\rm g}$ label) and choose one retained
completion for each image label. These representatives form a restriction of an
interlace of $R_{\rm temp}$ on the template rows, with column weight at
least $99/100$ (\cref{obs:representatives}).

We apply \cref{cl:separation} to the protocol inherited on these
representatives, without pruning its tree. We take $M=R_{\rm temp}$,
$d=H_2$, $u=s$, $x$ in place of its $\rho$, and row threshold $999/1000$ in
place of its $\xi$. Its hypotheses follow
from the deletion bounds $\RP{R_{\rm temp}}{x}{H_2}$, the depth
$C-1=H_2+\log s$, and the numerical inequalities
$999/1000>2sx$ and $999/1000-2(s-1)x>998/1000$, with $99/100>18/25$.
By that theorem, every binary history of length below $\log s$ is reached by
template inputs, and every such node is Alice's. They are actual nodes of the
original protocol, so Alice sends the first $\log s$ bits on every retained
input pair, including pairs with item rows or $\mathcal D$-inputs. At the resulting $s$
nodes, the respective template bins retain sets $A_1,\ldots,A_s\subseteq S$,
each of weight greater than $998/1000$.
We needed the representatives only to establish this prefix. From now on we use
all originally retained columns, with $C-1-\log s=H_2$ bits left.

We next fix the item assignment before choosing an index. We set aside
items whose rows retain less than half of their base row weight in the
restriction. By the row weights in \cref{def:source-matrix}, each item type has
total weight \(1/512\), uniform over \(4sn\) vertex or \(4s(s-3)\) auxiliary
items. Each set-aside item loses more than half its weight, and at most \(\eta\)
is deleted, so we set aside fewer than \(4096\eta sn\) vertex
items and fewer than \(4096\eta s(s-3)\) auxiliary items. These numbers are below
the deletion thresholds of \cref{src:item-witness}, because \eqref{eq:width-margins}
gives \(4096\eta sn\le\delta_v n/16\) and \(4096\eta s(s-3)<s/2\), as
checked in the proof of \cref{lem:common-schedule}. By averaging over these \(s\) nodes, we can assign every remaining item to
a node that receives at least \(1/(2s)\) of its base weight. The parameter choices give
\(4s^2\rho\le1\), so
\[
\frac1{2s}\ge2s\rho=x_{\rm cap}.
\]
Thus the assigned part retains at least \(x_{\rm cap}\) of the item's base row
weight. This assignment is fixed before the index is chosen. By
\cref{src:item-witness}, at least \(\zeta q_2\) coordinates contain two active
items and an inactive item in one bin. We assign each such coordinate to one node \(j\in[s]\)
at which its collision occurs and write \(U_j\subseteq[q_2]\) for the resulting
disjoint sets.

For each \(j\), we discard two kinds of indices \(\alpha\): those at which
the template labels in \(A_j\) project onto too little base row weight, and
those at which too little $\mathcal D$-weight is retained. First, we discard
\(\alpha\) for \(A_j\) if one of the
\(q_1\) blocks of the \(J\)-projection of \(A_j\) at \(\alpha\) retains base row
weight less than \(x_{\rm cap}\). Since \(x_{\rm cap}\le1/8\) by
\eqref{eq:width-margins} and each block has weight \(1/q_1\) in \(J\), the
\(J\)-projection then misses more than \(7/(8q_1)\) of the weight of \(J\).
If there were at least $8q_1$ such indices, we could choose $8q_1$ of them and let $Z$
count those at which $z_\alpha$ lies outside the projection at $\alpha$, for $z\in S$
drawn from its weights. Then $Z$ would satisfy $\mathbb EZ>7$ and, by pairwise independence,
$\operatorname{Var}Z\le\mathbb EZ$. Every label in $A_j$ has $Z=0$, so
Chebyshev's inequality would give $\Pr(A_j)\le\Pr(Z=0)<1/7$, contrary to
$\Pr(A_j)>998/1000$. Thus fewer than \(8q_1\) indices are discarded for \(A_j\).
Second, we discard \(\alpha\) if the $\mathcal D$-inputs with index \(\alpha\)
retain less than \(3/4\) of their weight. The $\mathcal D$-inputs have total
weight \(\mu\), and their index is uniform on \([q_2]\). Each such \(\alpha\)
therefore loses more than \(\mu/(4q_2)\) of the column weight. Since the
restriction deletes at most \(\eta\) of the column weight, we discard fewer
than \(4\eta q_2/\mu\) such indices. Let \(B_j\) be the union of these two
sets of discarded indices and put \(b_j:=|B_j|\). The inequalities in
\eqref{eq:width-margins} give
\[
\sum_{j=1}^s b_j<8sq_1+4s\eta q_2/\mu<\zeta q_2/4.
\]
Since \(\sum_j|U_j|\ge\zeta q_2\), some \(j\) satisfies \(|U_j|>4b_j\). Fix this
\(j\), write \(b:=b_j\), and define \(w_0:=|U_j\setminus B_j|\). Then
\(w_0\ge|U_j|-b>3b\).

Fix a collision index \(\alpha\in U_j\setminus B_j\). Let
\(i_1,i_2\) be two active items and \(i_3\) an inactive item in one bin at
\(\alpha\), as given by \cref{src:item-witness}. We choose retained template
rows for the first \(q_1\) projected blocks and retained rows of \(i_1,i_2,i_3\).
By the entry rule in \cref{def:source-matrix}, against a
$\mathcal D$-input with label \((\alpha,\gamma)\) these rows read the
following entries, where \(u\) is the base row of the respective row:
\[
R(u,\gamma_\ell)\quad(\ell\in[q_1]),\qquad
R(u,\gamma_{q_1+\pi_\alpha(i_1)}),\qquad
R(u,\gamma_{q_1+\pi_\alpha(i_2)}),\qquad R(u,\gamma_{q_1+s+1}).
\]
The three tail positions are distinct by the collision definition and the tail
convention in \cref{def:source-matrix}. At each of these positions, the selected rows have base row weight at
least \(x_{\rm cap}\), for template positions because \(\alpha\notin B_j\), and for
item positions because of the threshold in the item assignment above. The selected rows depend only on the
projected labels and not on \(\gamma\), so they form a restriction of the
interlace of \(R\) on \(\Gamma\) in \cref{src:profiles}. The collision bound of
\cref{src:profiles} therefore requires \(H_1+1\) bits whenever the retained
$\mathcal D$-inputs at index \(\alpha\) have weight greater than \(1/2\) in \(\Gamma\).

We now apply \cref{lem:collision-tracking} with $q=q_2$ and $H=H_1$.
Its $q_2-b$ indices are those outside $B_j$, and its $w_0>3b$ marked sets are
those at the indices in $U_j\setminus B_j$. The numerical bounds
\eqref{eq:width-margins} give $q_2x\le10^{-4}$, hence $2q_2x\le1/8$.
We now verify its hypotheses (i), (ii) and (iii).

Until Alice speaks again, the set $A_j$ of template labels is unchanged. Its projection onto
each coordinate $\alpha$ has weight at least $\Pr(A_j)>998/1000>3/5$ in $J$, so (i)
follows from $\RP{A}{x}{H_1}$ (\cref{obs:representatives}).
At the first Alice bit, one child retains a part of $A_j$ of weight greater than
$499/1000>9/25$. After transposition, the entries of that child at indices with
retained list sets are $A(\gamma,z_\alpha)$, so they form the interlace of $A$ with
those sets of lists as row blocks and tuple list $S$. Since $S$ is exactly
$8$-wise independent and this child retains template-label weight greater
than $9/25$, \cref{det:exact} with $M=A$, $d=H_1$ and $2^a=u$ gives cost at
least $H_1+\log u$ for any $u/2+1$ retained index sets of weight at least
$ux$. This gives (ii) for the template labels retained at that child. The collision bound
established above gives (iii), since Alice's inputs are unchanged before that bit. Hence
\cref{lem:collision-tracking} excludes depth $C-1$.
The upper-bound protocol of \cref{src:upper} has depth $C$, which proves
equality.
\end{proof}

\subsection{Lower bounds after restricting tuple inputs}
\label{sec:retained-tuples}

Restricting the tuple inputs can destroy the independence among their
coordinates (\cref{sec:overview}), so we use independence only under the original
tuple weights. For $0<v,\tau\le1$, the next lemma shows that, after
conditioning on any event of weight at least $v$, every joint marginal on at most eight coordinates
is within total variation distance $\varepsilon(v,\tau)$ of the product
weights, provided these coordinates avoid a set of at most $e(v,\tau)$
exceptional coordinates. Set
\[
\begin{aligned}
\varepsilon(v,\tau)&:=\min\{\tau,2^{-30}v^{1/256}\},\\
e(v,\tau)&:=8\ceil{2^{20}\varepsilon(v,\tau)^{-2}v^{-1/512}},
\qquad e(v):=e(v,1).
\end{aligned}
\]
We use independence order $8192=8\cdot1024$ so that any $1024$ events on
pairwise disjoint sets of at most eight coordinates are independent under
the original weights, as a $1024$th-moment bound requires.

\begin{lemma}[Conditioning outside exceptional coordinates]\label{det:conditioning}
Let $\nu$ be the normalized weights of an indexed exactly $8192$-wise independent tuple list
(\cref{def:exact-tuples}) on $Z^q$ with marginal $\mu$, and let $0<v,\tau\le1$. For every event
$Y$ of labels with $\nu(Y)\ge v$, there is a set $E\subseteq[q]$, with
$|E|\le e(v,\tau)$, such that
\[
d_{\rm TV}\bigl((\nu\mid Y)_J,\mu^J\bigr)\le\varepsilon(v,\tau)
\qquad
(J\subseteq[q]\setminus E,\ |J|\le8).
\]
Here total variation is the maximum absolute difference between the two
distributions' probabilities of an event. The reduction need not compute the
event or the exceptional set.
\end{lemma}

The proof is in \cref{sec:conditioning-moments}.

Combined with \cref{obs:representatives}, \cref{det:conditioning} bounds the column weight of
projected restrictions after conditioning. If $K\subseteq Y$ and $J$
consists of at most eight coordinates outside $E$, then the inverse image of
$\pi_JK$ contains $K$, so
\[
\mu^J(\pi_JK)
\ge(\nu\mid Y)_J(\pi_JK)-\varepsilon(v,\tau)
\ge(\nu\mid Y)(K)-\varepsilon(v,\tau).
\]
Thus the projected restriction of \cref{obs:representatives} has column weight
at least $(\nu\mid Y)(K)-\varepsilon(v,\tau)$, and we need no product assumption on
$\nu\mid Y$.

The next two results apply to full products and to explicit lists. Both
are proved in \cref{sec:surviving-block-proofs}
(p.~\pageref{proof:det:exact}).

\begin{lemma}[Surviving blocks in a tuple list]\label{det:exact}
Suppose $d\ge2$ and $\RP{M}{x}{d}$. Let an interlace of $M$ use a tuple list
that is exactly $8$-wise independent with the column weights of $M$ as marginal. If an integer
$a\ge1$ satisfies $2^ax\le1$, and a restriction retains at least
$2^{a-1}+1$ blocks, each of base row weight at least $2^ax$, together with a
label set of total weight at least $9/25$, then the restricted
interlace has cost at least $d+a$.
\end{lemma}

After a protocol prefix, the retained tuple labels may weigh less than the
$9/25$ required by \cref{det:exact}. In that case we condition on the event $Y$
of retained labels and discard its exceptional coordinates (\cref{det:conditioning}).

\begin{theorem}[Surviving blocks after conditioning]\label{det:rareboost}
Suppose $d\ge2$ and $\RP{M}{x}{d}$, and let an interlace of $M$ use an exactly $8192$-wise independent
tuple list with weights $\nu$ and the column weights of $M$ as marginal. Let $\nu(Y)\ge v>0$,
choose the exceptional set from \cref{det:conditioning} with any error bound
$0<\tau\le1$, and let an integer $a\ge1$ satisfy $2^ax\le1$. If a restriction
retains at least $2^{a-1}+1$ blocks indexed outside that exceptional set, each of base row weight at least
$2^ax$, and a column set $K\subseteq Y$ with $(\nu\mid Y)(K)\ge1/2$, then
its cost is at least $d+a$.
\end{theorem}

\subsection{Where displaced inputs can go}\label{sec:internal-interlace}

We now prove a lower bound for restrictions of $U=\Int{w}{N}$ that keep
almost all rows but few tuple inputs, following the
copies of $N$ through a hypothetical short protocol as in \cref{sec:overview}.
The argument applies to both constructions. Each copy has row weight $1/w$
in $U$, but we measure row weight in units of one complete copy, and losses
among a copy's $\Omega$-rows in weights normalized within $\Omega$. Recall that
$\Omega$ has weight $\lambda=255/256$ and $\mathcal D$ has weight $\mu=1/256$.

\paragraph{Common hypotheses.}
In this section we use only the following hypotheses, where $H_1,H_2,C$ are as in
\eqref{eq:source-costs} and $H_1,H_2\ge2$.
\begin{enumerate}
\item[(H1)] Each copy satisfies the deletion bounds $\RP{A}{x}{H_1}$,
$\RP{R_{\rm temp}}{x}{H_2}$ and $\RP{N_{\rm ref}}{x}{C-1}$.
\item[(H2)] The matrix $N$ has exact cost $C$, including on every restriction
retaining $1-\eta$ of each input set's total weight.
\item[(H3)] The weights on $S$ give exact $8192$-wise independence across its
$J$-valued coordinates, and the bin tuple in $\Omega$ has the same
independence order and the prescribed $J_2$ weights as marginal.
The tuple weights $\nu$ of $U$ have
the original $\mathcal K$ weights at each coordinate and arise from three mutually independent
label sequences, namely type/bin/item labels, base labels with the $X$ weights, and
template labels with the $S$ weights. Each sequence has a value at
every copy and is exactly $8192$-wise independent across copies. At copy $i$,
with template label $z_i$ and base label $u_i$, the decoder outputs
the template row $(p,z_i)$ if the first sequence gives a template type with bin $p$,
and the item row $(a,u_i)$ if it gives item $a$.
Thus, on any set of at most $8192$ copies, the decoded labels have the product of the original $\mathcal K$ weights.
The tuple coordinates \(z_{i,\alpha}\), indexed by copy $i$ and coordinate
$\alpha$, are also exactly $8192$-wise independent: group at most
$8192$ coordinates by copy, then use independence between copies and among
the tuple coordinates within each template input. By independence, fixing some of the three complete
sequences leaves the remaining ones with their original weights. The proof of
\cref{res:incoming} fixes the first two sequences by averaging and then
conditions on the retained template labels.
\item[(H4)] The widths and deletion thresholds satisfy the numerical bounds in
\eqref{eq:width-margins}: each quantity in the first column is below the
corresponding bound in the last column. These bounds provide enough coordinates
to encode the graph and ensure that, after deleting inputs and setting aside
exceptional coordinates, enough inputs remain for the lower-bound arguments.
\end{enumerate}
In both cases, (H1) follows from \cref{src:profiles} and (H2) from
\cref{src:near-full}. \Cref{lem:common-schedule} verifies (H4) for the parameter
choices of \cref{def:amplification-parameters}. For (H3) with full products, we take all
complete sequences, including values ignored by the decoder, with full product
weights. Decoding then induces the full product of the original
weights on $\mathcal K^w$, so pulling back along the decoder preserves the weights of retained
sets and the entries and depth of protocols. This representation is used
only in the analysis and is never enumerated. In the explicit case the complete sequences are the
actual indexed input labels (\cref{det:construction,det:flat}).

By \cref{det:conditioning} with $v=\theta$ and $\tau=\eta/2$, conditioning on
a tuple event of weight at least $\theta$ makes at most $e(\theta,\eta/2)$
copies exceptional. Write $E_{\rm exc}=2e(\theta,\eta/2)$, so that at most
$E_{\rm exc}/2$ copies are exceptional. The reduction need not identify them.

We stop each root-to-leaf path of the protocol just before Bob's first bit or just after Alice's
$(\log w)$-th bit, whichever comes first. Up to these
\emph{stopping nodes} only Alice has spoken, so each row follows the same path
for every remaining column. By following the $\Omega$-rows, we assign distinct
stopping nodes to all but at most $E_{\rm exc}$ copies (\cref{res:destinations}).

\begin{lemma}\label{res:destinations}
Under the common hypotheses above, suppose a restriction of $U$ retains row
weight at least $1-\delta_U$ and a tuple event $Y'$ of original weight at least
$\theta$, and has a protocol of depth at most $C+\log w-1$.
Apply \cref{det:conditioning} to $Y'$ with $v=\theta$ and error bound $\eta/2$, and set aside
the at most $E_{\rm exc}/2$ exceptional copies. At an Alice node of depth $r$, put
$\kappa=w2^{-r}$. Follow each tracked set of $\Omega$-rows of one remaining copy,
of normalized weight at least $2\kappa x$, into every child in which it
retains normalized weight at least $\kappa x$. An unsplit row set
contributes one. Once a row set first splits into two children meeting the
threshold, each descendant contributes one-half. We add these
contributions at each node. They count row sets, not input weight.

Stop at the first node where Bob speaks or at depth $\log w$, whichever comes first.
The latter stopping condition applies regardless of the next speaker.
Paths that end earlier carry no row sets, and we extend them to depth
$\log w$ by fixed Alice bits. At a node where Bob
first speaks with $\kappa\ge2$, the total contribution is at most $\kappa/2$.
A node at depth $\log w$ has at most one row set. At least $w-E_{\rm exc}$
nodes at that depth are reached by unsplit row sets from distinct copies. Assign each
such node to the copy of its row set. Over all other stopping nodes,
the sum of $\kappa$ is at most $E_{\rm exc}$.
\end{lemma}
\begin{proof}
Since $\delta_U=\eta/(16w)$, initial deletion removes at most
$w\delta_U=\eta/16$ in copy units, hence at most $\eta/16$ of any
single copy, so every copy starts with normalized weight on $\Omega$ at least
$1-\eta/(16\lambda)>2wx$, by \eqref{eq:width-margins}. A row set always has a child retaining at
least half its input weight. As with the marks in the tracking rule after
\cref{lem:kraft-count}, the first split of a row set preserves the total count,
since $1=\frac12+\frac12$, and further splits can only increase it. The descendants
of one row set lie in disjoint subtrees, so each node holds at most one row set
from each copy. Each of the at least $w-E_{\rm exc}/2$ copies outside the
exceptional set starts with an unsplit row set, so the total count at the
stopping nodes is at least $w-E_{\rm exc}/2$.

Until Bob speaks, the tuple set is the fixed $Y'$. For every non-exceptional copy $i$,
the projection of $Y'$ onto coordinate $i$ has original $\mathcal K$ weight at least $1-\eta/2>3/5$
(the display after \cref{det:conditioning} with $K=Y=Y'$), and together with
the $\Omega$-rows of copy $i$ it gives a restriction of $N_{\rm ref}$
(\cref{obs:representatives}). Since a row set
has row weight $\ge2x$, the bound
$\RP{N_{\rm ref}}{x}{C-1}$ therefore requires at least $C-1>0$ further bits,
and no row set reaches a leaf.
After we pad paths without row sets, every root-to-leaf path has one
stopping node, so the sum of $\kappa$ over these nodes is $w$.

At a node where Bob first speaks with $\kappa\ge2$, suppose there were at
least $\kappa/2+1$ row sets. They come from distinct copies whose coordinates
lie outside the exceptional set for the same event $Y'$. The Bob child retaining more column weight has conditional weight at least one-half.
We apply \cref{det:rareboost} with $M=N_{\rm ref}$, $d=C-1$, $a=\log\kappa$,
and $K$ the column set of this child. Here $a$ is an integer since $w$ is
a power of two, $\kappa x\le wx\le1$, and
each row set has weight at least $2\kappa x\ge\kappa x$. The child requires
$C-1+\log\kappa$ bits, but only $C+\log\kappa-2$ remain.
Consequently there are at most $\kappa/2$ row sets, and their total
contribution is at most $\kappa/2$.

At a node at depth $\log w$, two row sets of weight at least $2x$ would require $C$ bits by the same theorem
with $a=1$ and $K=Y'$, but only $C-1$ bits remain.
Thus a node at depth $\log w$ has at most one row set. If the set is unsplit, we
assign the node to the copy supplying it. Otherwise, the set's
contribution is at most one-half. Let $c$ be the number of assigned nodes.
The total contribution at every other stopping node is at most half its value
of $\kappa$. Since each assigned node has $\kappa=1$ and the values of $\kappa$ sum to $w$,
\[
 w-E_{\rm exc}/2\le c+\tfrac12(w-c),\qquad\text{hence}\qquad w-c\le E_{\rm exc}.
\]
An unsplit row set follows only one path,
so the assigned nodes correspond to distinct copies. The sum of $\kappa$
over all other stopping nodes is $w-c\le E_{\rm exc}$.
\end{proof}

A surviving row of an assigned copy that misses its copy's assigned node
reaches either an unassigned stopping node or a node assigned to another copy.
\Cref{res:incoming} bounds the weight arriving at unassigned nodes, and
\cref{res:foreign} the weight arriving at nodes assigned to other copies.
Their proofs are in
\cref{sec:incoming-bound-proof,sec:foreign-bound-proof}, respectively.

\begin{lemma}\label{res:incoming}
For the protocol and stopping nodes of \cref{res:destinations}, the total
$\mathcal D$-weight, over all copies, reaching a stopping node of depth $r$ where Bob first
speaks before depth $\log w$ is less than $2s\kappa$, where $\kappa=w2^{-r}$.
At a stopping node at depth $\log w$ this weight is less than $2s$.
Weight is measured in units of one complete copy of $N$.
\end{lemma}

\begin{lemma}\label{res:foreign}
In the same protocol, for each copy $i$ with an assigned node, let
$\varepsilon_i$ be the normalized weight of copy $i$'s $\Omega$-rows missing
at that node. For every
other copy $j$ with an assigned node, the $\mathcal D$-rows of copy $j$ reaching
the node assigned to copy $i$ have total weight at most $\mu\delta_i$, where
\[
 \delta_i=2q_2x+4\varepsilon_i.
\]
\end{lemma}

The proof of \cref{res:alignment} combines these bounds into the contradiction shown in
\cref{fig:reference-destinations}, in which $P_i$ is the $\mathcal D$-weight of
copy $i$ surviving outside its assigned node and $a_i$ is its initially
deleted $\mathcal D$-weight.

\begin{figure}[H]
\centering
\begin{tikzpicture}[
  >=Stealth,font=\footnotesize,
  box/.style={draw,rounded corners=2pt,align=center,inner sep=4pt,
    text width=33mm,minimum height=13mm},
  flow/.style={->,semithick}]
\node[box] (origin) at (0,0) {Rows from $\mathcal D$\\in copies with assigned nodes};
\node[box,draw=black!55] (deleted) at (0,-2)
  {Initially deleted\\$\sum_i a_i\le\eta/16$};
\node[box,draw=black!55] (own) at (4.9,1)
  {Stay at their assigned node,\\not counted in $P_i$};
\node[box,fill=aliceblue!6,draw=aliceblue] (outgoing) at (4.9,-1)
  {Reach other nodes\\[3pt] $\sum_iP_i>\eta w/2$};
\node[box] (other) at (9.8,0)
  {Other assigned nodes\\total $\le\eta w/32$};
\node[box] (exceptional) at (9.8,-2)
  {Nodes not assigned to a copy\\total $\le\eta w/16$};
\coordinate (firstsplit) at (2.45,0);
\draw[semithick] (origin.east) -- (firstsplit);
\draw[flow,black!55] (origin.south) -- (deleted.north);
\draw[flow,black!55] (firstsplit) |- (own.west);
\draw[flow] (firstsplit) |- (outgoing.west);
\coordinate (secondsplit) at (7.35,-1);
\draw[semithick] (outgoing.east) -- (secondsplit);
\draw[flow] (secondsplit) |- (other.west);
\draw[flow] (secondsplit) |- (exceptional.west);
\node[draw,rounded corners=2pt,align=center,inner sep=4pt,fill=black!4,
  text width=131mm] at (4.9,-3.75)
  {\textbf{There is nowhere for all the outgoing weight to go.}\\[3pt]
   Required outgoing weight $>\eta w/2
   \;>\;3\eta w/32=\eta w/32+\eta w/16$
   (total receiving bound).};
\end{tikzpicture}
\caption{Destinations of the displaced $\mathcal D$-weight. Every surviving $\mathcal D$-row outside the node assigned to its copy must reach one of the two
groups of stopping nodes on the right, whose combined bound on arriving weight
is smaller than the outgoing weight. The lower group contains the nodes not assigned to copies, either
where Bob first speaks before depth $\log w$ or at that depth. Rows deleted
initially and rows staying at the node assigned to their copy do not count toward $P_i$.
All weights are in
complete-copy units, and arrow widths are not to scale.}
\label{fig:reference-destinations}
\end{figure}

The deletion bounds for $W$ in the next theorem are one bit below its full cost,
and the next subsection recovers that bit.

\begin{theorem}[Lower bounds for $W$]\label{res:alignment}
Under the common hypotheses of \cref{sec:internal-interlace},
\[
 \D(W)=C+\log w,\qquad \RP{W}{\theta}{C+\log w-1}.
\]
Every restriction of $W$ retaining row weight at least $\theta$ and column
weight at least $1-\delta_U$ has cost at least $C+\log w$.
\end{theorem}
\begin{proof}
Suppose a restriction of $W$ with row weight $\ge\theta$ and column
weight $\ge1-\delta_U$ has a protocol of depth $\le C+\log w-1$.
Transposing gives a restriction of $U$ (recall $W=U^{\mathsf T}$) with a protocol of the same depth,
retained row weight at least $1-\delta_U$, and a retained tuple event $Y'$ of weight at least $\theta$.
By \cref{res:destinations} and $E_{\rm exc}<w/4$ from \cref{lem:common-schedule},
at least $w-E_{\rm exc}\ge3w/4$ copies
have assigned nodes, each with only $C-1$ bits left. These copies lie outside
the exceptional set, so conditioning on $Y'$ with
error bound $\eta/2$ gives each of them a column projection of original
weight at least $1-\eta/2$, which gives a restriction of $N$
(\cref{obs:representatives}). By \cref{src:near-full}, the assigned node must miss more than
$\eta$ of the row weight in its assigned copy.

An unsplit set of $\Omega$-rows discards less than
$\sum_{r=0}^{\log w-1}wx2^{-r}=2(w-1)x$ on its path, since at depth $r$
the discarded child retains less than $\kappa x=wx2^{-r}$.
Initial deletion removes at most $\eta/16$ complete copies. Summing over copies
with assigned nodes and using $w^2x/\eta\le1/(32\lambda)$ from
\eqref{eq:width-margins}, we obtain
\[
 \sum_i\varepsilon_i\le\frac{\eta}{16\lambda}+2w(w-1)x
 \le\frac{\eta}{8\lambda}.
\]
For these copies, the missing $\Omega$-rows, initially
deleted $\mathcal D$-rows, and $\mathcal D$-rows surviving elsewhere partition
the part absent from their own assigned node. Thus
$P_i+a_i+\lambda\varepsilon_i>\eta$, and
\[
\sum_iP_i>\frac{3\eta w}{4}-\frac{\eta}{16}-\frac{\eta}{8}
>\frac{\eta w}{2}.
\]
We now show that the other stopping nodes cannot receive this much weight. The sum of $\kappa$
over nodes not assigned to copies is at most $E_{\rm exc}$, so
\cref{res:incoming} bounds their arriving
weight by $2sE_{\rm exc}\le\eta w/16$. By \cref{res:foreign}, other assigned nodes receive
at most
\[
\mu w\sum_i(2q_2x+4\varepsilon_i)
\le\eta w\left(\frac1{64}+\frac1{510}\right)<\frac{\eta w}{32},
\]
using $4\mu wq_2x\le\eta/32$ and the bound on losses within $\Omega$ above.
Since only Alice has spoken, every surviving row reaches one stopping node.
The total arriving weight is therefore less than $3\eta w/32$, contradicting
$\sum_iP_i>\eta w/2$. This proves the restriction bound.

For the deletion bounds $\RP{W}{\theta}{C+\log w-1}$, we fix $j\in\{0,1\}$ and transpose a restriction of $W$
with row weight at least $\theta$ and column weight at least
$\beta_j=(3/5)2^{-j}$. In $U$ its tuple event $Y$ has weight at least
$\theta$. We discard the $\mathcal D$-rows, leaving a restriction of
the interlace of $N_{\rm ref}=N[\Omega,\mathcal K]$. Since the $\mathcal D$-rows
have weight $1-\lambda$, the retained weight on $\Omega$, normalized within $\Omega$
and averaged over the copies, is at least
$\beta'_j=(\beta_j-(1-\lambda))/\lambda$.
We put $\kappa=w2^{-j}$ and apply \cref{det:conditioning} to $Y$ with $v=\theta$ and
$\tau=1$, which leaves at most $e(\theta)$ exceptional copies. Let $m$ be the number of
non-exceptional copies retaining normalized weight on $\Omega$ at least $\kappa x$.
The block count of \cref{sec:models} gives $w\beta'_j\le m+e(\theta)+w\kappa x$, and
$wx\le1/256$ and $4e(\theta)/w<1/32$ (\cref{lem:common-schedule}) then give
\[
 \frac m\kappa\ge2^j\beta'_j-wx-\frac{4e(\theta)}w
 \ge\frac{(3/5)-4(1-\lambda)}\lambda-\frac1{256}-\frac1{32}
 =\frac{44}{75}-\frac1{256}-\frac1{32}>\frac12.
\]
Since $\kappa/2$ is an integer, at least $\kappa/2+1$ blocks indexed outside the exceptional set remain. We apply
\cref{det:rareboost} to $N_{\rm ref}$ with $a=\log\kappa$ and $K=Y$,
which has conditional weight one. The parameter choices give $\kappa\ge2$ and
$\kappa x\le1$. The lower bound is
$C-1+\log\kappa=C+\log w-1-j$, which proves the deletion bounds for both values of $j$ after transposition.
Applied to the whole matrix, the restriction bound proved first gives $\D(W)\ge C+\log w$.
Conversely, the player holding the copy index sends it in $\log w$ bits, and
the players then run the cost-$C$ protocol for $N$, giving equality.
\end{proof}

\subsection{Restoring the bound for the next round}

To use $R^+$ as the next base matrix, we need deletion bounds at its exact
cost, whereas \cref{res:alignment} gives them for $W$ one bit lower. For full
products, \cref{cl:bridge} recovers this bit from the full-cost bound of
\cref{res:alignment} for column weight at least $1-\delta_U$. Its proof
(p.~\pageref{proof:cl:bridge}) uses the induction of \cref{sec:interlace-induction}.

\begin{theorem}[Restoring the deletion bounds]\label{cl:bridge}
Let $W$ have $m\ge1$ rows, $n\ge1$ columns and $\D(W)\ge3$. Let $\theta>0$ and $0<\delta\le1/64$. Assume that every
restriction retaining row weight at least $\theta$ and column weight at least
$1-\delta$ has cost at least $\D(W)$, and that $\RP{W}{\theta}{\D(W)-1}$ holds.
Let $b\ge1$ be an integer and $Q=2^r$ a power of two with
$Q\delta\ge80b$ and $Q\theta\le1/20$. Then $R^+=(\Int{Q}{W})^{\mathsf T}$
satisfies
\[
\D(R^+)=\D(W)+r,\qquad \RP{R^+}{2^{-b}}{\D(W)+r}.
\]
Independently of the lower-bound hypotheses, $\D(W)\le y$ implies
$\D(R^+)\le y+r$. The matrix $R^+$ has exactly $n^Q$ rows
and $Qm$ columns.
\end{theorem}

The sample size in \cref{thm:compression} grows with the rank of the matrix.
The next lemma bounds the rank increase of one round in both cases.

\begin{lemma}\label{lem:amplifier-rank}
In the construction of \cref{def:source-matrix}, let
$h=2\log s+\log q_2+\log w+\log Q$ and suppose
$|\mathcal I|\le2sq_2$. Then $\rank(R^+)\le2^h\rank(R)$.
The same bound holds if any product set of tuples is replaced by an
indexed list of tuples from that set, or by a product of such lists.
\end{lemma}
\begin{proof}
A width-$t$ interlace stacks $t$ row blocks, each obtained by selecting or
repeating columns of its base matrix, so its rank is at most $t$ times the
base rank.
The template rows form nested interlaces of widths $q_1$, $q_2$ and $s$,
so the template part has rank at most $sq_1q_2\rank(R)$.
For each item $i$, the rows $\{i\}\times X$ have entries $R(u,\psi_i(Z))$, where
the column $\psi_i(Z)$, namely $\xi_{\pi_\alpha(i)}$ or $\xi_{s+1}$, is independent of $u$, so
they have rank at most $\rank(R)$.
After the outer interlaces, rank subadditivity gives
\[
\rank(R^+)\le wQ(sq_1q_2+|\mathcal I|)\rank(R)
\le wQsq_2(q_1+2)\rank(R)=2^h\rank(R),
\]
using $q_1+2=s$. On indexed lists every block still selects or repeats
base columns, and transposition preserves rank, so the count is unchanged.
\end{proof}

\begin{proof}[Proof of \cref{thm:amplifier}]
\phantomsection\label{proof:thm:amplifier}
As in \cref{src:profiles}, we suppress the subscript $\rm g$ in the explicit
case. In both cases $h=2\log s+\log q_2+\log w+\log Q$. For $n$ beyond the
effective cutoff of \cref{lem:full-schedule}, that lemma gives $h=L+O_{\Delta,\gamma}(\log L)$ in the
full-product case, and \cref{lem:generated-schedule} gives
$h=40\log s=10\log\mathsf N$ in the explicit case.

We first prove (ii), so let the graph be NO, $\D(R)=d\ge3$ and
$\RP{R}{\rho}{d}$. By \cref{src:profiles}, the deletion bounds for $A$,
$R_{\rm temp}$ and $N_{\rm ref}$ and the collision bound hold. The construction has
the exact marginals required by \cref{src:near-full} (in the explicit case,
those of $S_{\rm g}$ and $\mathcal J_{\rm g}$ in \cref{det:construction}), and
\cref{lem:common-schedule} gives the numerical bounds \eqref{eq:width-margins}.
Hence \cref{src:near-full} applies, and $N$ has cost $C$ even after we delete at most
$\eta$ of each input set's weight. This gives (H1) and (H2) of
\cref{sec:internal-interlace}, and \cref{lem:common-schedule} gives (H4). For full
products, (H3) is verified in \cref{sec:internal-interlace}. In the
explicit case, \cref{det:construction} provides the three independent
sequences. \Cref{det:flat} shows that decoding the nested template labels gives
the exact $8192$-wise independence of (H3), and that fixing the other two
sequences leaves it unchanged.
Thus, by (H1)--(H4), \cref{res:alignment} gives $\D(W)=C+\log w$ and
$\RP{W}{\theta}{C+\log w-1}$, and cost $\ge C+\log w$ on every restriction
with row weight $\ge\theta$ and column weight $\ge1-\delta_U$.
The two cases differ only in the restoration step.
\begin{itemize}
\item \emph{Full-product case.} By \cref{lem:full-schedule},
\(0<\delta_U\le1/64\), \(Q\delta_U\ge80L\), and \(Q\theta=1/512<1/20\), so
\cref{cl:bridge} applies with \(b=L\) and \(\delta=\delta_U\), at threshold
$2^{-L}=\rho$.
\item \emph{Explicit case.} \Cref{det:restore} applies with
$(\phi,\delta,Q)=(\rho,\delta_U,Q)$: its hypotheses on $W$ are the bounds above,
\cref{lem:generated-schedule} verifies its numerical hypotheses, and
\cref{det:construction} uses an exactly $8192$-wise independent outer list with the
column weights of $W$.
\end{itemize}
In both cases we restore the deletion bounds at threshold $\rho$ and cost
\[
C+\log w+\log Q
=d+\underbrace{2\log s+\log q_2+\log w+\log Q}_{h}+1,
\]
so $\D(R^+)=d+h+1$ and $\RP{R^+}{\rho}{d+h+1}$.

For (i), on either promise, at each of the two outer interlaces the player
holding the copy index sends it, in $\log w+\log Q$ bits, and the
players then run the protocol of \cref{src:upper}, of cost
$\D(R)+2\log s+\log q_2+1$. This gives \(\D(R^+)\le\D(R)+h+1\). On a YES graph
\cref{src:upper} omits the final $+1$, giving \(\D(R^+)\le\D(R)+h\). In the explicit
case each input decodes locally to a full-product input with the same entry,
and the protocol uses only decoded inputs, so it applies unchanged.

For (iii), the bound $q_2\ge4s|E|$ from \eqref{eq:width-margins} and
$|E|\ge n/2$ imply $|\mathcal I|=4s(n+s-3)\le2sq_2$, so
\cref{lem:amplifier-rank} gives the rank bound in both cases.
\end{proof}


\section{Interlacing on small input lists}
\label{sec:small-lists}
\subsection{Constructing exact marginals}

The deterministic reduction replaces each full product $Y^q$ by an explicit tuple
list (\cref{def:exact-tuples}) whose induced weights on every set of at most $8192$
coordinates equal the product weights.

We build the tuples one coordinate at a time. To each tuple we attach the
zero--one vector indicating, for each set $J$ of at most $8192$ coordinates and assignment $a$ on $J$, whether the tuple agrees with $a$.
After each extension, we move weight along linear dependences
among these vectors. Each move removes at least one tuple, keeps all weights
nonnegative and leaves the joint weights on these coordinate sets unchanged.
The moves continue until no more tuples remain than there are pairs $(J,a)$. This is the sequential
construction idea of \citet[Section~3]{koller1994sample}. The proof of the
next lemma, including the implementation and bit bound, is in
\cref{sec:support-algorithm}. We store the rational weights
explicitly and use them to measure retained input sets.

\begin{lemma}[Small rational tuple lists]\label{det:code}
Fix an integer independence order $\mathsf D\ge1$. Given normalized rational weights $\mu$ on an alphabet of
size $m$ and an integer $q\ge1$, a deterministic algorithm constructs an exactly
$\mathsf D$-wise independent tuple list $\mathsf{Code}_{\mathsf D}(q,\mu)$ (\cref{def:exact-tuples}) with
a label set of size at most $(1+qm)^{\mathsf D}$.
For fixed $\mathsf D$, the construction time, total output length and
rational-weight bit lengths are polynomial in $q$, $m$ and the binary encoding
length of $\mu$.
\end{lemma}

Below we always use $\mathsf{Code}_{8192}$.
Keeping the weights as binary-encoded rational numbers avoids expanding them
into repeated labels. A denominator with polynomially many bits could require
far too many copies to print in polynomial time.

\subsection{Restoring the deletion bounds}

The matrix $W$ of \cref{res:alignment} satisfies the deletion bounds one bit below
its cost, and has full cost on restrictions retaining almost all column weight. One more interlace
recovers the missing bit. The proof distinguishes whether Bob speaks in a protocol prefix. If not, the bound for almost all columns applies. Otherwise
\cref{det:suffix} gives the missing bit.

\begin{theorem}\label{det:restore}
Let $0<\phi\le1$, $0<\delta\le1/64$ and $\theta>0$. Suppose $W$, with input weights
$\mu_X,\mu_Y$, satisfies $\D(W)=C\ge3$, $\RP{W}{\theta}{C-1}$, and
\[
\D(W[A,B])\ge C \quad\text{whenever }\mu_X(A)\ge\theta
\text{ and }\mu_Y(B)\ge1-\delta.
\]
Let $Q$ be a power of two satisfying
\[
Q\ge128\bigl(e(\phi,\delta/2)+1\bigr), \qquad Q\theta\le1/20,
\]
where $e(\cdot,\cdot)$, defined before \cref{det:conditioning}, bounds the
number of exceptional coordinates in that lemma.
Form the $Q$-fold interlace of $W$ on any exactly $8192$-wise independent tuple list with
marginal $\mu_Y$, and transpose it to obtain $R^+$. Then
\[
\D(R^+)=C+\log Q, \qquad \RP{R^+}{\phi}{C+\log Q}.
\]
Independently of the lower-bound hypotheses, a protocol of depth $U$ for $W$
yields one of depth $U+\log Q$ for $R^+$.
\end{theorem}
\begin{proof}
To prove the lower bounds, we fix $j\in\{0,1\}$ and a restriction of $R^+$ retaining row weight
$\ge\phi$ and column weight $\ge(3/5)2^{-j}$. After undoing the transpose, the retained
columns form a label event $Y$ of tuple-list weight $\nu(Y)\ge\phi$, and the within-block
weights of the retained rows average at least $(3/5)2^{-j}$ over the $Q$ blocks. We apply
\cref{det:conditioning} to $Y$ with $v=\phi$ and error bound $\delta/2$ and put
\[
e=e(\phi,\delta/2),\qquad k=Q2^{-j},\qquad a=\log k.
\]
Since $Q$ is a power of two with $Q\ge128(e+1)>2$, $a$ is a positive integer. More than $k/2$ blocks
are indexed outside the exceptional set and retain base-row weight
at least $k\theta$. Otherwise the
block-counting bound of \cref{sec:models} with $e$ blocks set aside gives
$(3/5)Q2^{-j}\le k/2+e+Qk\theta$, and hence
\[
\frac35\le\frac12+\frac{2^je}{Q}+Q\theta
\le\frac12+\frac1{64}+\frac1{20}<\frac35.
\]
We retain these blocks and all of $Y$.

Suppose the restriction to these blocks and $Y$ has a protocol of depth at most $C+a-1$.
We follow the protocol for $a$
messages, starting from conditional column weight one, by the weight-halving
walk from the proof of \cref{det:rareboost}, which there runs for $a-1$ messages. At each Alice
message, we assign every current block to a child retaining at least half its row
weight and keep a child assigned at least half the blocks. At each Bob message, we
keep a child retaining at least half the current column weight under $\nu\mid Y$.
If the protocol ends earlier, we pad with dummy Alice messages. If $\alpha$ messages are Alice's and $\beta$
are Bob's, then $\alpha+\beta=a$, and the surviving restriction has a column set $K'\subseteq Y$ and
\[
\begin{gathered}
\text{more than }2^{\beta-1}\text{ blocks outside the exceptional set},\qquad
\text{row weight per block at least }2^\beta\theta,\\
(\nu\mid Y)(K')\ge2^{-\beta},\qquad
\text{at most }C-1\text{ bits remaining}.
\end{gathered}
\]
If $\beta=0$, the columns are still all of $Y$. At least one block indexed
outside the exceptional set remains, and its row weight is at least $\theta$. The projection of $Y$ onto its coordinate has base
weight at least $1-\varepsilon(\phi,\delta/2)\ge1-\delta$, by
\cref{det:conditioning}. Choosing retained representatives
(\cref{obs:representatives}), we obtain a restriction of $W$ that requires $C$ bits by the hypothesis for row weight $\ge\theta$ and column weight $\ge1-\delta$.

If $\beta\ge1$, we apply the short-continuation bound \cref{det:suffix} with
$d=C-1$, $b=\beta-1$, and $x=\theta$, under hypothesis (C) with $v=\phi$ and $\tau=\delta/2$.
Its hypotheses hold. The bound $\RP{W}{\theta}{C-1}$ is assumed, the blocks avoid the exceptional set,
and the number of blocks is an integer exceeding
$2^{\beta-1}$, so it is at least $2^b+1$. Each block has row
weight at least $2^{b+1}x$, and $(\nu\mid Y)(K')\ge2^{-b-1}$. Also $d\ge2$ and
$2^{b+1}x=2^\beta\theta\le k\theta\le1$. The lemma therefore requires $d+1=C$ bits.
In both cases this exceeds the $C-1$ remaining bits.

Thus the restriction to these blocks and $Y$ requires at least $C+a=C+\log Q-j$ bits. Transposing
back proves both deletion bounds and, for $j=0$, the lower bound on the whole matrix. For the
upper bound, Alice sends the block index $i$ in $\log Q$ bits, Bob decodes coordinate $i$ of his tuple,
and they run the depth-$U$ protocol for $W$. Exchanging the players gives the
same cost for $R^+$. Taking $U=C$ proves equality.
\end{proof}

\section{Deterministic gap tables}\label{sec:deterministic}
\label{sec:deterministic_amplification}

The proof of \cref{thm:amplifier} covers explicit lists once their
marginal weights and independence properties are verified. We now construct
these lists and bound their encoding length, making each round a deterministic
polynomial-time operation. \Cref{det:reduction} gives the resulting gap and
encoding bound after \(k\) rounds.

The subscript $\rm g$ marks parameters and matrices of the explicit-list
construction. We choose a
power of two $s_{\rm g}$ above the fixed cutoff $s_0$ of
\cref{lem:generated-schedule}, with $\mathsf N:=s_{\rm g}^4\ge|V(G)|$ for the
source graph $G$.
Set $L_{\rm g}=\log\mathsf N$ and $\rho_{\rm g}=1/\mathsf N$. The choices in
\cref{def:amplification-parameters} then fix every other parameter as a
function of $s_{\rm g}$. In particular, $q_{2,\rm g}=s_{\rm g}^2/\rho_{\rm g}=s_{\rm g}^6$ since $\rho_{\rm g}=s_{\rm g}^{-4}$,
$w_{\rm g}=s_{\rm g}^8$, $Q_{\rm g}=s_{\rm g}^{24}$, and the block threshold is
$x_{\rm g}=\rho_{\rm g}^8$. These values are the same in every round, although the base
matrix $R$ changes from round to round.

\begin{corollary}[Explicit case]\label{det:amplifier}
For the construction \(R\mapsto R_{\rm g}^{+}\) of \cref{det:construction} with the
parameters of \cref{lem:generated-schedule}, parts (i) and (ii) of
\cref{thm:amplifier} hold with \(L=L_{\rm g}\), so that \(2^{-L}=\rho_{\rm g}\), and with
\[
h=h_{\rm g}:=2\log s_{\rm g}+\log q_{2,\rm g}+\log w_{\rm g}+\log Q_{\rm g}
=40\log s_{\rm g}=10\log\mathsf N.
\]
The construction is deterministic and runs in time polynomial in \(\mathsf N\) and the
explicit rational encoding length of \(R\). Every row and column label of \(R_{\rm g}^{+}\)
decodes to one of the full-product matrix \(R^+\), with the same entry at each decoded pair.
\end{corollary}

After defining the lists below, we prove this corollary in
\cref{sec:deterministic-endpoints}.

\subsection{Replacing products by explicit lists}

A coordinate of a tuple in one small list can itself be a label of another
small list. The next lemma shows that decoding such nested labels preserves
the product weights on every set of at most $8192$ coordinates. It also shows that
fixing independently weighted labels of other lists does not change these weights.
As in \cref{sec:models}, inputs with the same decoded value keep separate
labels and weights.

\begin{lemma}\label{det:flat}
Let an inner indexed list (\cref{def:exact-tuples}) on \(Z^{q_{\rm in}}\) be exactly \(8192\)-wise independent
with marginal \(\mu_Z\). Let an outer list on \(q_{\rm out}\)-tuples of inner labels be
exactly \(8192\)-wise independent with the inner label weights as marginal.
Then the decoded coordinates, indexed by pairs in \([q_{\rm out}]\times[q_{\rm in}]\), are exactly
\(8192\)-wise independent with marginal \(\mu_Z\). If the outer-list label
has product weights with the tuple of labels from finitely many other lists,
fixing those other labels at any joint value of positive weight leaves the
outer-label weights, and hence the conclusion above, unchanged.
\end{lemma}

\begin{proof}
Given at most $8192$ prescribed decoded coordinates, we group them by
outer position. Outer independence factors their joint weight over
these groups, and inner independence factors each group into its prescribed
coordinate marginals. Each factorization involves at most $8192$ coordinates,
as both independence hypotheses require.
By the product form, fixing the other labels leaves the outer-label weights,
which determine the decoded weights, unchanged.
\end{proof}

\begin{definition}\label{det:construction}
Let \(R:X\times Y\to\{0,1\}\) be a rationally weighted base matrix. With the
parameter choices of \cref{lem:generated-schedule}, construct the items
and the indicators \(v_i(\alpha)\) as in \cref{def:source-matrix}.
Replace the product alphabets \(\Gamma\), \(S\), and \(J_2^s\) of that definition by the indexed
lists \(\Gamma_{\rm g}\), \(S_{\rm g}\), and \(\mathcal J_{\rm g}\) of
\cref{det:code}, defined by
\[
\Gamma_{\rm g}=\mathsf{Code}_{8192}(2s_{\rm g}-1,\mu_Y),\qquad J_{\rm
g}=[q_{1,\rm g}]\times X,\qquad S_{\rm g}=\mathsf{Code}_{8192}(q_{2,\rm
g},\mu_{J_{\rm g}}),
\]
\[
J_{2,\rm g}=[q_{2,\rm g}]\times\Gamma_{\rm g},\qquad \mathcal J_{\rm
g}=\mathsf{Code}_{8192}(s_{\rm g},\mu_{J_{2,\rm g}}).
\]
Integer factors are uniform, and all other factors have their base or indexed
weights. On decoded labels set
\[
A_{\rm g}(\gamma,(r,u))=R(u,\gamma_r),\qquad R_{\rm temp,g}(z,(\alpha,\gamma))=A_{\rm g}(\gamma,z_\alpha).
\]
Use the raw entry rules, family tags, and type weights of
\cref{def:source-matrix}. For the family corresponding to $\Omega$, the
column weights are the product of the weights of a $\mathcal J_{\rm g}$ label
and the weights induced on the tail by $\Gamma_{\rm g}$. For the family
corresponding to $\mathcal D$, the weights are induced from $J_{2,\rm g}$ under
the map $j=(\alpha,\gamma)\mapsto(\operatorname{tail}(\gamma),j,\ldots,j)$ defining
$\mathcal D$. These rules define $B_{\rm raw,g}$, its transpose $N_{\rm g}$, and its
normalized restriction $N_{\rm ref,g}$ to inputs with the $\Omega$ family tag.

The lower-bound proof fixes the type/bin/item sequence and the base-label
sequence before restricting the template labels, which must then still have
their original weights, as required by (H3) of \cref{sec:internal-interlace}.
To interlace \(w_{\rm g}\) copies of \(N_{\rm g}\), we therefore combine three
tuple lists whose labels have product weights. Each list has
tuples of length \(w_{\rm g}\) and is exactly \(8192\)-wise independent,
with the following marginals.
Type/bin/item labels have the weights induced by the raw-row weights,
base labels have weights \(\mu_X\), and template labels have the \(S_{\rm g}\)
weights. We enumerate the Cartesian product of these three label sets.
At each position, the type label in the first sequence determines whether the
decoded row is a template row \((p,z)\), from the bin and template labels, or a row
\((i,u)\) indexed by an item, from the item and base labels. Since the three sequences
are independent and each is exactly \(8192\)-wise independent, their
coordinatewise combination gives an exactly \(8192\)-wise
independent tuple list with the raw-row marginal. Interlacing \(N_{\rm g}\) on
this combined list defines \(U_{\rm g}\). Set \(W_{\rm g}=U_{\rm
g}^{\mathsf T}\). We interlace \(W_{\rm g}\) at width \(Q_{\rm g}\) on
\(\mathsf{Code}_{8192}\) with the column weights of \(W_{\rm g}\) and then transpose to obtain \(R_{\rm
g}^{+}\). We store all input labels and rational weights explicitly, keeping inputs with
equal decoded values separate.
\end{definition}

As in \cref{obs:representatives}, we obtain projected restrictions from
chosen representatives. The per-copy deletion bounds and the collision
bound are the explicit case of \cref{src:profiles}. The proof of the deletion bounds uses
\cref{det:conditioning,det:rareboost} to handle retained events of weight
as small as $x_{\rm g}$, and the collision bound follows from \cref{det:exact}.
With these bounds and the exact independence of \cref{det:flat}, the collision and stopping-node
arguments of \cref{src:near-full,sec:internal-interlace} apply to the
indexed construction.

\subsection{Iteration, encoding size and deterministic running time}
\label{sec:deterministic-endpoints}

To prove \cref{det:amplifier}, we need only add the encoding bound to the
explicit case of \cref{thm:amplifier}.

\begin{proof}[Proof of \cref{det:amplifier}]
\phantomsection\label{proof:det:amplifier}
We suppress the subscript $\rm g$. Each width $q_1,q_2,w,Q$ is a fixed polynomial
in $\mathsf N$ (\cref{lem:generated-schedule}). Therefore the fixed number of calls to \cref{det:code}, all of order
$8192$, and enumeration of the Cartesian products of their label sets take time bounded by a fixed polynomial
in $\mathsf N$ and the encoding length of $R$. The indexed labels and rational
weights needed by the next round have encoding length within the same bound.

Parts (i) and (ii) follow from the proof of \cref{thm:amplifier}, with
\cref{det:restore} used in place of \cref{cl:bridge}. We apply \cref{det:restore}
with $(\phi,\delta,Q)=(\rho,\delta_U,Q)$, and \cref{lem:generated-schedule}
verifies its numerical hypotheses. The upper-bound
protocol uses only the decoded inputs and not their weights, so it also works
for the indexed construction.
\end{proof}

\begin{theorem}\label{det:reduction}
There are effective absolute constants \(\mathsf A\ge2\) and \(C_0>0\) and a
deterministic algorithm with the following guarantee. Given a YES or NO
source graph \(G\) from \cref{src:gap-colouring}, an integer \(k\ge1\), and a power of two
\(s_{\rm g}\) such that \(\mathsf N=s_{\rm g}^4\ge|V(G)|\) meets the parameter conditions
of \cref{lem:generated-schedule}, it produces an explicit total Boolean matrix
\(F_{G,k}\) and a public integer \(B_k^{\rm g}\) such that
\[
\text{YES: }\D(F_{G,k})\le B_k^{\rm g},\qquad \text{NO: }\D(F_{G,k})=B_k^{\rm
g}+k.
\]
Its running time and full
binary output length are at most \(\mathsf N^{C_0\mathsf A^k}\). The constants
depend only on the fixed source reduction and construction. The output is
square with power-of-two side length at least \(\mathsf N\). The public threshold is
\[
B_k^{\rm g}=d_{0,\rm g}+kh_{\rm g},
\]
where \(d_{0,\rm g}\) is the cost of the initial matrix in \cref{cl:seed}
with \(L=L_{\rm g}\):
\[
d_{0,\rm g}=\log\operatorname{pow}_2(64(L_{\rm g}+1))+1,\qquad L_{\rm
g}=\log\mathsf N.
\]
\end{theorem}

\begin{proof}
We use the parameters of \cref{lem:generated-schedule} for the given $s_{\rm g}$ and let $R_0$ be the
initial matrix of \cref{cl:seed} with $L=L_{\rm g}=\log\mathsf N$. We apply
\cref{det:construction} $k$ times, using each output as the next base matrix
and storing each indexed table with its rational
input weights. Finally, we forget the weights
and duplicate inputs to make a square table with power-of-two side at least $\mathsf N$.

By \cref{cl:seed}, the initial matrix has exact cost $d_{0,\rm g}\ge3$ and
satisfies the deletion bounds at threshold $\rho_{\rm g}=2^{-L_{\rm g}}$.
\Cref{prop:iteration}, in the explicit case (\cref{det:amplifier}), therefore gives
\[
\D(R_k)\le d_{0,\rm g}+kh_{\rm g}\quad\text{on YES},\qquad
\D(R_k)=d_{0,\rm g}+k(h_{\rm g}+1)\quad\text{on NO}
\]
for the matrix $R_k$ after $k$ rounds. Forgetting weights and duplicating rows
and columns to make the table square change neither cost (\cref{sec:models}). This proves the gap
with $B_k^{\rm g}=d_{0,\rm g}+kh_{\rm g}$.

For size and time, let $Z_i$ be $\mathsf N+2$ plus the complete encoding
length after round $i$, including labels and normalized rational weights. The polynomial
bound for the initial matrix and \cref{det:amplifier} give effective constants $C_1$ and a
power of two $\mathsf A\ge2$, independent of $i$ and $k$, such that, for all $i\ge0$,
\[
Z_0\le\mathsf N^{C_1},\qquad
Z_{i+1}\le Z_i^{\mathsf A},\qquad \text{time for round }i+1\le Z_i^{\mathsf A},
\]
and hence $Z_i\le\mathsf N^{C_1\mathsf A^i}$.
The padded side is below $2Z_k$ (\cref{sec:models}). After enlarging the effective constant $C_0$ if necessary, we bound the sum of
the $k$ round times and the final square padding by $\mathsf N^{C_0\mathsf A^k}$.
Computing the public threshold and the parameters takes time within the same bound.
\end{proof}

We now derive \cref{intro:deterministic} by taking a fixed number $k$ of rounds,
and \cref{det:time} by taking $k=\Theta(\log\log N)$ rounds for a sparse formula on
$N$ variables.

\begin{proof}[Proof of \cref{intro:deterministic}]
\phantomsection\label{proof:intro:deterministic}
For fixed $k$, we apply the graph reduction of \cref{src:gap-colouring} and then
\cref{det:reduction}. The least value $\mathsf N=s_{\rm g}^4$ meeting the parameter conditions obeys
$\mathsf N\le16\max\{|V(G)|,\mathsf N_0\}$ for the effective constant
$\mathsf N_0=s_0^4$, by \cref{lem:generated-schedule}. Since the graph reduction runs in
polynomial time, $|V(G)|$ is polynomial in the formula length, and since $k$ is fixed, so is
$\mathsf N^{C_0\mathsf A^k}$.
Thus $B=B_k^{\rm g}$ separates cost $\le B$ from
cost exactly $B+k$, on a square table with power-of-two side.
\end{proof}

\begin{proof}[Proof of \cref{det:time}]
\phantomsection\label{proof:det:time}
We choose the number $k$ of rounds so that the gap $k$ exceeds twice the
allowed error $c_*\log\log T$. At the same time, $k$ must be small enough that the
construction and the estimator run in quasipolynomial time in the size of the sparse formula.
Let $\mathsf A$ be the effective power of two from \cref{det:reduction}, and
set $c_*=1/(8\log\mathsf A)$. Suppose such an estimator exists.
For a sparse formula on $N$ variables and the graph $G$ of \cref{src:gap-colouring},
the least power of two $s_{\rm g}\ge\max\{N^{1/4},|V(G)|^{1/4},s_0\}$ meets the
conditions of \cref{lem:generated-schedule}, and
$\max\{N,|V(G)|\}\le\mathsf N=s_{\rm g}^4<16\max\{N,|V(G)|,\mathsf N_0\}$.
Since $|V(G)|\le N(\log(N+2))^p$ for the fixed $p$ of
\cref{src:gap-colouring}, we have $\mathsf N=N^{O(1)}$.
For sufficiently large $N$, we choose
\[
k=\floor{\frac{\log\log N}{\log\mathsf A}}.
\]
Then $k=\Theta(\log\log N)$ and $\mathsf A^k\le\log N$. The square output of
\cref{det:reduction} has threshold $B_k^{\rm g}$ and $T$ entries, where
\[
N^2\le T\le\mathsf N^{C_0\mathsf A^k},\qquad
\log\log T\le k\log\mathsf A+O(\log\log N).
\]
For sufficiently large $N$, the second bound is at most $3k\log\mathsf A$:
indeed it is at most $2\log\log N+O(1)$, whereas
$k\log\mathsf A=\log\log N+O(1)$.
Consequently
\[
2c_*\log\log T\le 3k/4<k.
\]
The estimator's guarantee holds on all sufficiently large tables, and $T\ge N^2$.
Hence, for sufficiently large $N$, comparing the estimate with the public midpoint
$B_k^{\rm g}+k/2$ decides satisfiability.
The construction time is at most $\mathsf N^{C_0\mathsf A^k}$, and the estimator's
running time is a fixed polynomial in $T\le\mathsf N^{C_0\mathsf A^k}$. Since
$\mathsf N=N^{O(1)}$ and $\mathsf A^k\le\log N$, both are bounded by
\[
2^{O((\log N)^2)}.
\]
For the finitely many remaining values of $N$, including those with $k<1$,
exhaustive search over assignments decides satisfiability.
This proves the claimed deterministic running-time implication.
\end{proof}


\section{Final sampling and growing error}
\label{sec:quantitative}

In the full-product case of \cref{prop:iteration}, the iterated matrix $R_k$
has cost at most $B_k$ on YES graphs and exactly $H_k=B_k+k$ on NO graphs, but it is too large to print, so we replace it by a sampled
submatrix. A single sample must be hard for every protocol of depth below the NO-case
bound. A small sample suffices because the rank bound $\rank(R_k)\le r_k$
of \cref{prop:iteration} limits the number of column sets on which such
protocols can be correct (\cref{thm:rank-count}).

\subsection{Counting column sets on which short protocols are correct}

For the union bound in \cref{sec:sampling}, we count the sets of columns on
which a short protocol can be correct on a retained row set. One bound
covers all row sets at once. We fix the tree, Alice's message functions and the leaf
outputs, and let Bob's bits depend on the column.

\begin{definition}[Columns on which Bob can complete the protocol]\label{def:compatibility}
Let a depth-$t$ tree $P$ specify a complete binary tree of
depth $t$, the speaker at each internal node, all Alice maps (from her row to
her bit at each of her nodes), and all leaf outputs. Bob's maps are unspecified. For a Boolean matrix
$F:X\times Y\to\{0,1\}$ and $U\subseteq X$, let $G_P(U)$ be the set of columns
for which some assignment of bits to Bob's nodes makes the tree correct on every
row in $U$. The assignment may depend on the column, but not on Alice's row.
\end{definition}

\begin{observation}[Lemma~3.2 in \cite{mackenzie2026condensation}]\label{obs:compatible-protocol}
For retained sets $U\subseteq X$, $V\subseteq Y$ and an integer $t\ge0$, with correctness on an empty restriction understood
vacuously,
\begin{equation}\label{eq:compatibility}
 \D(F[U,V])\le t
 \quad\Longleftrightarrow\quad
 V\subseteq G_P(U)\text{ for some depth-}t\text{ protocol tree }P\text{ of this kind}.
\end{equation}
\end{observation}

\Cref{fig:compatible-trees} uses two one-bit trees to show why leaf
requirements combine by intersection at an Alice node and by union at a Bob
node. We label each leaf by the columns on which its output is correct for
the rows reaching it.

\begin{figure}[H]
\centering
{\small
The same matrix in both trees: $F(x,y)=x_y$, the $y$th bit of $x$.

\smallskip
\begin{tabular}{c@{\qquad}*{3}{>{\centering\arraybackslash}p{8mm}}}
\toprule
&\multicolumn{3}{c}{Bob's column $y$}\\
\cmidrule(l){2-4}
Alice's row $x$&$1$&$2$&$3$\\
\midrule
$001$&$0$&$0$&$1$\\
$011$&$0$&$1$&$1$\\
\bottomrule
\end{tabular}
\par}
\medskip
\captionsetup[subfigure]{justification=centering}
\begin{subfigure}[t]{.48\linewidth}
\centering
\begin{tikzpicture}[font=\small,>=Stealth,
 box/.style={draw,rounded corners=2pt,align=center,inner sep=4pt},
 leaf/.style={box,text width=2.55cm}]
\node[box,draw=aliceblue] (a) at (0,0) {Alice sends $x_2$};
\node[leaf,anchor=north] (a0) at (-1.65,-1.55) {row $001$\\output $0$\\columns $\{1,2\}$};
\node[leaf,anchor=north] (a1) at (1.65,-1.55) {row $011$\\output $1$\\columns $\{2,3\}$};
\coordinate (am) at (0,-.95);
\coordinate (al) at (-1.65,-.95);
\coordinate (ar) at (1.65,-.95);
\draw[aliceblue] (a.south) -- (am);
\draw[aliceblue] (am) -- node[midway,above=3pt,inner sep=1pt] {$0$} (al);
\draw[aliceblue] (am) -- node[midway,above=3pt,inner sep=1pt] {$1$} (ar);
\draw[->,aliceblue] (al) -- (a0.north);
\draw[->,aliceblue] (ar) -- (a1.north);
\end{tikzpicture}
\caption{\textbf{Both required: intersection.}\\
$\{1,2\}\cap\{2,3\}=\{2\}$.}
\end{subfigure}\hfill
\begin{subfigure}[t]{.48\linewidth}
\centering
\begin{tikzpicture}[font=\small,>=Stealth,
 box/.style={draw,rounded corners=2pt,align=center,inner sep=4pt},
 leaf/.style={box,text width=2.55cm}]
\node[box,draw=boborange] (b) at (0,0) {Bob chooses using $y$};
\node[leaf,anchor=north] (b0) at (-1.65,-1.55) {rows $001,011$\\output $0$\\columns $\{1\}$};
\node[leaf,anchor=north] (b1) at (1.65,-1.55) {rows $001,011$\\output $1$\\columns $\{3\}$};
\coordinate (bm) at (0,-.95);
\coordinate (bl) at (-1.65,-.95);
\coordinate (br) at (1.65,-.95);
\draw[boborange] (b.south) -- (bm);
\draw[boborange] (bm) -- node[midway,above=3pt,inner sep=1pt] {$0$} (bl);
\draw[boborange] (bm) -- node[midway,above=3pt,inner sep=1pt] {$1$} (br);
\draw[->,boborange] (bl) -- (b0.north);
\draw[->,boborange] (br) -- (b1.north);
\end{tikzpicture}
\caption{\textbf{Either works: union.}\\
$\{1\}\cup\{3\}=\{1,3\}$.}
\end{subfigure}
\caption{Alice's branch depends on the row, so a column on which the tree is correct must
satisfy both leaf conditions. Bob's branch depends on the column, so either
leaf may work, but his chosen leaf must answer both rows correctly.}
\label{fig:compatible-trees}
\end{figure}

Mackenzie proves the count in \cref{thm:rank-count} with the row-sum argument of Oum, S{\ae}ther and
Vatshelle~\cite[Theorem~4.2]{oum2014faster}: the columns on which selected
Boolean rows all vanish are exactly the zero coordinates of the integer sum of
those rows. Counting these sums instead of the row subsets gives a bound that depends
only on the rank and not on the number of rows, which keeps the final sample small.

\begin{theorem}[Lemma~4.3 in \cite{mackenzie2026condensation}]\label{thm:rank-count}
Let $r\ge1$ and $t\ge0$ be integers. If a Boolean matrix $F$ has real rank at most
$r$, then, over all depth-at-most-$t$ trees $P$ specified as in
\cref{def:compatibility} and all retained row subsets $U$, there are at most
\begin{equation}\label{eq:rank-count}
 2^{(r+1)\ceil{\log(r+2)}\,2^{t+1}}
\end{equation}
distinct sets $G_P(U)$. Repeated row and column labels are allowed.
\end{theorem}

\Cref{thm:rank-count} follows from Mackenzie's bound $2^{B(r)2^t-1}$ for depth exactly $t$, where
$B(r)=(r+1)\ceil{\log(r+1)}+2$. Since $B(r)\le2(r+1)\ceil{\log(r+2)}$ for
$r\ge1$, and padding turns a tree of depth $t'<t$ into a depth-$t$ tree with the
same sets $G_P(U)$, this gives \eqref{eq:rank-count}. Repeated labels are
handled by passing to distinct values~\cite[Section~2]{mackenzie2026condensation}.

\subsection{Sampling columns and then rows}
\label{sec:sampling}

We sample the same number $m$ of rows and columns, and we take $m$ to be a
power of two, so the output has the square power-of-two form of
\cref{thm:quantitative}. In the proof, we first show that, with high
probability, the restriction to the sampled columns still has cost at least $H$ on every
row set of original weight $\ge3/5$. We then analyze the row sample, which is independent of the columns.

\begin{theorem}[Sampling rows and columns]\label{thm:compression}
Let $F:X\times Y\to\{0,1\}$ have real rank at most $r$, where $r\ge1$ is an
integer, and let $\mu_X,\mu_Y$ be input distributions on nonempty finite indexed sets.
Suppose an integer $H\ge1$ satisfies, for all $U\subseteq X$ and $V\subseteq Y$,
\[
\mu_X(U)\ge3/5,\quad\mu_Y(V)\ge3/5
\quad\Longrightarrow\quad \D(F[U,V])\ge H.
\]
Put
\[
m=\operatorname{pow}_2\bigl(16(r+1)\ceil{\log(r+2)}\,2^H\bigr),
\]
where $\operatorname{pow}_2(u)$ is the least power of two at least $u$.
Draw $m$ rows from $\mu_X$ and $m$ columns from $\mu_Y$, all independently, retaining separate indices for
repeated draws. The resulting $m\times m$ matrix has cost at least $H$ except
with probability less than $1/100$. Every protocol for $F$ also computes the
sampled matrix, so the sampled matrix has cost at most $\D(F)$ on every sample outcome.
\end{theorem}
\begin{proof}
Separate indices for repeated draws do not change costs. Each player can decode an index to its label, and conversely a protocol on
the indexed matrix restricts to one occurrence of each label.

For every $U$ with $\mu_X(U)\ge3/5$ and every depth-$(H-1)$ tree
$P$ specified as in \cref{def:compatibility}, we have $\mu_Y(G_P(U))<3/5$: otherwise
\cref{obs:compatible-protocol} with $V=G_P(U)$ gives $\D(F[U,V])\le H-1$,
contradicting the restriction hypothesis. As $U$ and
$P$ vary, these sets form a family of size at most
$2^{m/16}$ by \cref{thm:rank-count}, since $m\ge16(r+1)\ceil{\log(r+2)}\,2^H$. By a union bound, the probability that all sampled
columns lie in one of them is at most $2^{m/16}(3/5)^m$.
Outside this event, the sampled-column matrix
$F_Y(x,j)=F(x,y_j)$ has cost $\ge H$ on every original row set of weight
$\ge3/5$.

We condition on such a column outcome. The transpose $F_Y^{\mathsf T}$ still has
rank at most $r$. Consider the sets $G_P(U)$ for this transpose, where $U=[m]$ is
its full row set and $P$ has depth $H-1$. Each has $\mu_X(G_P(U))<3/5$: otherwise
\cref{obs:compatible-protocol} and transposition give
$\D(F_Y[G_P(U),[m]])\le H-1$, contradicting the preceding guarantee. There are again at most $2^{m/16}$ such sets.
The row draws remain independent with distribution $\mu_X$, so the probability that
they all lie in one of these sets is at most $2^{m/16}(3/5)^m$. By \cref{obs:compatible-protocol} for the transpose of the final matrix, a protocol of depth at most
$H-1$ on the final matrix would put all sampled rows in one of these sets. Since $r,H\ge1$ give $m\ge32$,
the total failure probability is at most
\[
2\cdot2^{m/16}(3/5)^m\le2^{1-7m/16}\le2^{-13}<1/100.
\]
The universal upper bound follows by decoding sampled indices and running a
protocol for $F$.
\end{proof}

We sample only once, after all amplification rounds on full products.
\Cref{thm:compression} bounds the failure probability but gives no test
certifying that a particular sample succeeds.

\subsection{Printing the sample and comparing errors}
\label{sec:reduction}
\label{sec:randomised_reduction}

We generate the sampled row and column labels and evaluate $R_k$ only on
the pairs they form.

We take a source graph on $n$ vertices above the fixed effective cutoff in
\cref{lem:full-schedule}, use the parameter choices of that lemma at every
round, and write
\[
L=\ceil{\log n},\qquad B_k=d_0+kh,\qquad H_k=B_k+k,\qquad r_k=2^{B_k-1}
\]
as in \cref{prop:iteration}, which gives $\rank(R_k)\le r_k$. The reduction
uses only this bound and never computes a rank. A \emph{random tape} is the sequence of random bits used
by the algorithm. The next lemma bounds the running time on every tape, not
only on likely ones.

\begin{lemma}\label{lem:sample-access}
Let $R_k$ be the matrix of \cref{prop:iteration} in the full-product case, and let $m$ be the sample size of
\cref{thm:compression} with $r=r_k$ and $H=H_k$. An algorithm generates $m$ row and $m$ column labels of $R_k$
and evaluates $R_k$ on all their pairs, using $n^{O(k+1)}$
time and random bits on every tape, with an absolute implied constant
independent of $k$. Every output label is an actual input of $R_k$, and the
sample counts and tape length do not depend on the tape. The joint distribution of the generated labels is within total
variation less than $1/100$ of the ideal independent sample of \cref{thm:compression}.
\end{lemma}

Here total variation is the maximum, over events, of the absolute difference between the probabilities
that two distributions assign to the event. The proof in
\cref{sec:sampling-algorithm} gives the recursive construction with a fixed
number of random bits. The label distribution is only close to the ideal one,
but every printed entry is an exact entry of $R_k$. Combining the lemma with
\cref{thm:compression}, we obtain the following reduction.

\begin{theorem}[Randomized reduction]\label{thm:uniform-reduction}
There is a randomized algorithm that, given a source graph on $n$ vertices above the fixed cutoff in \cref{lem:full-schedule} and
an integer $k\ge1$, outputs a total Boolean matrix $M$ and an integer $B$.
The table is square with power-of-two side length. On every tape the algorithm
uses $n^{O(k+1)}$ time and random bits, with an absolute implied
constant independent of $k$. For a YES source, $\D(M)\le B$ on every tape.
For a NO source,
\[
\Pr[\D(M)=B+k]\ge49/50.
\]
Here $B=B_k$ and the tape-independent entry count is
\[
m=\operatorname{pow}_2\bigl(16(r_k+1)\ceil{\log(r_k+2)}\,2^{H_k}\bigr),
\qquad T=m^2.
\]
\end{theorem}

\begin{proof}
The algorithm computes the parameters of \cref{lem:full-schedule} and the
initial matrix of \cref{cl:seed}, then $B_k,H_k,r_k,m$. It generates $m$ columns
and $m$ rows with \cref{lem:sample-access}, evaluates all $m^2$ row--column
pairs (keeping repeated labels as distinct indices), and outputs $(M,B_k)$.
The time and random-bit bounds follow from \cref{lem:sample-access}, and the
output is $m\times m$ with $m$ a power of two.
Every generated label is an input of $R_k$. On a YES source, decoding the
labels and running a protocol for $R_k$ of cost at most $B_k$ (\cref{prop:iteration}) shows that every sample has cost at most $B_k$.

On a NO source, \cref{prop:iteration} gives
\[
\D(R_k)=H_k,\qquad \RP{R_k}{2^{-L}}{H_k},\qquad \rank(R_k)\le r_k.
\]
Because $2^{-L}\le1/2<3/5$, the $j=0$ lower bound applies whenever both retained
weights are at least $3/5$. Thus, by \cref{thm:compression}, an ideal sample has cost below $H_k$
with probability less than $1/100$. By \cref{lem:sample-access}, using the implemented
sampler instead of the ideal independent sample adds less than $1/100$ to this
failure probability. Therefore, except with probability less than $1/50$, the
output has cost $\ge H_k$. Decoding the inputs and using $\D(R_k)\le H_k$ (\cref{prop:iteration}) gives cost $\le H_k$ on every
tape, so the cost is then exactly $B_k+k$.
\end{proof}

\subsection{From an approximation algorithm to a $3$-SAT algorithm}
\label{sec:time-tradeoff}

We now combine the two constructions with an estimator, proving the general
implication stated in \cref{thm:time-tradeoff}. Each construction produces
a known threshold $B$ and a gap $k$. The size bound of the construction determines which
approximation errors are small enough to distinguish the two cases.

\begin{theorem}[From approximation algorithms to $3$-SAT algorithms]
\label{thm:time-tradeoff}
There is an effective absolute integer $\mathsf A\ge2$ and, for each fixed
clause-density bound $c_s$, an effective integer $C\ge1$ with the following
property. Suppose a deterministic or
bounded-error randomized polynomial-time estimator approximates $\D(M)$
with additive error at most a nondecreasing function $a(T)$ on every
sufficiently large explicit total Boolean table with $T$ entries.
Let the gap $k=k(N)\ge1$ be an integer function computable in time polynomial in
$N+k(N)$. Choose either construction and its size bound:
\[
\begin{array}{c|c}
\text{Construction} & S(N,k)\\ \hline
\text{Randomized} & (N+2)^{C(k+1)}\\[2pt]
\text{Deterministic} & (N+2)^{C\mathsf A^k}
\end{array}
\]
If, for all sufficiently large $N$,
\[
2a\bigl(S(N,k(N))\bigr)<k(N),
\]
then $3$-SAT on $N$ variables and at most $c_sN$ clauses has an algorithm
running in time $S(N,k(N))^{O(1)}$. It is deterministic when both the
construction and the estimator are deterministic, and bounded-error
randomized otherwise. The exponent may depend on the estimator and the
algorithm computing $k$. It suffices for the estimator to work on square
tables whose side length is a power of two.
\end{theorem}

\begin{proof}[Proof of \cref{thm:time-tradeoff}]
\phantomsection\label{proof:thm:time-tradeoff}
Apply \cref{src:gap-colouring}, then \cref{thm:uniform-reduction} in the
randomized case or \cref{det:reduction} in the deterministic case. Since
the source graph and the size parameter $\mathsf N$ of \cref{det:reduction} are $N^{O(1)}$, a
fixed $C$ bounds both the table size $T$ and construction time by the
chosen $S(N,k)$. The constant $\mathsf A$ is the absolute constant from
\cref{det:reduction}. The table sizes $T$ tend to infinity with $N$, so the estimator's guarantee applies for all sufficiently large $N$.

Because $a$ is nondecreasing, $a(T)\le a(S(N,k))<k/2$. An accurate
estimate is therefore below the midpoint $B+k/2$ on a YES output and
above it on a successful NO output. If the estimator is randomized,
repeat this comparison independently a constant number of times and take
the majority answer, reducing its error below $1/100$. The randomized
construction adds failure probability at most $1/50$, and the deterministic
construction adds none. Thus the combined algorithm has error less than
$1/3$ whenever randomness is used, and is exact otherwise.

Construction, computing $k$, estimation, and comparison with the binary
rational midpoint all take time $S(N,k)^{O(1)}$. In particular, the assumed polynomial cost of computing $k$
is $S(N,k)^{O(1)}$ because $S(N,k)\ge N+k$.
Exhaustive search handles the finitely many smaller sizes.
\end{proof}

To read this in terms of a proposed $3$-SAT running time $R(N)=2^{g(N)}$,
suppose $g(N)/\log(N+2)$ grows without bound and the chosen number of rounds
can be computed within the stated time. A running time $2^{O(g(N))}$ allows
\[
\begin{aligned}
k&=\Theta\!\left(\frac{g(N)}{\log(N+2)}\right)
&&\text{in the randomized construction},\\
k&=\Theta\!\left(\log\frac{g(N)}{\log(N+2)}\right)
&&\text{in the deterministic construction}.
\end{aligned}
\]
The constant in the second choice is chosen so that
$\mathsf A^k\log(N+2)=O(g(N))$.
Additive error below $k/2$ on tables of size $S(N,k)$
therefore yields a $3$-SAT algorithm running in time $2^{O(g(N))}$. If such a $3$-SAT
algorithm is excluded by an explicit hypothesis, the same statement gives
inapproximability. To quantify a further improvement over a known algorithm,
choose a smaller target running time $R$. The theorem does not assume that the best
known running time is optimal.

For \cref{thm:quantitative}, we use the exact sample size rather than
the coarser bound $S(N,k)$. This gives the coefficient $1/8$ in the
randomized error bound, and the resulting algorithm is the same.

\begin{proof}[Proof of \cref{thm:quantitative}]
\phantomsection\label{proof:thm:quantitative}
Suppose a bounded-error randomized polynomial-time estimator achieves additive
error $a(T)=c\log T/\log\log T$ for some fixed $0\le c<1/8$ on every
sufficiently large square table of power-of-two side length. Fix
$8c<\alpha<1$. We obtain a bounded-error $2^{O(N^\alpha)}$ algorithm for
sparse $3$-SAT.

For a formula with $N$ variables, \cref{src:gap-colouring} produces a source
graph with $n$ vertices, where
\[
2N/\Delta\le n\le N(\log(N+2))^p
\]
for the fixed integer $p$ of that theorem. The lower bound follows from
$N\le|E|\le\Delta n/2$, also given by that theorem.
Thus the graph is above the fixed cutoff for every sufficiently large $N$, and
\[
L=\ceil{\log n}=\log N+O(\log\log N).
\]
Choose a fixed rational $\beta$ with $8c<\beta<\alpha$, and put
$k=\floor{N^\beta/\ceil{\log N}}$. This integer is computable in polynomial
time, using an integer root for the fixed rational power. By
\cref{lem:full-schedule,cl:seed}, $h=L+O(\log L)$ and $d_0=O(\log L)$, so
\[
B_k=d_0+kh=(1+o(1))k\log N.
\]
Substituting $r_k=2^{B_k-1}$ and $H_k=B_k+k$ into
\cref{thm:uniform-reduction} gives
$\log m=2B_k+k+O(\log(B_k+1))$. Hence the printed size satisfies
\begin{align}
\log T&=2\log m=4B_k+2k+O(\log(B_k+1))
             =(4+o(1))k\log N,\label{eq:quantitative-size}\\
\log\log T&=(\beta+o(1))\log N,
\qquad k=\left(\frac\beta4+o(1)\right)\frac{\log T}{\log\log T}.
\end{align}
Consequently $k>2a(T)$ for all sufficiently large $N$. The table already has
the square power-of-two form required by the assumed estimator.

For each estimator run, we test whether its output is below the rational
midpoint $B_k+k/2$. An accurate estimate answers this test correctly on every
YES output and every successful NO output. We repeat the run independently a
constant number of times and take the majority answer, which reduces the
decision error on each such table below $1/100$. This uses only comparison with a binary rational, which the
estimator model supports (\cref{def:table-estimation}). The reduction
fails with probability at most $1/50$, so the decision error is less than
$3/100<1/3$.

Both construction time and random-bit usage are
$n^{O(k+1)}=2^{O(N^\beta)}$ on every tape. The estimator also takes
$2^{O(N^\beta)}$ time because $\log T=O(N^\beta)$.
For all sufficiently large $N$, we have $k\ge1$, the graph is above the fixed cutoff, and $T$ meets the estimator's size threshold.
The finitely many smaller values of $N$ are decided by exhaustive search within
a fixed bound. Since $\beta<\alpha$, the total running time is
$2^{O(N^\alpha)}$, as claimed.
\end{proof}

\section{Strictly alternating protocols}\label{sec:alternation}
\label{sec:alternating}

We now prove the result that settles Yao's original question. If we simulate
the multi-bit messages of the unrestricted construction by inserting the missing
turns, the YES-case upper bound can double while the known NO-case lower
bound stays the same, so the additive gap is lost. We therefore use a
separate matrix designed for strict alternation. The end of \cref{sec:overview} sketches
this construction. In the YES case, a balanced coloring of the source hypergraph gives a short alternating protocol
(\cref{prop:alternating-upper}). In the NO case, a short protocol
colors each vertex by Alice's most common response rules for her early turns. A monochromatic edge then gives many
array positions on which her remaining messages produce too few answer patterns
(\cref{yao:small-bias-escape}). With an explicit array list (\cref{yao:array-list}),
the table has polynomial size for each fixed gap, and no repeated amplification is
needed.

\subsection{Starter, output and matrix inputs}

Changing the starting player or the output convention changes the cost by at
most two bits, so the gap transfers between the conventions of \cref{yao:model}.

\begin{lemma}\label{yao:comparison}
On nonempty input sets, each cost
$C\in\{D_A^{\rm alt},D_B^{\rm alt},D_{\rm free}^{\rm alt},Y_{\rm free}\}$
satisfies
\[
|C(M)-D_A^{\rm alt}(M)|\le2.
\]
\end{lemma}
\begin{proof}
Prepending one fixed bit sent by the new starter changes the starter, so
$|D_A^{\rm alt}-D_B^{\rm alt}|\le1$ and
\[
D_A^{\mathrm{alt}}-1\le D_{\mathrm{free}}^{\mathrm{alt}}\le D_A^{\mathrm{alt}}.
\]
Public output also gives receiver output. Conversely, at a nonzero-depth
terminal transcript the receiver is the next scheduled speaker and can announce
the answer in one bit. Public stopping tells both players when to make this
announcement. At depth zero, choose the designated output player as the starter
and let that player announce the answer. Thus
\[
Y_{\mathrm{free}}\le D_{\mathrm{free}}^{\mathrm{alt}}\le Y_{\mathrm{free}}+1,
\]
and the claimed two-bit bound follows. Each of these conversions preserves public
stopping and uses an input-independent starter.
\end{proof}

A hypergraph coloring is proper if no edge is monochromatic. In the YES case
the source hypergraph has a balanced coloring with few colors. In the NO case it has no
proper coloring even with the larger number of colors that a short protocol would induce. We use the balanced
hypergraph coloring reduction of Guruswami and Lee~\cite{guruswami2018balanced},
modified so that every edge has distinct vertices.

\begin{proposition}[Hypergraph coloring promise]\label{yao:source}
Fix an integer $t\ge4$, and set
\[
s_{\rm a}=2^t,\qquad K_{\rm a}=s_{\rm a}^2,\qquad S_{\rm a}=2^{s_{\rm a}}.
\]
There is a deterministic polynomial-time reduction from $3$-SAT to
$K_{\rm a}$-uniform hypergraphs with nonempty vertex sets and distinct vertices
in every edge such that:
\begin{enumerate}
\item a satisfiable formula has an $s_{\rm a}$-coloring with at least $s_{\rm a}-1$ vertices of every color in every edge;
\item an unsatisfiable formula has no proper $S_{\rm a}$-coloring.
\end{enumerate}
The polynomial may depend on $t$.
\end{proposition}

The proof in \cref{app:hypergraph-source} derives this promise from the cited
result and ensures that every edge has distinct vertices.

Alice holds a vertex and a tuple of addresses, each indexing the first
coordinate of Bob's array. Bob holds the array, a coordinate selecting one of Alice's addresses,
and an edge index. If the vertex
is outside the selected edge, or the edge index is unused, the answer is zero.

\begin{definition}\label{def:alternating-table}
Let $\mathcal H=(V,E)$ be a $K_{\rm a}$-uniform source hypergraph as in
\cref{yao:source}, with a fixed ordering of the vertices in each edge. Set
\[
q_{\rm a}=2^{t-1}=s_{\rm a}/2, \qquad H_{\rm a}=2^b=\operatorname{pow}_2(\max\{1,|E|\}).
\]
Index the
edges by distinct $\alpha\in[H_{\rm a}]$, and let $\iota_\alpha(v)$ be the
position of $v$ in that edge, or zero if $v$ is absent or the index is unused.
Alice's address tuples are all of
\[
\mathcal C=[H_{\rm a}]^{q_{\rm a}},
\]
each used once.
For any labeled Boolean array list $\mathcal W=(w^\ell)$ on
\[
\Omega_{\rm a}=[H_{\rm a}]\times[K_{\rm a}],
\]
define
\[
M_{\rm a}((v,c),(g,\alpha,\ell)):=
\begin{cases}
w^\ell(c_g,\iota_\alpha(v)),&\iota_\alpha(v)>0,\\
0,&\iota_\alpha(v)=0.
\end{cases}
\]
Rows are indexed by $(v,c)\in V\times\mathcal C$. Column labels lie in
$[q_{\rm a}]\times[H_{\rm a}]\times[|\mathcal W|]$. Equal arrays retain their
separate labels.
\end{definition}

In \cref{fig:yao-entry}, \(g\) selects an address from Alice's tuple, which
gives the row of Bob's array, and the position of her vertex in Bob's edge gives
the column.

\begin{figure}[H]
\centering
\begin{tikzpicture}[>=Stealth,every node/.style={font=\small},
 data/.style={draw,rounded corners=2pt,align=left,inner sep=5pt}]
\matrix (bits) at (8,0) [matrix of math nodes,
 nodes={minimum width=8mm,minimum height=6mm,inner sep=1pt},
 column sep=0pt,row sep=0pt,
 row 1/.style={nodes={font=\footnotesize}},
 column 1/.style={nodes={font=\footnotesize}}] {
  &1&2&3&4\\
 1&1&0&1&0\\
 2&0&1&0&0\\
 3&0&1&1&1\\
 4&1&0&0&1\\
};
\begin{scope}[on background layer]
  \node[fill=aliceblue!8,fit=(bits-2-2)(bits-2-5),inner sep=0pt] {};
  \node[fill=boborange!8,fit=(bits-2-4)(bits-5-4),inner sep=0pt] {};
\end{scope}
\node[draw=aliceblue,fit=(bits-2-1),inner sep=0pt] {};
\node[draw=boborange,fit=(bits-1-4),inner sep=0pt] {};
\node[draw=answergreen,very thick,fit=(bits-2-4),inner sep=0pt] {};
\node[data,draw=aliceblue,fill=aliceblue!5,text width=39mm,anchor=east]
 (address) at ([xshift=-10mm]bits-2-1.west)
 {Alice's tuple: $c=(3,\boxed{1})$\\
  Bob selects $g=2$\\[2pt]
  Address: $c_g=1$};
\node[data,draw=boborange,fill=boborange!5,text width=45mm,anchor=south]
 (position) at ([yshift=7mm]bits-1-4.north)
 {Bob's edge: $e_\alpha=(u,v,\boxed{z},r)$\\
  Alice's vertex: $z$\\[2pt]
  Position: $\iota_\alpha(z)=3$};
\draw[->,aliceblue,semithick] (address.east) -- (bits-2-1.west);
\draw[->,boborange,semithick] (position.south) -- (bits-1-4.north);
\node[below=5mm of bits,align=center]
 {Bob's array $w^\ell$\\[2pt]
  \textcolor{answergreen}{Requested bit: $w^\ell(1,3)=1$}};
\node[data,text width=29mm,anchor=north west]
 (outside) at ([yshift=-12mm]address.south west)
 {Vertex outside\\the selected edge};
\node[data,anchor=west] (zero) at ([xshift=9mm]outside.east) {output $0$};
\draw[->,semithick] (outside.east) -- (zero.west);
\end{tikzpicture}
\caption{The address selects a row and the vertex's edge position selects a
column, giving \(w^\ell(1,3)=1\). A vertex outside the edge gives zero without
reading the array. The indices are illustrative, and the arrows show how the
entry is selected, not a protocol's messages.}
\label{fig:yao-entry}
\end{figure}

\subsection{The protocol given by a coloring}

Once Bob has sent the edge, a balanced coloring lets Alice name her
position among the at most $2s_{\rm a}-1$ positions of her color in that edge. The schedule in the proof of
\cref{prop:alternating-upper} exchanges color and coordinate, then edge and
address, then this position, and ends with Bob's answer bit. It costs
\[
B_{\rm a}:=(2t-1)+2b+2(t+1)+1=2b+4t+2.
\]

\begin{proposition}\label{prop:alternating-upper}
For the table of \cref{def:alternating-table}, an $s_{\rm a}$-coloring with at
least $s_{\rm a}-1$ vertices of every color in every edge gives
\[
D_A^{\rm alt}(M_{\rm a})\le B_{\rm a}.
\]
\end{proposition}
\begin{proof}
Fix a row $(v,c)$, a column $(g,\alpha,\ell)$ and a balanced coloring $\chi$.
Every edge contains at least
$s_{\rm a}-1$ vertices of each of the $s_{\rm a}-1$ colors other than $\chi(v)$. Consequently
the vertices whose color is $\chi(v)$ occupy at most
\[
K_{\rm a}-(s_{\rm a}-1)^2=2s_{\rm a}-1
\]
positions.
Alice's message value is the rank of her position among these, or one
extra value for ``outside the edge.'' There are therefore at most
$2s_{\rm a}=2^{t+1}$ message values, encoded in $t+1$ bits. The ranking and the encoding are fixed publicly
for every color and edge index. An unused edge index has only the value for ``outside the edge.''
Alice determines her message value from $v$, $\alpha$ and $\chi(v)$.

First Alice sends the $t$ bits of $\chi(v)$, while Bob sends
the $t-1$ bits of $g$ between consecutive Alice bits. This uses $2t-1$ bits and
leaves Alice knowing $g$. Next Bob sends the $b$ bits of $\alpha$, while Alice
sends the $b$ bits of $c_g$, alternating. This uses $2b$ bits, after which
Alice knows $\alpha$ and Bob knows $c_g$. Since Bob speaks next, he sends $t+1$ dummy bits while Alice sends
the $t+1$ bits encoding her message value, after which Bob sends one answer bit. He answers zero for
``outside the edge.'' Otherwise he decodes, from $\chi(v)$, $\alpha$ and the rank, the position
$\iota_\alpha(v)$ and answers $w^\ell(c_g,\iota_\alpha(v))$. The three
stages use $B_{\rm a}$ bits in total.
\end{proof}

\subsection{Few response patterns cannot cover many array positions}

Bob's early messages may depend on the whole array, so the transcript on one
column does not determine how Alice behaves on other columns. To analyze an arbitrary
protocol, we therefore also consider Bob histories that do not occur on a given column.
For a fixed row, we record Alice's $r$th bit after every possible history of
$r-1$ Bob bits, including hypothetical histories. Her own preceding bits
are then determined. For this count, we pad branches that stop early with
fixed bits in the alternating order. For example, her first
two turns require three response bits: the initial bit and one response to each
possible first Bob bit.
In general, her first $t$ turns require
$\sum_{r=1}^t2^{r-1}=2^t-1$ response bits, giving at most
$2^{2^t-1}$ possible response rules for these turns.

Among all Boolean arrays, the fraction agreeing with prescribed bits on
$k$ distinct positions is $2^{-k}$. We need only an upper bound of
$2^{-k}+\varepsilon$ on this fraction for a small $\varepsilon>0$, and a short explicit list already
satisfies this weaker bound (\cref{yao:array-list}). Here we count all array labels equally, keeping
repeated arrays as separate labels. For a list
$\mathcal W=(w^\ell)_{\ell\in\mathcal L}$ on a finite position set $\Omega$,
with $\mathcal L$ finite and nonempty, we require
\begin{equation}\label{eq:alternating-pattern}
\frac{|\{\ell\in\mathcal L:w^\ell|_I=z\}|}{|\mathcal L|}
\le2^{-|I|}+\varepsilon
\qquad
(I\subseteq\Omega,\ z\in\{0,1\}^I).
\end{equation}
Adding these bounds over $P$ patterns on the same set $I$, at most a
$P(2^{-|I|}+\varepsilon)$ fraction of all labels matches one of them. Every fraction below is measured
against the original list size $|\mathcal L|$, even after a transcript has
removed some columns.

\Cref{yao:small-bias-escape} rules out a short protocol when many requested
positions share Alice's early response rules. Her early messages cannot distinguish
those positions, so many array labels remain after them. The remaining
messages produce only a few answer patterns, and by \cref{eq:alternating-pattern}
these match fewer labels than survive. In the lemma, $m$
is the total number of Alice turns, and $t$ is the number of initial Alice
turns during which her response rules agree across the rows for every possible
Bob history.

\begin{lemma}[Lower bound for alternating protocols]\label{yao:small-bias-escape}
Let $m,t$ be integers with $m\ge2t\ge2$. Suppose the nonempty labeled
array list $\mathcal W$ satisfies \cref{eq:alternating-pattern} with
$\varepsilon\le2^{-4m}$. If $J\subseteq\Omega$ has at least
$2^{m-t+1}$ positions, then the matrix
\[
L_J(j,\ell):=w^\ell(j)
\]
has no Alice-first alternating protocol of depth at most $2m-1$ in which all
rows $j\in J$ give Alice the same response rules for her first $t$ turns on every possible Bob history.
\end{lemma}
\begin{proof}
Suppose such a protocol exists. We pad every shorter branch to depth $2m-1$
with fixed bits in the alternating order. At the final Alice node, her row and the transcript
determine the old output leaf, so we replace her last bit by that output. This
uses no extra bit and, since $m>t$, preserves her first $t$ responses.

We stop with about $\log m$ turns per player remaining, so that Bob's
remaining response rules require fewer bits to describe than there are
surviving array positions.
Put $h=\ceil{\log m}\le m-t$ and $Q=2^h$. The inequality follows from
$m\le2(m-t)\le2^{m-t}$. We follow the first $2(m-h)-1$ messages from the root.
After each Bob message we keep the child with more column labels, and after
each Alice message the child with more rows.
Alice's first $t$ turns discard no rows because these response rules agree.
Each of her remaining $m-h-t$ turns retains at least half the rows,
and each of the $m-h-1$ Bob turns retains at least half the columns.
The resulting rectangle $X\times Y$, with $Y\subseteq\mathcal L$, therefore satisfies
\[
|X|\ge2^{h+1}=2Q,\qquad
\frac{|Y|}{|\mathcal L|}\ge2^{-(m-h-1)}.
\]
Exactly $h$ Bob--Alice pairs remain. We record Bob's next bit after every
hypothetical history of earlier Alice bits, which gives
$1+2+\cdots+2^{h-1}=Q-1$ bits. Fixing these bits and Alice's response rules determines
her final output on every retained row. On any $2Q$ of those rows, there are
therefore at most $2^{Q-1}$ output patterns (vectors of outputs on these rows).

Since $m\le Q<2m$, we have $\varepsilon\le2^{-2Q}$. Every column in $Y$
must match one of these patterns. Applying \cref{eq:alternating-pattern}
to each pattern and adding the bounds gives
\[
\frac{|Y|}{|\mathcal L|}\le2^{Q-1}(2^{-2Q}+\varepsilon)\le2^{-Q}.
\]
This contradicts the lower bound on the surviving fraction because
$Q\ge m>m-h-1$.
\end{proof}

Since \cref{eq:alternating-pattern} holds for every set of positions, one
array list works against every protocol.
We use the accuracy
$\epsilon_{\rm a}=2^{-4(b+3t)}=2^{-12t}H_{\rm a}^{-4}$, which is inverse
polynomial in $H_{\rm a}$ for each fixed $t$. At this accuracy, the
finite-field construction of \cref{sec:alternating_implementation} gives such a
list with polynomially many arrays in $H_{\rm a}$, and we use it in place of the
set of all Boolean arrays in $M_{\rm a}$.

\begin{corollary}\label{yao:array-list}
For every fixed $t\ge4$, a nonempty labeled array list on $\Omega_{\rm a}$
satisfying \cref{eq:alternating-pattern} with
\[
\epsilon_{\rm a}:=2^{-4(b+3t)}
\]
can be generated in deterministic time polynomial in $H_{\rm a}$.
\end{corollary}

The construction and proof are in \cref{sec:arrays}.

\subsection{From a hypergraph to an explicit gap}

Given a short protocol, we color each vertex $v$ by Alice's most common response
rules for her first $t$ turns among the rows $(v,c)$ with $c\in\mathcal C$. A monochromatic edge then gives the many distinct
array positions that \cref{yao:small-bias-escape} requires.

\begin{proposition}\label{yao:lift}
Use the complete address list $\mathcal C$ and the explicit array list of
\cref{yao:array-list} in $M_{\rm a}$. A balanced coloring of $\mathcal H$ as in
\cref{prop:alternating-upper} gives
\[
D_A^{\rm alt}(M_{\rm a})\le B_{\rm a},
\]
whereas the absence of a proper $S_{\rm a}$-coloring gives
\[
D_A^{\rm alt}(M_{\rm a})\ge B_{\rm a}+2t-8.
\]
For each fixed $t$, the full table can be printed in deterministic time
polynomial in $|V|+H_{\rm a}$.
\end{proposition}
\begin{proof}
The upper bound is \cref{prop:alternating-upper}. For the lower bound, we put
$m=b+3t-3$, so $2m-1=B_{\rm a}+2t-9$, and suppose that an Alice-first protocol
has depth at most $2m-1$. We pad it to that depth and, for each vertex $v$,
partition its address tuples by Alice's response rules on her first $t$ turns.
By the count of response bits above, there are at most
$2^{2^t-1}\le S_{\rm a}$ choices of these rules. A largest class $C_v$ therefore has
\[
|C_v|\ge H_{\rm a}^{q_{\rm a}}/S_{\rm a}.
\]
We color $v$ by this class's response rules. Since the hypergraph has no proper
$S_{\rm a}$-coloring, some edge $e_\alpha$ is monochromatic, that is, all its
vertices receive the same response rules.

For $v\in e_\alpha$ and $g\in[q_{\rm a}]$, let
$J_{v,g}=\{c_g:c\in C_v\}$ be the set of addresses occurring in coordinate
$g$. Since $C_v$ is contained in the product of these projections,
\[
\prod_{g=1}^{q_{\rm a}}|J_{v,g}|\ge |C_v|
\ge H_{\rm a}^{q_{\rm a}}/S_{\rm a}.
\]
A column $(g,\alpha,\ell)$ reads only coordinate $g$ of Alice's tuple, so we
need one coordinate $g$ that is good for all vertices of the edge. Multiplying
the last display over the $K_{\rm a}$ vertices of $e_\alpha$ and choosing $g$ to
maximize $\prod_{v\in e_\alpha}|J_{v,g}|$ gives
$\prod_{v\in e_\alpha}|J_{v,g}|\ge(H_{\rm a}S_{\rm a}^{-1/q_{\rm a}})^{K_{\rm a}}$.
By AM--GM, this coordinate satisfies
\[
\begin{aligned}
N:=\sum_{v\in e_\alpha}|J_{v,g}|
&\ge K_{\rm a}H_{\rm a}S_{\rm a}^{-1/q_{\rm a}}\\
&=\frac{K_{\rm a}H_{\rm a}}4=2^{m-t+1},
\end{aligned}
\]
where $S_{\rm a}=4^{q_{\rm a}}$.
For each $v\in e_\alpha$ and $u\in J_{v,g}$, we choose a tuple $c(v,u)\in C_v$
with $c(v,u)_g=u$. We retain the corresponding rows and all columns
$(g,\alpha,\ell)$. Their entries are
\[
M_{\rm a}((v,c(v,u)),(g,\alpha,\ell))
=w^\ell(u,\iota_\alpha(v)).
\]
The positions $(u,\iota_\alpha(v))$ are $N$ distinct elements of $\Omega_{\rm a}$, because different vertices occupy different
positions in the edge and different addresses give different first coordinates.
All retained rows give Alice the same response rules for her first $t$ turns. Since $m=b+3t-3\ge2t$ for $t\ge4$
and $\epsilon_{\rm a}\le2^{-4m}$, the restriction to these rows and columns contradicts
\cref{yao:small-bias-escape}. Integer depths give
$D_A^{\rm alt}(M_{\rm a})\ge2m=B_{\rm a}+2t-8$.

There are $|V|H_{\rm a}^{q_{\rm a}}$ rows and
$q_{\rm a}H_{\rm a}|\mathcal W|$ columns. For fixed $t$, the number
$q_{\rm a}$ is constant and \cref{yao:array-list} constructs the arrays in
polynomial time. Enumerating all tuples and evaluating the displayed entry rule
therefore prints the entire table in polynomial time.
\end{proof}

To finish, we choose $t$ so that the gap $2t-8$ of \cref{yao:lift} is at least
the requested gap. We then transfer the resulting hardness to the other starter
and output conventions with \cref{yao:comparison}.

\begin{proof}[Proof of \cref{yao:main,yao:bpp}]
\phantomsection\label{proof:yao:main}\label{proof:yao:bpp}
Given a fixed requested gap $k\ge1$, we choose $t=\ceil{k/2}+4$. Then
$2t-8=2\ceil{k/2}\ge k$.
Composing the reduction of \cref{yao:source} with \cref{yao:lift}, we obtain,
in deterministic polynomial time, a table whose cost $D_A^{\rm alt}$ is at
most $B_{\rm a}$ in the YES case and at least $B_{\rm a}+k$ in the NO case, proving \cref{yao:main}.

An additive-$u$ estimate of any cost in \cref{yao:comparison} is, by the
triangle inequality, an additive-$(u+2)$ estimate of $D_A^{\rm alt}$ with the
same success probability. We choose any fixed integer $k>2(u+2)$ and compare
the estimate with $B_{\rm a}+k/2$. This gives
$\mathsf P=\NP$ for a deterministic estimator and $\NP\subseteq\BPP$ for a
bounded-error randomized estimator, proving \cref{yao:bpp}.
\end{proof}

The address length $q_{\rm a}=2^{t-1}$ and the degree of the polynomial
bounding the running time of the source reduction may both grow with $t$. Hence the construction is polynomial only for each
fixed gap, and this argument gives no growing-error result for strict
alternation.

\section{Remaining approximation questions}
\label{sec:conclusion}

We obtained fixed-error NP-hardness in both models and algorithmic implications
for growing errors in the unrestricted model. The quantitative consequences
depend on how fast the explicit table grows with the requested gap. Smaller
tables would give faster SAT algorithms from the same approximation guarantee,
and stronger inapproximability under assumptions excluding those SAT algorithms.

For unrestricted protocols, our deterministic construction outputs a table
of size at most $N^{2^{O(k)}}$ for an $N$-variable sparse formula and gap $k$
(\cref{det:reduction}). Can smaller lists preserve the restriction bounds needed for the
next round, or can another construction improve the $\log\log T$ error scale?

For strict alternation, a construction whose size and running time depend
more mildly on the gap could let the gap grow with the table size, giving
analogous approximation-to-SAT implications in Yao's model.

\appendix
\section{Interlacing estimates and the initial matrix}\label{sec:interlacing-details}
\label{sec:interlacing_details}

We first prove the projection inequality and the small-block bounds used in
\cref{sec:surviving-block-proofs}. The surviving-block bound proved there
(\cref{det:exact}) is then used in the separation theorem and the column-loss induction below.
We use these estimates to prove \cref{lem:many-product-blocks,cl:bridge},
and finish with the initial matrix and template-layer bounds
(\cref{cl:seed,src:profiles}).

\subsection{Projecting column tuples and retained weight}

After a protocol prefix, a restriction of an interlace may retain many blocks
but only a small fraction of its column weight. We select a few blocks whose
projected column tuples still have enough weight for a lower bound. We measure the
loss by the logarithm $\Delta_p$ below, which increases by one each time the
retained column weight is halved.

Fix a Boolean matrix $M:X\times Y\to\{0,1\}$ with positive normalized row weights $\mu$ and
column weights $\nu$. Every weight below uses these original weights or their products.
For a nonempty $K\subseteq Y^p$, define
$\Delta_p(K)=\log(1/\nu^p(K))$, where $\nu^p$ is the product weight on $Y^p$. The projection onto no coordinates has loss $\Delta_0=0$, since $Y^0$ is
a single point of weight one.

We show that projection cannot increase the loss, and that projecting onto a
suitable set of $l$ of the $p$ coordinates leaves at most an $l/p$ fraction of it.

\begin{lemma}\label{cl:projection}
Let $K\subseteq Y^p$ be nonempty, and write $\pi_JK$ for its projection onto the
coordinates in $J\subseteq[p]$. Then
\begin{align*}
 \Delta_{\card J}(\pi_JK)&\le\Delta_p(K),\\
 \Delta_{\card{J_0}}(\pi_{J_0}K)+\Delta_{\card{J_1}}(\pi_{J_1}K)
 &\le\Delta_p(K) &&(J_0\sqcup J_1=[p]).
\end{align*}
For every integer $1\le l\le p$, some $l$-element set $J$ satisfies
\[
\Delta_l(\pi_JK)\le\frac lp\Delta_p(K).
\]
Viewing $K$ as a column set of $\Int{p}{M}$, the interlace on the blocks in
$J$, restricted to the columns $\pi_JK$ and to any retained rows of these
blocks, is a submatrix of the restriction of $\Int{p}{M}$ to these rows and the
columns $K$, and each block keeps its row weights (\cref{obs:representatives}).
\end{lemma}

\begin{proof}
The first two inequalities follow from
$K\subseteq\pi_J^{-1}(\pi_JK)$ and
$K\subseteq\pi_{J_0}K\times\pi_{J_1}K$, using the product weights.

For the averaged bound, we regard $\nu^p$ as a distribution, let $Z$ have distribution $Q=\nu^p(\cdot\mid K)$, and
write $Q_J$ for the distribution of $Z_J$.
For a distribution $P$ and a positive distribution $P'$ on a finite set, their relative entropy is
$D_{\rm KL}(P\|P')=\sum_zP(z)\log(P(z)/P'(z))$, and entropy is
$\mathsf H(P)=-\sum_zP(z)\log P(z)$, where zero terms are omitted.
Since $Q(z)/\nu^p(z)=1/\nu^p(K)$ for $z\in K$,
\[
D_{\rm KL}(Q\|\nu^p)=\Delta_p(K).
\]
Since $Q_J$ is supported on $\pi_JK$, nonnegativity of relative entropy gives
\[
\Delta_{|J|}(\pi_JK)\le D_{\rm KL}(Q_J\|\nu^J).
\]
For a uniformly chosen $l$-subset $J$, the chain rule and the fact that
conditioning does not increase entropy give
\[
\mathbb E_J\mathsf H(Z_J)
\ge\mathbb E_J\sum_{i\in J}\mathsf H(Z_i\mid Z_1,\ldots,Z_{i-1})
=\frac lp\mathsf H(Z).
\]
Also, since $\nu^J$ is a product, each coordinate contributes to
$-\mathbb E\log\nu^J(Z_J)$ exactly when it lies in $J$, which happens with probability $l/p$. Since
$D_{\rm KL}(Q_J\|\nu^J)=-\mathsf H(Z_J)-\mathbb E\log\nu^J(Z_J)$, subtracting the entropy
inequality yields
\[
\mathbb E_J\Delta_l(\pi_JK)
\le\mathbb E_JD_{\rm KL}(Q_J\|\nu^J)
\le\frac lpD_{\rm KL}(Q\|\nu^p)=\frac lp\Delta_p(K).
\]
Hence some $J$ satisfies the desired bound. The last assertion is \cref{obs:representatives}.
\end{proof}

\subsection{Two or three surviving blocks}

Later arguments, such as the proof of \cref{det:exact}, reduce a large
collection of surviving blocks to two or three. We therefore first prove lower
bounds for these small collections. Each proof splits into cases according
to whether Alice or Bob speaks first and uses both column-weight thresholds of \cref{def:profile}.

\begin{lemma}[Small block bounds]\label{lem:small-blocks}
Suppose $\RP{M}{x}{H}$ with $H\ge2$. A restriction of a full-product interlace of $M$
retaining the listed number of blocks, each of at least the listed base row weight,
and a column set of at least the listed product weight has at least the listed cost
whenever the listed row weight is at most one.
\[
\begin{array}{c|c|c|c}
\text{blocks}&\text{row weight per block}&\text{product column weight}&\text{cost}\\
\hline
2&2x&9/50&H\\
2&2x&9/25&H+1\\
3&4x&9/50&H+1
\end{array}
\]
\end{lemma}
\begin{proof}
For the first row, \cref{cl:projection} gives a singleton column projection
of weight $\ge\sqrt{9/50}>3/10$. Applying $\RP{M}{x}{H}$ with $j=1$ to
that block gives cost $\ge H-1\ge1$,
so the restriction is nonconstant. If Alice speaks first, we follow that block
into a child retaining base row weight $\ge x$. The retained columns are
unchanged, so the case $j=1$ of $\RP{M}{x}{H}$ again gives suffix cost at least $H-1$.
If Bob speaks first, his denser child has column weight $\ge9/100$
and hence a singleton projection of weight $\ge3/10$. The suffix at that
child again costs at least $H-1$. Including the first bit gives cost at least $H$.

For the second row, the first row gives cost at least $H\ge2$, hence nonconstancy. If Alice
speaks, some singleton projection has column weight $\ge\sqrt{9/25}=3/5$, and by $j=0$
in $\RP{M}{x}{H}$, a child retaining half the row weight of that block costs at least $H$. If Bob speaks, his denser
child has column weight $\ge9/50$, so the first row gives suffix
cost at least $H$. Including the first bit gives cost at least $H+1$.

For the third row, the projection onto any two blocks has product column weight
$\ge9/50$ and row weights $\ge2x$, so the first row shows that the restriction is not constant. If Alice speaks, we assign each block to a child
retaining at least half its row weight. One child receives two blocks of weight $\ge2x$,
and the first row gives suffix cost $H$. If Bob speaks, his denser child
retains column weight $\ge9/100$. The averaged projection inequality gives
two of its three coordinates with product weight at least
\[
(9/100)^{2/3}>9/50.
\]
Their row weights are still $\ge4x\ge2x$, so the first row again gives
suffix cost $H$. Including the first bit proves the claim. By
\cref{obs:representatives}, each projection is a restriction of the original matrix.
\end{proof}

The next theorem shows that a short protocol for a width-$u$ interlace must
begin with $\log u$ Alice bits, and that each node reached by these bits keeps most of
the row weight of one block and little of every other block. Its proof uses \cref{det:exact}, whose proof in
\cref{sec:conditioning-details} does not use the theorem.

\begin{theorem}\label{cl:separation}
Suppose $\RP{M}{\rho}{d}$ with $d\ge2$. Let $u\ge2$ be dyadic and
$1\ge\xi>2u\rho$. Interlace $M$ at width $u$ using an exactly $8$-wise independent
tuple list with the base-column marginal, as in \cref{def:exact-tuples}.
Every restriction retaining base row weight at least $\xi$ in every block
and tuple-label weight at least $18/25$ has the following property: every
protocol of depth at most $d+\log u$ for it begins with $\log u$ Alice bits
along every path.
The $u$ nodes at depth $\log u$ are in bijection with the blocks: each such
node retains more
than $\xi-2(u-1)\rho$ of its assigned block and less than $2\rho$ of every
other block. Full products are a special case, and fixed bits count as transmitted bits.
\end{theorem}

\begin{proof}
We maintain the following invariant at each protocol-tree node of depth
$r<\log u$: there are $\kappa=u2^{-r}$ assigned blocks, each retaining base row
weight $\ge\xi2^{-r}>2\kappa\rho$, and the retained column set is the original
one. The root has this property. By
\cref{obs:representatives}, the projection onto the coordinate of one assigned
block is a restriction of $M$ with column weight $\ge18/25>3/5$, so $\RP{M}{\rho}{d}$ gives cost at least $d\ge2$ and the
node is not a leaf. Since $1\ge\xi>2u\rho$, every row threshold $2^a\rho$ with
$1\le a\le\log u$ used below is less than one.

If Bob speaks, one child retains column weight $\ge9/25$ and all
$\kappa$ assigned blocks. By \cref{det:exact} with $a=\log\kappa$,
it needs $d+\log\kappa$ bits, but has only
$d+\log u-r-1=d+\log\kappa-1$ left. Thus Alice speaks. We assign each block
to a child retaining at least half its weight. If a child received more than
$\kappa/2$ blocks, it would contain at least $\kappa/2+1$ blocks of weight
$\xi2^{-r-1}>\kappa\rho$, and \cref{det:exact} with $a=\log\kappa$ would again
require $d+\log\kappa$ bits, exceeding the $d+\log\kappa-1$ left.
Both children therefore receive exactly $\kappa/2$ blocks. This proves the
invariant, and hence that Alice sends the first $\log u$ bits. In particular, a fixed Alice
bit is impossible, since it would assign every block to one child.

At depth $\log u$, each node has one assigned block of retained base row weight at least
$\xi/u>2\rho$. If such a node retained two blocks of weight at least $2\rho$, they
would require $d+1$ bits by
\cref{det:exact} with $a=1$, whereas only $d$ remain. Every other block thus has
weight less than $2\rho$. Each original block is assigned to exactly one node, and only Alice has spoken,
so its weight at the other $u-1$ nodes totals less than $2(u-1)\rho$, proving the final bound.
\end{proof}

\subsection{Induction on column loss}
\label{sec:interlace-induction}

For a restriction of $\Int{p}{M}$ with surviving columns $K$, Bob's messages
shrink $K$ and may increase the loss $\Delta_p(K)$. The following induction gives lower bounds
that tolerate this loss, and it uses only lower bounds on the base matrix.

\begin{lemma}\label{cl:defect-induction}
Let $C\ge3$, $0<\theta\le1/4$, and $0<\delta\le1/64$.
Suppose $\RP{M}{\theta}{C-1}$ (\cref{def:profile}) and every restriction of $M$ retaining row
weight at least $\theta$ and column weight at least $1-\delta$ has cost at least $C$.
Let $e\in\{0,1\}$, let $a\ge2e$ be an integer with $2^a\theta\le1$, and let
$p>2^{a-1}$ be an integer. If a restriction $G$ of $\Int{p}{M}$ retains base
row weight at least $2^a\theta$ in every block and its nonempty column set $K$ satisfies
\[
\Delta_p(K)\le(p-2^{a-1})\delta+\frac{4e}{3},
\]
then $\D(G)\ge C+a-e$. Here $C$ is only a lower bound, and we do not assume
$\D(M)=C$.
\end{lemma}

\begin{proof}
Say that $G$ is in state $(a,e)$, or $(a,e,p)$, when it satisfies the
hypotheses for these values. We prove the lower bound by induction on
\[
\mathfrak m(a,e):=2a+1-e.
\]
An Alice bit halves the row thresholds and lowers $a$.
A Bob bit increases the loss by at most one. We cover this increase either
by raising $e$ to one, which adds $4/3$ to the allowed loss, or by projecting
onto about half the blocks with \cref{cl:projection}. Each such step lowers $\mathfrak m$.

Since $C-1\ge2$, $4\theta\le1$ and $\RP{M}{\theta}{C-1}$ hold, we may apply
\cref{det:exact} with $d=C-1$ and $x=\theta$ to full-product interlaces of $M$. At column weight $9/25$, its cases $a=1,2$
give cost at least $C$ for two blocks of row weight $2\theta$, and at least
$C+1$ for three blocks of row weight $4\theta$. We call these the two-copy and
three-copy bounds. Every projection below with loss $\le3\delta+4/3$
has column weight $>9/25$, because
\[
3\delta+\frac43\le\frac{265}{192}<\frac75<-\log(9/25).
\]
The last inequality follows from \(2^7\cdot9^5<25^5\).

We first verify the base states, where an Alice bit would leave the range \(a\ge2e\). For \((a,e)=(0,0)\), the averaged
projection bound in \cref{cl:projection} gives a singleton \(J\subseteq[p]\)
with
\[
\Delta_1(\pi_JK)\le\frac{p-1/2}{p}\delta<\delta.
\]
Since \(2^{-v}\ge1-v\) for \(0\le v\le1\), applied with \(v=\Delta_1(\pi_JK)\), the projected column weight is
greater than \(1-\delta\). The block in \(J\) retains row weight
\(\ge\theta\), so the hypothesis gives cost at least \(C\).

For \((a,e)=(2,1)\), the loss bound and \cref{cl:projection} give a
three-coordinate projection with loss at most
\[
\frac3p\left((p-2)\delta+\frac43\right)
\le3\delta+\frac43<-\log(9/25).
\]
Its column weight exceeds \(9/25\), and each block retains weight
\(\ge4\theta\). The three-copy bound above gives cost at least
\(C+1=C+2-1\).
There is one further base case, \((a,e,p)=(3,1,5)\), since the Bob step below needs \(p\ge6\). Here \(\Delta_5(K)\le\delta+4/3<-\log(9/25)\), and every block retains weight \(\ge8\theta\).
Its column weight therefore exceeds \(9/25\), and \cref{det:exact} with $a=3$
applies to these five blocks, giving $C+2=C+3-1$ bits.

We also handle the Bob-first case of \((a,e)=(1,0)\) now, since \((1,1)\) is not a
valid state. The general step below covers its Alice-first case.
Some singleton projection has loss
\(\le(p-1)\delta/p<\delta\), so its column weight exceeds \(1-\delta\).
Its block retains row weight \(\ge2\theta\ge\theta\), so the hypothesis gives it
cost at least \(C\ge3\), and the restriction is nonconstant. After a Bob bit,
the denser child has a two-coordinate projection of loss at most
\[
\frac2p\left((p-1)\delta+1\right) \le2\delta+1<3\delta+\frac43,
\]
so the two-copy bound gives cost at least \(C\) after that bit. Every other
valid state has \(a\ge2\), \(p\ge3\), and base row weight \(\ge4\theta\). By \cref{cl:projection}, some
three-coordinate projection has loss \(\le3\delta+4/3\), so by the
three-copy bound the restriction is not constant.

Now consider any case not yet treated, and the first bit of a protocol for its restriction. If Alice
speaks, we assign every block to a child retaining at least half its row weight. For
child \(i\in\{0,1\}\), let \(p_i\) be the number of assigned blocks and let
\(L_i\) be the loss after projecting the retained column set onto those
blocks. Then \(p_0+p_1=p\), and the partition inequality in \cref{cl:projection}
gives \(L_0+L_1\le\Delta_p(K)\). Put
\[
s_i=\max\{p_i-2^{a-2},0\}.
\]
Since \(s_0+s_1\ge p-2^{a-1}>0\), at least one \(s_i\) is positive. The
loss bounds required of the children with \(s_i>0\) satisfy
\[
 \begin{aligned}
\sum_{i:s_i>0}\left(s_i\delta+\frac{4e}{3}\right)
 &\ge (p-2^{a-1})\delta+\frac{4e}{3}\\
&\ge\sum_{i:s_i>0}L_i.
 \end{aligned}
\]
At least one eligible child contributes the constant \(4e/3\), and the
final inequality uses the assumed loss bound on \(K\) and
\(L_0+L_1\le\Delta_p(K)\). Consequently some child with \(s_i>0\) has
\(L_i\le s_i\delta+4e/3\). Each of its assigned blocks retains weight \(\ge2^{a-1}\theta\), and
there are \(p_i>2^{a-2}\) such blocks. Since \((a,e)\ne(2,1)\), we have \(a-1\ge2e\), so the
projection of this child onto its assigned blocks, a submatrix of the child by \cref{cl:projection},
is in state \((a-1,e,p_i)\). The induction hypothesis gives cost at least
\(C+a-1-e\) in that child, and Alice's bit gives \(C+a-e\). For \((a,e)=(1,0)\),
this is the step to state \((0,0)\).

If Bob speaks, one child has at least half the current column weight, so its loss
is \(\le\Delta_p(K)+1\). When \(e=0\) and \(a\ge2\), this child satisfies
state \((a,1)\), because
\[
\Delta_p(K)+1\le(p-2^{a-1})\delta+\frac43.
\]
Its induction cost is at least \(C+a-1\), and adding Bob's bit proves the claim.

It remains to treat a Bob bit when \(e=1\) and \(a\ge3\). The extra term
\(4/3\) has then already been used, so we instead project the denser child onto
about half the blocks and show that the projection is in state \((a-1,1)\). The five-block base case already
handles \(p=5\), so \(p\ge6\). We keep \(p'=\ceil{p/2}\) blocks, chosen below by \cref{cl:projection}. Then
\[
\frac12\le\frac{p'}p\le\frac47,\qquad
\frac{p'}p(p-2^{a-1})\le p'-2^{a-2},\qquad
\frac47\left(\frac43+1\right)=\frac43.
\]
By the averaged projection bound in \cref{cl:projection}, some \(p'\)
coordinates of the denser child therefore have loss at most
\[
\frac{p'}p\left((p-2^{a-1})\delta+\frac43+1\right)
\le(p'-2^{a-2})\delta+\frac43.
\]
The projected child retains base row weight \(\ge2^{a-1}\theta\) in each
of its \(p'>2^{a-2}\) blocks, so it is state \((a-1,1)\). Its induction cost is at
least \(C+a-2\), and adding Bob's bit gives \(C+a-1\), as required.
\end{proof}

We prove \cref{lem:many-product-blocks} by applying \cref{cl:defect-induction}
to $p=9u/16$ blocks. The induction needs more than $u/2$ blocks, and the
extra $u/16$ blocks give slack $(9u/16-u/2)\delta=u/1024$ at $\delta=1/64$ in the
loss bound. The retained column weight may therefore be as small as $2^{-u/1024}$.

\begin{proof}[Proof of \cref{lem:many-product-blocks}]
\phantomsection\label{proof:lem:many-product-blocks}
Let $K\subseteq Y^q$ be the retained column set, where $q$ is the width of the
interlace. Since $u\ge16$ is dyadic, $9u/16$ is an integer. We select a set
$J$ of exactly $p=9u/16$ blocks retaining base row weight $\ge ux$ and project
$K$ onto them. The hypothesis $\nu^q(K)\ge2^{-u/1024}$ and \cref{cl:projection} give
$\Delta_p(\pi_JK)\le\Delta_q(K)\le u/1024$, and the projected interlace is a
restriction of the original one.
Apply \cref{cl:defect-induction} with
\[
a=\log u,\quad \theta=x,\quad \delta=1/64,\quad C=H,\quad e=0.
\]
Here $x\le1/16$, $p>2^{a-1}=u/2$, and $2^ax=ux\le1$.
The hypothesis $\RP{M}{x}{H}$ implies the weaker bounds $\RP{M}{x}{H-1}$
and, since $63/64\ge3/5$, cost $\ge H$ at column weight $63/64$. These are the
two lower-bound hypotheses of \cref{cl:defect-induction}. Finally,
\[
\Delta_p(\pi_JK)\le u/1024
=(9u/16-u/2)/64=(p-2^{a-1})\delta.
\]
The induction gives cost at least $C+a=H+\log u$.
\end{proof}

We use \cref{cl:defect-induction} to restore both deletion bounds at full cost for
$R^+=(\Int{Q}{W})^{\mathsf T}$.

\begin{proof}[Proof of \cref{cl:bridge}]
\phantomsection\label{proof:cl:bridge}
Put $C=\D(W)$ and $f=\ceil{b/\delta}$. The assumptions $b\ge1$, $\delta\le1/64$ and $Q\delta\ge80b$ imply
\[
f\delta\ge b,\qquad (f-1)/Q<1/80,\qquad f<Q/40,\qquad Q\ge5120.
\]
Indeed, $b/\delta\ge64$, $Q\delta\ge80b$, and
$f\le b/\delta+1<2b/\delta$. Fix $j\in\{0,1\}$ and put
\[
a_j=r-j,\qquad p_j=2^{a_j-1}+f,\qquad t_j=2^{a_j}\theta.
\]
These choices give $a_j\ge0$, $2^{a_j-1}<p_j<Q$, and $t_j<1$.

Consider a restriction of $R^+$ to which $\RP{R^+}{2^{-b}}{C+r}$ applies with
this $j$, and transpose it to a restriction of $\Int{Q}{W}$. By
\cref{def:profile}, the latter retains tuple
weight $\ge2^{-b}$ and total block-row weight $\ge(3/5)2^{-j}$.
If fewer than $p_j$ blocks had base row weight at least $t_j$, then, as each block
has weight $1/Q$, this total would be less than
\[
\frac{p_j-1}{Q}+t_j
<2^{-j-1}+\frac1{80}+\frac1{20\cdot2^j}
\le\frac35\,2^{-j},
\]
a contradiction, where the last inequality uses $j\le1$. We select a set $J$ of $p_j$ blocks meeting the row-weight threshold. For the retained tuple set $K$, \cref{cl:projection} gives
\[
\Delta_{p_j}(\pi_JK)\le\Delta_Q(K)\le b
\le f\delta=(p_j-2^{a_j-1})\delta.
\]
The projected restriction therefore satisfies the hypotheses of
\cref{cl:defect-induction} with $e=0$, $a=a_j$, $p=p_j$, and row threshold $2^{a_j}\theta=t_j$.
Its two lower-bound hypotheses are those assumed for $W$, and $\theta<1/4$ follows from $Q\theta\le1/20$.
The lemma gives cost at least $C+a_j=C+r-j$, so \cref{obs:representatives} and
transposition prove $\RP{R^+}{2^{-b}}{C+r}$.

Sending the copy index and then running a protocol for $W$ costs
$r+\D(W)$ bits. This gives $\D(R^+)=\D(W)+r$, and it turns any
upper bound $y$ on $\D(W)$ into the bound $y+r$ on $\D(R^+)$. By definition the
transpose has $n^Q$ rows and $Qm$ columns.
\end{proof}

\subsection{The initial matrix and template layers}
\label{sec:initial-template-proofs}

For the initial matrix $R_0(z,j)=z_j$ on $\{0,1\}^{p_0}\times[p_0]$, we use
real rank. Every restriction tested by the deletion bounds still has large
rank, and large rank forces large depth.

\begin{proof}[Proof of \cref{cl:seed}]
\phantomsection\label{proof:cl:seed}

Since \(p_0\) is a power of two, \(a=\log p_0\) is an integer. Fix
\(j\in\{0,1\}\), a retained row set
\(V\subseteq\{0,1\}^{p_0}\) with \(\card V\ge2^{p_0-L}\), and a retained column
set \(J\subseteq[p_0]\) with \(\card J\ge\ceil{(3/5)2^{-j}p_0}\). Under uniform
weights, these are the restrictions tested by \(\RP{R_0}{2^{-L}}{d_0}\). Let \(v\) be
the dimension of the real span of \(V\). The matrix with row set \(V\) has
rank \(v\) (with all \(p_0\) columns), so by \citet[Lemma~4.1(1)]{mackenzie2026condensation} it has at most
\(2^v\) distinct rows, that is, \(\card V\le2^v\). The
lower bound on \(\card V\) therefore implies \(v\ge p_0-L\).

Deleting the \(p_0-\card J\) coordinates outside \(J\) can lower the row rank by
at most \(p_0-\card J\). Thus the restricted initial matrix has rank \(r\)
satisfying
\[
r\ge \card J-L \ge \frac35\,2^{-j}p_0-L >2^{a-j-1}.
\]
The strict inequality follows from
\[
\frac35\,2^{-j}p_0-2^{a-j-1} =\frac{p_0}{10\cdot2^j}\ge\frac{p_0}{40}>L,
\]
where \(p_0/40-L>0\) follows from \(p_0\ge64(L+1)\).

By \citet[Lemma~4.1(2)]{mackenzie2026condensation}, a matrix of
depth at most \(a-j\ge1\) has rank at most \(2^{a-j-1}\). Hence the restriction has
depth at least \(a+1-j=d_0-j\), proving \(\RP{R_0}{2^{-L}}{d_0}\). At \(j=0\), this gives \(\D(R_0)\ge a+1\). Conversely,
Bob sends the column index in \(a\) bits and Alice sends the corresponding bit, so
\(\D(R_0)\le a+1\), and hence \(\D(R_0)=a+1=d_0\). The unit-vector rows form the
\(p_0\times p_0\) identity submatrix, so \(\rank(R_0)=p_0=2^{d_0-1}\).

By minimality of \(p_0=\operatorname{pow}_2(64(L+1))\), \(p_0<128(L+1)\). If \(L=\ceil{\log n}\) and
\(n\ge4\), then \(L+1\le\log n+2\le2\log n\), so \(p_0<256\log n\) and
\(2^{p_0}<n^{256}\). Enumerating the \(2^{p_0}\) bit strings and the \(p_0\) column
indices therefore prints the table deterministically in polynomial time.
\end{proof}

\begin{proof}[Proof of \cref{src:profiles}]
\phantomsection\label{proof:src:profiles}
We treat the full-product and explicit cases together, except for the step that turns many blocks
above the threshold into a lower bound. Fix $j\in\{0,1\}$, put
$\beta_j=(3/5)2^{-j}$, and let $\lambda=255/256$ be the total weight of the
template rows (\cref{def:source-matrix}). After transposition, each of the three deletion
bounds concerns a restriction of an interlace with the following data. Here $q$ is the interlace
width, $u$ is the block parameter to which we apply
\cref{lem:many-product-blocks} or \cref{det:rareboost}, and the last column is
a lower bound on the average block retention.
\[
\begin{array}{c|c|c|c|c}
\text{target}&\text{base}&q&u&\text{average block retention}\\ \hline
A&R&q_1=s-2&s2^{-j}&\beta_j\\
R_{\rm temp}&A&q_2&q_2 2^{-j}&\beta_j\\
N_{\rm ref}&R_{\rm temp}&s&s2^{-j}&(\beta_j-(1-\lambda))/\lambda
\end{array}
\]
In each case the retained tuple-label event $Y$ has weight $\ge x$. For
$N_{\rm ref}$, we discard the non-template rows. They have total weight
$1-\lambda$, which gives the displayed average retention. We then map each retained $\Omega$
label to its bin part, the tuple $(j_1,\ldots,j_s)$ in the full-product case
and its $\mathcal J_{\rm g}$ label in the explicit case, forgetting the tail
$\xi$, which template entries ignore. This map induces the original
weights on its target labels, so its image still has weight $\ge x$, and choosing one retained
label for each image label gives the template entries (\cref{obs:representatives}).

For the three rows, let $m$ be the number of blocks retaining base row weight at least the
threshold $t=u\rho$, $ux$, and $ux$, respectively. The block-counting bound
$q\beta\le m+qt$ of \cref{sec:models}, with $\beta$ the tabulated retention, gives the following lower bounds for $m/u$:
\[
\frac35(1-2/s)-q_1\rho,\qquad
\frac35-q_2x,\qquad
\frac{2^j(\beta_j-(1-\lambda))}{\lambda}-sx.
\]
All three exceed $9/16$ by the numerical bounds of \cref{lem:common-schedule}, which
both parameter schedules satisfy. They also give the row thresholds
$u\rho\le s\rho\le1$ and $ux\le\max\{q_2x,sx\}\le1$, and $u\ge s/2\ge16$
since $s\ge256$.

\emph{Full-product case.} Here $x=2^{-8L}\ge2^{-u/1024}$ by
\cref{lem:full-schedule}, so \cref{lem:many-product-blocks} applies to the
more than $9u/16$ blocks above the threshold.

\emph{Explicit case.} We apply \cref{det:conditioning} to $Y$ with $v=x$ and
$\tau=1$. Outside at most $e(x)$ exceptional coordinates, the marginals of
$\nu\mid Y$ on at most eight coordinates stay close to the product weights, and
$e(x)\le s/128\le u/32$ by \cref{lem:generated-schedule}. Discarding the blocks
at these coordinates leaves more than $9u/16-u/32>u/2$ blocks above the
threshold. Since $u$ is dyadic, at least $u/2+1$ of them remain. These blocks
and $K=Y$, for which $(\nu\mid Y)(K)=1$, meet the hypotheses of
\cref{det:rareboost} with $a=\log u$, which gives the costs computed below.

In both cases the base bound is $\RP{R}{\rho}{d}$ for the first row and, for
the other two rows, the deletion bounds just established for $A$ and
$R_{\rm temp}$. The required costs are therefore
\[
d+\log s-j=H_1-j,\qquad
H_1+\log q_2-j=H_2-j,\qquad
H_2+\log s-j=C-1-j,
\]
which are the three asserted pairs of deletion bounds.

It remains to prove the collision bound for the interlace on $\Gamma$.
The first $q_1$ positions and three distinct tail positions give
$q_1+3=s+1=(2s)/2+1$ blocks. By hypothesis, each retains base row weight
$\ge x_{\rm cap}=2s\rho$, and the retained list labels have weight greater
than $1/2>9/25$. In either case the list $\Gamma$ is exactly $8$-wise
independent with the base-column marginal, so \cref{det:exact} with
$a=\log(2s)$ gives $d+\log(2s)=H_1+1$ bits.
\end{proof}

\section{Bounds on transferred inputs}\label{sec:mass-details}
\label{sec:mass_bounds}

We prove \cref{res:incoming,res:foreign}, which bound the $\mathcal D$-weight
arriving at a stopping node (defined in \cref{res:destinations}) and the
weight sent to the stopping node assigned to another copy. Throughout, we
assume the common hypotheses (H1)--(H4) of \cref{sec:internal-interlace} and
drop the subscript $\rm g$ of the explicit construction. In each source copy,
the $\mathcal D$-inputs have total weight $\mu=1/256$, and the template rows
and the $\Omega$-inputs each have total weight $\lambda=255/256$.

Transferred weights below are in units of one complete copy, whereas the
fraction $\varepsilon_i$ of \cref{res:foreign} is normalized within copy $i$'s $\Omega$-inputs.
\Cref{det:conditioning} controls conditional marginals on at most eight
nonexceptional coordinates, and the projection inequality following it converts
this control into bounds in the original product weights.
By \cref{obs:representatives}, choosing one retained representative for each
projected label turns these projected tables into submatrices of the restriction.

\subsection{Two template inputs at one stopping node}

The next lemma uses the orientation of $B_{\rm raw}$, transposed relative to $N$. Its rows
are template-label pairs, and its two column families come from one
copy's $\Omega$-inputs and another copy's $\mathcal D$-inputs.

At the node assigned to copy $i$, an $\Omega$-input of copy $i$ reads the
template label at coordinate $i$, and a $\mathcal D$-input of a copy $j\ne i$
reads the template label at coordinate $j$. By (H3), these two labels are
independent under the original weights. The next lemma shows that if enough of both
families is retained, the protocol needs one bit more than the cost $H_2$
given by $\RP{R_{\rm temp}}{x}{H_2}$, even if the two families have the
same column labels in $J_2$.

\begin{lemma}[Two template inputs]\label{res:paired}
Let $A:\Gamma\times J\to\{0,1\}$ satisfy $\RP{A}{x}{H_1}$ with $H_1\ge2$,
and let $q_2$ be dyadic with $2q_2x\le1/8$. Take two weighted lists $S_o,S_f$ of words in $J^{q_2}$, each
exactly $8$-wise independent with the column weights of $A$ as marginal, and use
product weights on $S_o\times S_f$. Put $H_2=H_1+\log q_2$ and give
$J_2=[q_2]\times\Gamma$ the product of uniform index weights and the row weights of $A$ on $\Gamma$.
Rows are pairs of template labels $(z_o,z_f)\in S_o\times S_f$. The columns form two separately tagged copies of $J_2$, with entries
\[
A(\gamma,(z_o)_\alpha)\quad\text{and}\quad
A(\gamma,(z_f)_\alpha),
\]
respectively. The tags remain distinct even when the inner indices coincide.
Retain a set of pairs of original product weight at least $49/64$, columns reading $z_o$
of normalized weight at least $1-\varepsilon$, and columns reading $z_f$ of
normalized weight at least $\delta$, each tagged copy of $J_2$ having total weight one, where
\[
\varepsilon\ge0,\qquad 0<\delta\le1,\qquad
\delta>2q_2x+\varepsilon.
\]
Then the retained matrix requires at least $H_2+1$ bits.
\end{lemma}

\begin{proof}
Put $t=2q_2x$. We group columns by their tag and index $\alpha$.
These $2q_2$ groups have total retained $\Gamma$-weight at least
\[
q_2(1-\varepsilon+\delta)>q_2(1+t).
\]
If at most $q_2$ groups had weight at least $t$, this sum would be at most
$q_2+q_2t$. Thus at least $q_2+1$ groups retain weight $\ge t$.

We transpose the matrix. These groups are now distinct row blocks of an
interlace of $A$. Concatenating $S_o$ and $S_f$ with product weights gives
an exactly $8$-wise independent tuple list on $2q_2$ tagged coordinates,
since any set of at most eight coordinates splits between the two independent
lists, and each list has the required marginals on its part. We keep the surviving pair
labels as tuple labels, whose original weight is $\ge49/64>9/25$, and
apply \cref{det:exact} with the integer $a=\log(2q_2)$ and row threshold
$2q_2x\le1/8$. This gives $H_1+\log(2q_2)=H_2+1$ bits, as required.
\end{proof}

In the application (the proof of \cref{res:foreign}), we fix a bin at each
of two copies $i$ and $j$, and the pairs $(z_o,z_f)$ occur only in the bin fixed at
copy $i$, which we call the \emph{distinguished bin}. In the other bins of copy $i$, we know
nothing about the entries in the columns of copy $j$, so the next claim
extends \cref{res:paired} to allow arbitrary entries there.

\begin{claim}\label{cl:two-stream}
Use the source weights of \cref{def:source-matrix} and the common hypotheses (H1)--(H4) of \cref{sec:internal-interlace}, and put
$\lambda_0=2q_2x$, $\tau_0=2sx$, $C=H_2+\log s+1$, and
\[
a_{\min}=\frac{99/100+\lambda^2-1}{\lambda^2}.
\]
Retain $\Omega$-inputs of normalized weight at least
$1-\varepsilon$, and separately tagged $\mathcal D$-inputs of another copy whose
weight-preserving map $(\operatorname{tail}(\gamma),j,\ldots,j)\mapsto j$ to $J_2$
has retained image weight at least $\delta$.
Number the bins so that the distinguished bin is bin one. This bin retains a set
$P_*\subseteq S^2$ of pairs $(z_o,z_f)$ of product weight at least $a_{\min}$, and every other bin retains
a set of labels of $S$-weight at least $\tau_0$. As in \cref{def:source-matrix}, write a column in $\Omega$ as
$Z=(\xi,j_1,\ldots,j_s)$, where $\xi\in Y^{s+1}$ and $j_p=(\alpha_p,\gamma_p)\in J_2$. Require entries
\[
\begin{array}{c|cc}
 & Z\in\Omega&\text{other-copy image }(\alpha,\gamma)\\ \hline
(z_o,z_f)\in P_*&A(\gamma_1,(z_o)_{\alpha_1})&A(\gamma,(z_f)_\alpha)\\
(p,z),\ p\ne1&A(\gamma_p,z_{\alpha_p})&\text{arbitrary}.
\end{array}
\]
If $\varepsilon\ge0$, $0<\delta\le1$, and $\delta>\lambda_0+4\varepsilon$, then every such matrix
requires at least $C$ bits.
\end{claim}

\begin{proof}
Suppose this matrix has a protocol of depth at most $C-1$. In each bin, we
follow the retained set of template labels through Alice's bits until Bob
first speaks, using the tracking rule stated before
\cref{lem:collision-tracking} with $t=\tau_0$. The only marked set is the set
$P_*$ of pairs in the distinguished bin. For the thresholds, we measure the
weight of $P_*$ and of its unsplit descendants by their projections onto $z_o$, and a
split replaces the two parts by their projections onto $z_o$, which are sets
of labels from $S$. A pair set weighs no more than its projection onto $z_o$,
so by the tracking rule an unsplit pair set loses less than $\tau_0$ of its original weight.
The numerical bounds of \cref{lem:common-schedule} give
$\tau_0\le10^{-3}$ and $a_{\min}-\tau_0>49/64$. An unsplit set of pairs therefore
retains weight $>49/64$, and every retained set at depth $r$ has original
weight $\ge\tau_0 2^{-r}$.

The hypothesis on $\delta$ implies $\varepsilon<1/4$, so the columns from $\Omega$
retain normalized weight $>3/4$, and so does their projection onto any bin.
Since $H_2>0$, the bounds $\RP{R_{\rm temp}}{x}{H_2}$ rule out a leaf
containing a retained set of weight $\ge x$. A retained set also cannot reach depth $\log s+1$:
it would have weight at least $x$, but only $H_2-1$ bits would remain.
Thus every retained set reaches a Bob node by depth $\log s$.

At each stopping node $v$, let $h_v$ be the number of tracked sets and $z_v$
the number of unsplit pair sets among them. By the tracking rule,
$\sum_v(h_v+z_v)$ equals $s+1$, the $s$ initial sets plus the one initial
mark. Thus the stopping nodes satisfy the hypotheses of
\cref{lem:kraft-count} with $q=s$.
Let $v$, at depth $r$, be the node given by that lemma, and write $\kappa=\kappa_v=s2^{-r}$,
$h=h_v$, $z=z_v$. The sets at $v$ have distinct bin coordinates, since
descendants in different branches never rejoin, and each has original weight
$\ge\tau_0 2^{-r}=2\kappa x$. By \cref{lem:kraft-count},
$h\ge\kappa/2+1$ if $\kappa\ge2$, while for $\kappa=1$, either $h\ge2$ or $h=z=1$.

We first treat the case $\kappa=1$, $h=z=1$, before restricting columns or projecting the pair.
Since only Alice has spoken, both tagged column families are still unchanged. We project columns from $\Omega$
onto $j_1$ and other-copy columns onto their $J_2$ images, choosing one surviving
column per image label. These maps preserve the original weights, so the projected weights
are at least $1-\varepsilon$ and $\delta$. The set of pairs has weight
$>49/64$. Thus \cref{res:paired} requires $H_2+1$ bits, while the stopping
node has only $C-1-r=H_2$ bits left. The original weights on pairs of template
labels are the product of their weights on $S$, as required by that lemma.

In all other cases, at Bob's bit we choose the child with more weight from
$\Omega$, which retains weight $>3/8>9/25$. We replace any set of pairs by its
projection onto $z_o$, which has at least its weight. The template entries at
the chosen child now form a restriction of an interlace of $R_{\rm temp}$ with one
block for each tracked bin. The original column list of this interlace is the bin part
$(j_1,\ldots,j_s)$ of $\Omega$, which by (H3) has exactly
$8192$-wise independent bin coordinates under the original weights in either
construction. We forget the tail $\xi$, which template entries ignore, but keep
the whole tuple $(j_1,\ldots,j_s)$ in the
full-product case and its whole list index in the explicit case. The retained
image has original weight $>3/8$, and choosing one surviving row or column
for each projected label gives the required entries (\cref{obs:representatives}). Thus
\cref{det:exact} applies in both constructions. For $\kappa\ge2$ we
take $a=\log\kappa$. The $\kappa/2+1$ blocks have row weight
$\ge2\kappa x\ge\kappa x$, so the lemma requires $H_2+\log\kappa$ bits, but
the child has only $H_2+\log\kappa-1$ bits left. In the remaining case $\kappa=1$, $h\ge2$, there are at least two
blocks of weight $\ge2x$, and $a=1$ gives the stronger bound $H_2+1$.
The conditions $2^ax\le1$ hold since $2^a\le s$ and $sx\le10^{-4}$ by \eqref{eq:width-margins}. This
contradiction does not use the other-copy entries outside the distinguished
bin, which proves the claim.
\end{proof}

\subsection{Total weight arriving at a stopping node}
\label{sec:incoming-bound-proof}

We first bound the total $\mathcal D$-weight arriving at one stopping node,
including weight from copies without an assigned node. Too much incoming
weight leaves many blocks of substantial row weight, and \cref{det:rareboost},
applied after conditioning, then requires more communication than remains.

\begin{proof}[Proof of \cref{res:incoming}]
\phantomsection\label{proof:res:incoming}
At a node at depth $\log w$ we put $\kappa=1$, so that both bounds read $2s\kappa$. Fix any collection of copies, and suppose
the weight of their inputs from $\mathcal D$ arriving at one stopping node is at least $2s\kappa$.
Let $d_i\in[0,1]$ be the fraction of copy $i$'s inputs from $\mathcal D$ reaching this node,
and set $d_i=0$ outside this collection. Since each such family has weight
$\mu=1/256$,
\[
D_0:=\sum_i d_i\ge512s\kappa.
\]
These fractions are fixed by the Alice-only transcript. At a node where Bob first speaks before depth $\log w$,
we choose a child whose column set $Y^*$ has original weight $\ge\theta/2$,
which is possible since the tuple event $Y'$ of \cref{res:destinations} has weight $\ge\theta$.
At a node at depth $\log w$ we put $Y^*=Y'$.

To apply the template lower bounds, we need retained column tuples in which
at least half of the incoming weight $D_0$ comes from copies $i$ at which the
tuple's label is a template row.
Let $T_i$ indicate that the label at coordinate $i$ of a column tuple is a template row. When the type, bin, and item labels are drawn according to their fixed weights, the centered variables $T_i-\lambda$ are $256$-wise independent.
The weighted form of \cref{lem:conditioning-moment} with $r=128$ and coefficients
$d_i$, followed by Markov's inequality, gives
\[
\Pr\left\{\sum_i d_iT_i<D_0/2\right\}
\le C_{128}D_0^{-128}\le C_{128}s^{-128}\le\theta/4.
\]
Here $C_{128}=(2^{14}/(\lambda-1/2)^2)^{128}$ is the fixed constant
of \eqref{eq:common-budgets}, and the last inequality is a row of \eqref{eq:width-margins}.
Thus $Y^*$ intersects the event $\sum_i d_iT_i\ge D_0/2$ in weight
$\ge\theta/4$.
This event depends only on the type/bin/item sequence. Averaging over that
sequence and the base-label sequence, we fix a
configuration whose retained set $Z$ of template-label sequences has original weight
$\ge\theta/4$ and whose template positions have total weight
$\sum_i d_iT_i\ge D_0/2$.
By (H3), fixing those two sequences leaves the template-label sequence
with its original weights and exactly $8192$-wise independent
$J$ coordinates $z_{i,\alpha}$. We apply
\cref{det:conditioning} to $Z$ with $v=\theta/4$ and $\tau=1$, which gives an exceptional set of at most
$e(\theta/4)\le sq_2$ coordinates $(i,\alpha)$ by \eqref{eq:width-margins}.

By \cref{def:source-matrix}, each retained input $(\alpha,\gamma)$ from $\mathcal D$ at a copy $i$ with $T_i=1$ reads
$A(\gamma,z_{i,\alpha})$, regardless of its fixed bin. We group these inputs by $(i,\alpha)$
and measure each group within its fiber, that is, by the fraction of the inputs
from $\mathcal D$ of copy $i$ with index $\alpha$ that it contains. Since the
index $\alpha$ is uniform, the $q_2$ groups of copy $i$ then have total weight $q_2d_i$,
so the groups with $T_i=1$ have total base-row weight $\ge q_2D_0/2$.
Set $K_0=2s\kappa q_2$ and $t=K_0x$. From $w^2q_2x\le1$ and $512s\kappa\le D_0\le w$,
\[
t\le\frac{2s\kappa}{w^2}\le\frac1{256w}<1.
\]
There are at most $wq_2$ groups, each of weight at most one. If $h$ groups have weight
at least $t$, the total weight of all groups is at most $h+wq_2t$, so
\[
h\ge\frac{q_2D_0}{2}-\frac{K_0}{w}
\ge K_0(128-1/w)>K_0\ge K_0/2+sq_2.
\]
After deleting the exceptional coordinates, more than $K_0/2$ blocks outside that exceptional set
remain, each of base row weight $\ge K_0x$. We retain $K_0/2+1$ of them and
all labels in $Z$. Distinct triples $(i,\alpha,\gamma)$ label distinct incoming
rows, so these rows and $Z$ form a restriction with the required entries. Row weights use the original $\Gamma$ weights, and the
retained column set is all of $Z$, so it has weight one under the conditional weights on $Z$.

Since $K_0$ is dyadic, we can apply \cref{det:rareboost} with base bounds
$\RP{A}{x}{H_1}$ and integer $a=\log K_0$.
It requires $H_1+\log K_0=C+\log\kappa$ bits.
Since the protocol has depth at most $C+\log w-1$, if Bob first speaks at a depth $r<\log w$, the chosen child has only $C+\log\kappa-2$ bits left, and a node at depth $\log w$
has only $C-1$. This proves the bound for every collection of copies.
\end{proof}

\subsection{Transfer to another copy's assigned node}
\label{sec:foreign-bound-proof}

We next bound the $\mathcal D$-weight that one copy can send to the node assigned to
another copy $i$, which already retains all but the normalized fraction $\varepsilon_i$ of
copy $i$'s $\Omega$-inputs. A large transfer would yield a restriction
of the form in \cref{cl:two-stream}, which needs more bits than remain at that node.

\begin{proof}[Proof of \cref{res:foreign}]
\phantomsection\label{proof:res:foreign}
Let $\varepsilon_i$ be the normalized $\Omega$-weight missing at the node assigned to copy $i$, and put
$\delta_i=2q_2x+4\varepsilon_i$. If $\delta_i\ge1$, the bound $\mu\delta_i$ is automatic, since
the inputs from $\mathcal D$ in one copy have weight only $\mu\le\mu\delta_i$. Otherwise, suppose the transfer
exceeds $\mu\delta_i$. Then the node receives a set of inputs labeled by $\mathcal D$ from some copy $j\ne i$ of actual
weight $\mu\delta$ for some $\delta>\delta_i$.

Both copies have assigned nodes, so their indices lie outside the exceptional set for $Y'$, and only Alice has spoken before
reaching the assigned node. By \cref{det:conditioning} with error bound $\eta/2$ and the display
after it, the projection of $Y'$ onto
coordinates $i$ and $j$ has original product weight $\ge1-\eta/2>99/100$.
We restrict to the event that both coordinates are template rows. This event
has weight $\lambda^2$, so the
retained relative weight is at least
\[
a_{\min}=\frac{99/100+\lambda^2-1}{\lambda^2}.
\]
We average over the two bin labels and fix a pair of bins attaining this relative
weight. The retained pairs of template labels in these two bins form a set $P_*\subseteq S^2$ of original
product weight $\ge a_{\min}$. The bin fixed at copy $i$, which we call $p_i$, is the distinguished bin. The
projection at coordinate $i$, which lies outside the exceptional set, loses at most $\eta/2$ of its row weight
by the same error bound, while each bin has weight $\lambda/s$,
so every other
bin retains weight at least
\[
1-\frac{\eta s}{2\lambda}>2sx.
\]
We choose one surviving completion in $Y'$ for every pair in $P_*$ and
for every retained template label in the other bins. The bin label at $i$ distinguishes
the representatives of the other bins from those of the pairs, and the projected labels distinguish the
representatives within each bin. Thus the chosen completions define a
restriction of the original table with the claimed entries.

After transposition, the columns labeled by $\Omega$ from copy $i$ have normalized weight
$1-\varepsilon_i$. The columns labeled by $\mathcal D$ from $j$ have $J_2$-image weight
$\delta>\delta_i$, since that image map preserves their original weights. A pair
$(z_o,z_f)$ reads $A(\gamma_{p_i},(z_o)_{\alpha_{p_i}})$ on columns labeled by $\Omega$ and
$A(\gamma,(z_f)_\alpha)$ on columns labeled by $\mathcal D$. Other bins have the same
template entries on the first family, and their entries on the second family are arbitrary. This is the
restriction in \cref{cl:two-stream}, after renaming the distinguished bin.
The claim requires $C$ bits because $\delta>2q_2x+4\varepsilon_i$, whereas the
assigned node has only $C-1$ bits left. This contradiction proves that the
transfer is at most $\mu\delta_i$, also when $\varepsilon_i=0$.
\end{proof}

\section{Conditioning and surviving blocks}
\label{sec:conditioning-details}

We prove \cref{det:conditioning} by bounding the number of exceptional
coordinates, and then show that a short protocol must leave two or three
blocks to which \cref{lem:small-blocks} still applies. For these estimates only,
we view a label drawn according to its fixed weights as a random variable.
Expectations then express weighted sums over the fixed tuple list. The
$1024$th-moment bounds use the original weights $\nu$, under which
independence is exact, whereas the fourth-moment bound under $\nu\mid Y$ in
\cref{lem:pair-projections} uses only the approximate marginals outside the
exceptional coordinates.

\subsection{A moment count and the exceptional coordinates}
\label{sec:conditioning-moments}

We use the following moment bound twice: to bound the number of exceptional
coordinates, and, in the proof of \cref{res:incoming}, to bound the probability
that too few template positions survive. In the second use the coefficients are
the fixed fractions $d_i$ of incoming weight, so the bound applies directly to
the incoming template weight $\sum_id_iT_i$.

\begin{lemma}[Moment bound]\label{lem:conditioning-moment}
Let $r\ge1$ be an integer. Suppose $X_1,\ldots,X_m$ have mean zero,
$|X_i|\le1$, and every subfamily of at most $2r$ variables is independent.
For fixed coefficients $a_i\in[0,1]$ with $W=\sum_i a_i\ge1$,
\[
\mathbb E\left|\sum_i a_iX_i\right|^{2r}\le(r^2W)^r.
\]
Unit coefficients give the same bound with $W=m$, and the empty sum is zero.
In particular, centered indicators with $1024$-wise independence satisfy
\[
\mathbb E\left|\sum_iX_i\right|^{1024}\le(2^{18}m)^{512},
\qquad
\mathbb E\left|\sum_iX_i\right|^{256}\le(2^{14}m)^{128}.
\]
\end{lemma}
\begin{proof}
We expand $(\sum_ia_iX_i)^{2r}$ and group the terms by the pattern of equal
indices among the $2r$ factors. A term in which some index appears once has
expectation zero. Every
remaining pattern has $k\le r$ index classes, each of size at least two.
Labeling the classes in order of first appearance shows that there are at
most $r^{2r}$ such patterns. If their sizes are $n_1,\ldots,n_k$, then,
since $|X_i|\le1$, the absolute contribution from a fixed pattern is at most
\[
\sum_{\substack{i_1,\ldots,i_k\\\text{distinct}}}
a_{i_1}^{n_1}\cdots a_{i_k}^{n_k}\le W^k\le W^r,
\]
because $a_i^{n_j}\le a_i$. Summing over patterns proves the bound.
Taking $r=512$ and $r=128$ gives the displayed special cases.
\end{proof}

We take the exceptional coordinates to be the union of a maximal disjoint
family of sets of at most eight coordinates with large marginal error. \Cref{lem:conditioning-moment} bounds the size of this family.

\begin{proof}[Proof of \cref{det:conditioning}]
\phantomsection\label{proof:det:conditioning}
Put $\varepsilon=\varepsilon(v,\tau)$. We choose a maximal disjoint family
$J_1,\ldots,J_m$ of sets of at most eight coordinates with
$d_{\rm TV}((\nu\mid Y)_{J_i},\mu^{J_i})>\varepsilon$.
If $m=0$, we delete nothing. Otherwise, since $\nu$ has marginal $\mu^{J_i}$ on $J_i$, the definition of total variation distance
gives an event indicator $f_i$ on each $J_i$ with
$\mathbb E_{\nu\mid Y}f_i-\mathbb E_\nu f_i>\varepsilon$ (take its complement
if necessary). Under the original weights $\nu$, any subfamily of at most $1024$
such indicators is independent, because its disjoint sets use at most
$8192$ coordinates.

Put $S=\sum_i(f_i-\mathbb E_\nu f_i)$, so $\mathbb E_{\nu\mid Y}S>m\varepsilon$.
Jensen's inequality and \cref{lem:conditioning-moment} give
\[
v(m\varepsilon)^{1024}
<\nu(Y)\mathbb E_{\nu\mid Y}|S|^{1024}
\le\mathbb E_\nu|S|^{1024}
\le(2^{18}m)^{512}.
\]
Thus $m<2^{18}\varepsilon^{-2}v^{-1/512}$.
We delete $E=\bigcup_iJ_i$, so $|E|\le8m\le e(v,\tau)$. Every remaining set of
at most eight coordinates has marginal within $\varepsilon$ of the product,
since otherwise it could be added to the disjoint family, contradicting maximality.
\end{proof}

\subsection{Some coordinate pair has a large projection}

When many blocks survive, we apply the two-block case of \cref{lem:small-blocks} to a pair of
coordinates on which the retained labels project with product weight at least
$9/25$. Such a pair exists because, if every one of many disjoint pairs had a
smaller projection, the retained set itself would be too small.

\begin{lemma}\label{lem:pair-projections}
Fix disjoint coordinate pairs $J_1,\ldots,J_m$ of a tuple list with marginal
$\mu$ and label weights $\nu$. Suppose a label set $K$ satisfies
$\mu^2(\pi_{J_i}K)<9/25$ for every $i\in[m]$.
For an exactly $8$-wise independent list,
\[
\nu(K)<\frac2{m^2}<\frac9{50m}\qquad(m\ge16).
\]
For an exactly $8192$-wise independent list, let $K\subseteq Y$ with
$\nu(Y)\ge v>0$, and choose the pairs outside the exceptional set from
\cref{det:conditioning}, with any error bound $0<\tau\le1$. Then
\[
(\nu\mid Y)(K)<\frac1{4m}\qquad(m\ge16).
\]
Thus a retained set of conditional weight at least $1/(4m)$ forces one of
these pairs to have product projection weight at least $9/25$.
\end{lemma}
\begin{proof}
Write $P_i=\pi_{J_i}K$ and put
\[
p_i=\mu^2(P_i)<9/25,\qquad
X_i=\ind[z_{J_i}\in P_i]-p_i,\qquad S=\sum_{i=1}^mX_i.
\]
On $K$ every indicator equals one, so $S>(16/25)m$. Under the full product
weights, the centered indicators are independent, and
$\mathbb EX_i^2,\mathbb EX_i^4\le1/4$. Only terms with no
singleton index survive the fourth-power expansion. Hence
\[
\mathbb E_{\rm prod}S^4
=\sum_i\mathbb EX_i^4+6\sum_{i<j}\mathbb EX_i^2\mathbb EX_j^2
\le\frac m4+\frac6{16}\binom m2\le\frac{m^2}4.
\]
Exact $8$-wise independence gives the same fourth moment under $\nu$,
since each term uses at most four pairs. Markov's inequality for $S^4$ gives
$\nu(K)<2/m^2$, because $1/[4(16/25)^4]<2$.
For $m\ge16$, $2/m^2\le1/(8m)<9/(50m)$.

Now suppose the pairs lie outside the exceptional set for the fixed event $Y$.
Every term in $S^4$
lies in $[-1,1]$ and depends on a set $J$ of at most eight coordinates outside that set.
Since $\nu$ has marginal $\mu^J$, \cref{det:conditioning} shows that conditioning on $Y$ changes
the term's expectation by at most $2\varepsilon$, where
$\varepsilon=\varepsilon(v,\tau)$. Hence $\mathbb E_{\nu\mid Y}S^4\le m^2/4+2\varepsilon m^4$, and Markov's inequality gives
\[
p:=(\nu\mid Y)(K)
\le\frac2{m^2}+\frac{2\varepsilon}{(16/25)^4}
\le\frac2{m^2}+16\varepsilon.
\]
The error term $16\varepsilon$ does not decrease with the number of pairs,
so for large $m$ we also use the original weights $\nu$. Any subfamily of at most $1024$ pair indicators involves at most $2048$
coordinates and is therefore independent under $\nu$. Since $K\subseteq Y$ and
$\nu(Y)\ge v$, \cref{lem:conditioning-moment}, Markov's inequality and
$(25/16)^2<4$ give
\[
p=\frac{\nu(K)}{\nu(Y)}
\le\frac{(2^{18}m)^{512}}{v((16/25)m)^{1024}}
\le\frac{(2^{20}/m)^{512}}v.
\]
Combining the bounds and using
$\min\{u,w\}\le u^{511/512}w^{1/512}$ yields
\[
\begin{aligned}
mp&\le\frac2m+
\min\{16\varepsilon m,2^{20\cdot512}v^{-1}m^{-511}\}
\\
&\le\frac2m+(16\varepsilon)^{511/512}2^{20}v^{-1/512}.
\end{aligned}
\]
The last term is at most
\[
2^{20-26\cdot511/512}v^{255/131072}<\frac1{32},
\]
by $\varepsilon\le2^{-30}v^{1/256}$ and $0<v\le1$. For $m\ge16$, this gives
$mp<1/8+1/32<1/4$, as required.
\end{proof}

\subsection{Small projections and short continuations}
\label{sec:surviving-block-proofs}

Under $\RP{M}{x}{d}$ with $d\ge2$, \cref{lem:small-blocks} with $H=d$ shows that
$d+1$ bits are needed for two blocks of row weight $2x$ with product column
weight $9/25$, and for three blocks of row weight $4x$ with product column
weight $9/50$, whenever this row threshold, $2x$ or $4x$, is at most one. We apply
\cref{lem:small-blocks} to the projections of \cref{obs:representatives}, whose
column weight $\mu^J(\pi_JK)$ is a product weight even when the labels have conditional weights. The proofs of \cref{det:exact} and
\cref{det:rareboost} follow a protocol for $a-1$ messages and then apply the next
lemma, which shows that enough surviving blocks and column weight still force
$d+1$ bits.

\begin{lemma}\label{det:suffix}
Suppose $d\ge2$ and $\RP{M}{x}{d}$, and let $b\ge0$ be an integer with
$2^{b+1}x\le1$. Consider an interlace of $M$ on a tuple list whose marginal is
the column weights of $M$, and a restriction of it that retains at least $2^b+1$ blocks, each
of base row weight at least $2^{b+1}x$, and a label set $K$. Assume one of the
following.
\begin{enumerate}
\item[(E)] The list is exactly $8$-wise independent with weights $\nu$, and
$\nu(K)\ge(9/25)2^{-b}$.
\item[(C)] The list is exactly $8192$-wise independent with weights $\nu$,
$\nu(Y)\ge v>0$, the exceptional set is chosen from \cref{det:conditioning}
with error bound $0<\tau\le1$, the retained blocks have indices outside that set,
$K\subseteq Y$, and $(\nu\mid Y)(K)\ge2^{-b-1}$.
\end{enumerate}
Then the restriction requires at least $d+1$ bits.
\end{lemma}
\begin{proof}
Let $J$ be any set of at most eight retained block indices. Under (E), exact
marginals give $\mu^J(\pi_JK)\ge\nu(K)\ge(9/25)2^{-b}$
(\cref{obs:representatives}). Under (C), the indices in $J$ lie outside the
exceptional set. With
$\varepsilon=\varepsilon(v,\tau)\le2^{-30}$, \cref{det:conditioning} gives
\[
\mu^J(\pi_JK)\ge(\nu\mid Y)(K)-\varepsilon
\ge2^{-b-1}-\varepsilon.
\]
For $b\le4$ this is again $\ge(9/25)2^{-b}$, since the smallest margin
is $(7/50)2^{-4}=7/800>2^{-30}$.

For $b\le4$ we select the blocks directly. If $b=0$, any two blocks meet the
$9/25$ threshold above, and if $b=1$, any three meet the $9/50$ threshold. If $b=2$, the projection onto five retained blocks has
product weight $\ge9/100$, so \cref{cl:projection} with $l=2$ gives a pair of weight
$\ge(9/100)^{2/5}>9/25$. If $b=3$ or $b=4$, the projection onto eight retained
blocks has weight $\ge9/400$, so \cref{cl:projection} gives a pair of weight
$\ge(9/400)^{1/4}>9/25$.

For $b\ge5$, we choose $m=2^{b-1}\ge16$ disjoint pairs from the at least $2m+1$
retained blocks. If every pair projection had product weight less than
$9/25$, \cref{lem:pair-projections} would give $\nu(K)<9/(50m)=(9/25)2^{-b}$
under (E), and $(\nu\mid Y)(K)<1/(4m)=2^{-b-1}$ under (C), a contradiction.

Each selected block has row weight $\ge2^{b+1}x$, hence $\ge2x$,
and $\ge4x$ in the triple at $b=1$. These thresholds are at most one by
$2^{b+1}x\le1$. Thus \cref{lem:small-blocks} gives $d+1$ bits in every case.
\end{proof}

\begin{proof}[Proof of \cref{det:exact,det:rareboost}]
\phantomsection\label{proof:det:exact}\label{proof:det:rareboost}
In \cref{det:exact} we measure column weight under $\nu$, and in
\cref{det:rareboost} under the fixed conditional weights $\nu\mid Y$, where
the event $Y$ and its exceptional set stay fixed. Suppose a protocol for the
restriction has depth at most $d+a-1$. We follow it for $a-1$ messages by a
weight-halving walk. At each Alice message, we assign every current block to
a child retaining at least half its row weight and keep a child assigned at
least half the blocks. At each Bob message, we keep a child retaining at least
half the current column weight. If the protocol ends earlier, we pad the leaf with dummy Alice messages, which leave the
restriction unchanged. If $A$ messages are Alice's and $B$ are Bob's, then
$A+B=a-1$, at most $d$ bits remain, and the surviving restriction, with label
set $K'$, has
\[
\begin{aligned}
\text{number of blocks}
&\ge\ceil{(2^{a-1}+1)/2^A}\ge2^B+1,\\
\text{row weight per block}&\ge2^{a-A}x=2^{B+1}x,\\
\text{column weight of }K'&\ge2^{-B}\cdot\text{initial column weight}.
\end{aligned}
\]
In \cref{det:rareboost} the surviving blocks still have indices outside the
exceptional set. The initial column weight is $\nu(K)\ge9/25$ in
\cref{det:exact} and $(\nu\mid Y)(K)\ge1/2$ in \cref{det:rareboost}, so $K'$
satisfies hypothesis (E), respectively (C), of \cref{det:suffix} with $b=B$.
Its row threshold is at most one because $B+1\le a$ and $2^ax\le1$. Hence
it requires $d+1$ further bits, while only $d$ remain.
\end{proof}

\section{Constructing small rational tuple lists}
\label{sec:support-algorithm}

The deterministic reduction prints its tuple lists together with their
rational weights, so \cref{det:code} must bound the encoding length and not
only the number of tuples. We use the sequential construction of
\citet[Section~3]{koller1994sample}: extend each word by every coordinate
value with its product weight, then reduce the list while preserving the
joint weights on all sets of at most $\mathsf D$ coordinates. We recall the
procedure and bound the bit length of the weights.

\begin{proof}[Proof of \cref{det:code}]
\phantomsection\label{proof:det:code}
Let $Y$ be the alphabet, $|Y|=m$, and put $P=(1+qm)^{\mathsf D}$. We start
with the empty word of weight one. At step $i\in[q]$, we append every
value $y\in Y$ to every retained word and multiply its weight by $\mu(y)$. On the resulting candidate
set $\mathcal U\subseteq Y^i$, the weights satisfy, by induction on $i$,
\[
\sum_{u\in\mathcal U}w_u\ind[u_J=a]
=\prod_{j\in J}\mu(a_j)
\qquad
(J\subseteq[i],\ |J|\le\mathsf D,\ a\in Y^J).
\tag{*}\label{eq:support-cylinders}
\]
There are at most $P$ equalities in \eqref{eq:support-cylinders}, including
the normalization equation for $J=\varnothing$, since
$\sum_{k\le\mathsf D}\binom ik m^k\le(1+qm)^{\mathsf D}$. The constraint column
$a(u)$ of a word $u$ has entry $\ind[u_J=a]$ in row $(J,a)$. The reduction
below removes linear dependences among these columns until at most
$P$ words remain, and it preserves the equalities and the nonnegativity of the
weights. Each stage therefore extends at most $P$ words to at most $mP$ candidates.
After $q$ stages the equalities give the required $\mathsf D$-wise independence.

Words of weight zero are deleted. A reduction step takes a nonzero integer
dependence $\sum_uc_ua(u)=0$, whose row $J=\varnothing$ gives $\sum_uc_u=0$,
so some $c_u>0$. It sets $w'_u=w_u-tc_u$, where
$t=\min_{c_u>0}w_u/c_u>0$, which keeps the weights nonnegative and
\eqref{eq:support-cylinders} valid, and removes at least one word.
With more than $P$ words, we take $r\le P$
columns forming a basis of their span, one further column, and $r$ rows on
which the basis columns are independent. The signed $r\times r$ minors of
this $r\times(r+1)$ zero--one matrix give a nonzero integer relation among these
columns, valid on these rows. This relation holds in every row, since the basis spans the further column with
coefficients determined by these rows. By Hadamard's bound, each
coefficient has absolute value at most $\max\{1,r^{r/2}\}$, hence $O(P\log(P+1))$ bits. Exact linear
algebra finds this relation in polynomial time, and fixed orderings of words and rows fix every choice.

Let $L_\mu$ be the total binary encoding length of the rationals $\mu(y)$. An extension
multiplies the common denominator of the weights by at most the product of the
denominators in $\mu$, adding at most $L_\mu$ bits. A reduction step multiplies
it by the coefficient $c_u>0$ attaining the minimum ratio, adding
$O(P\log(P+1))$ bits. A stage has at most $mP$ reduction steps, each deleting a word. Across $q$ extensions and at most $qmP$ reduction steps,
every denominator therefore has bit length
\[
O\bigl(qL_\mu+qmP^2\log(P+1)\bigr).
\]
Numerators obey the same bound because the weights lie in $[0,1]$. Hence
all arithmetic has polynomial cost for fixed $\mathsf D$, and so does the
output of at most $P$ weighted words of length $q$.
\end{proof}

\section{Printing a sample of the implicit table}\label{sec:sampling-algorithm}
\label{app:implementation}

The randomized reduction uses the implicit table only through sampled rows,
columns and their entries, which we generate without printing the table. We
first bound the length and generation cost of complete row and column labels,
then implement the elementary draws with fixed blocks of fair bits. Evaluating
one entry reads only one base entry per recursion level, but the sampler
outputs complete labels, including the coordinates that the evaluation never reads.

\begin{proof}[Proof of \cref{lem:sample-access}]
\phantomsection\label{proof:lem:sample-access}
We fix a source graph on $n$ vertices above the size cutoff in \cref{lem:full-schedule}, put $L=\ceil{\log n}$, and use the parameter choices of that lemma at
every level. Write $R:X\times Y\to\{0,1\}$ for the current base matrix.

\emph{Labels and entries.}
The two interlaces $U=\Int{w}{N}$, $W=U^{\mathsf T}$ and
$R^+=(\Int{Q}{W})^{\mathsf T}$ give the row and column labels of $R^+$ (the successor labels)
\[
\mathbf a=((i_t,Z_t))_{t\in[Q]},\qquad
\mathbf b=(j,(k_i)_{i\in[w]}),
\]
where $i_t\in[w]$ and $j\in[Q]$ are uniform, $Z_t$ are tagged columns of $B_{\rm raw}$,
and $k_i\in\mathcal K$ are rows of $B_{\rm raw}$. We draw all these coordinates independently, each according to its weight
in \cref{def:interlace,def:source-matrix}. The entry rule is
\[
R^+(\mathbf a,\mathbf b)=B_{\rm raw}(k_{i_j},Z_j).
\]
The entry rules in \cref{def:source-matrix} select exactly one entry of
$R$. To generate a row $k_i$ or column $Z_t$ of $B_{\rm raw}$, we sample according to the row-type and column-family weights from that
definition and draw every listed coordinate, including those not visited by
the eventual entry evaluation. A $\mathcal D$-input needs one draw of $j=(\alpha,\gamma)\in J_2$, since its
other positions copy $j$ or $\operatorname{tail}(\gamma)$.
We generate only actual input coordinates, not the auxiliary sequences used
only in the lower-bound proof. The numbers of rows and columns of $R$ drawn
for each label are
\[
\begin{array}{c|c|c}
\text{label}&\text{base-row draws}&\text{base-column draws}\\ \hline
\text{template row}&q_2&0\\
\text{row for an item}&1&0\\
\Omega\text{ column}&0&(s+1)+s(2s-1)=2s^2+1\\
\mathcal D\text{ column}&0&2s-1\\
\text{successor row}&0&\le Q(2s^2+1)\\
\text{successor column}&\le wq_2&0
\end{array}
\]
Apart from these recursive draws, generating indices and scanning activity
lists cost polynomial time in $n+s+q_2+w+Q+L+1$, copying a label costs time
linear in its length, and no coloring is required. Every width $s,q_1,q_2,w,Q$ is $2^{O(L)}$ by
\cref{lem:full-schedule}, and so is every draw count in the table. Hence we can choose common computable integer bounds $J_i$ for
the label length, generation time, elementary draw count and entry-evaluation
time at level $i$ such that
\[
J_0=O(L),\qquad J_{i+1}\le2^{c_0L}J_i,\qquad
J_k=2^{O((k+1)L)},
\]
where $c_0$ is fixed and effective. Here each elementary draw counts as one
operation, and we implement it with fair bits below. At the initial matrix $R_0$ of \cref{cl:seed},
a row consists of $p_0=O(L)$ independent uniform bits, a column is a uniform index in
$[p_0]$, and an entry can be read in $O(L)$ time. This gives $J_0=O(L)$.

Since $h=O(L)$ and $d_0=O(L)$ by \cref{lem:full-schedule}, we have $B_k=O((k+1)L)$. The direct sample
size in \cref{thm:compression} then gives $m=2^{O((k+1)L)}$.
We generate and store the $m$ row labels and $m$ column labels, then evaluate all
$m^2$ row--column entries.
Including the stored tuples and printed entries, this costs
\[
O((2m+m^2)J_k)=2^{O((k+1)L)}=n^{O(k+1)}.
\]
All constants are independent of $k$.

\emph{Fixed-bit implementation.}
Let $N_{\rm draw}\ge1$ be a computable bound on the elementary draws for
all $2m$ labels, so $N_{\rm draw}=2^{O((k+1)L)}$. On every tape, each draw
has at most
\[
A=\max\{512,4s(n+s),q_2,w,Q,p_0\}=2^{O(L)}
\]
outcomes, and the weights of each draw have a common denominator $\le A$.
This holds for type draws (whose weights in \cref{def:source-matrix} have
denominators $512$ and $256$), item indices, bins, template positions, activity indices,
interlace-copy indices and draws from the initial matrix. Rows and columns of
base matrices above $R_0$ are generated recursively, not drawn as single outcomes. We set $b=\ceil{\log(200A N_{\rm draw})}$ and reserve
$N_{\rm draw}$ blocks of $b$ fair bits. A uniform draw from $[a]$, $a\le A$, reads the next
block as a uniform integer $U\in\{0,\ldots,2^b-1\}$ and returns
\begin{equation}\label{eq:rounded-draw}
\operatorname{Draw}(a)=1+\floor{aU/2^b}.
\end{equation}
A draw whose weights have common denominator $a\le A$ maps consecutive
intervals of values of $\operatorname{Draw}(a)$, whose lengths are the
numerators of its weights, to its types. Unused blocks are ignored.

Each value in $[a]$ has $\floor{2^b/a}$ or $\ceil{2^b/a}$ preimages, so a
draw is within total variation $a/2^b\le1/(200N_{\rm draw})$ of the ideal
one, and mapping values to types cannot increase this distance. Because every
draw reads a fresh block, this bound holds conditionally on all earlier
outcomes, even when the alphabet depends on them.
We couple the implemented and ideal draws one at a time. By a union bound
over the at most $N_{\rm draw}$ draws, the implemented and ideal draws differ somewhere with probability at most
$N_{\rm draw}/(200N_{\rm draw})=1/200<1/100$.

Every tape thus produces exactly $m$ legal row labels and $m$ legal column
labels, and all $m^2$ entries are evaluated exactly. The tape has fixed length
$bN_{\rm draw}$, where $b=O((k+1)L)$. Arithmetic on $O(b+\log A)$-bit integers adds only
polynomial overhead. Hence the running time and the number of random bits are
$n^{O(k+1)}$ on every tape.
\end{proof}

\section{One parameter recipe for both constructions}
\label{sec:parameters}

We first verify the numerical requirements of the recipe in
\cref{def:amplification-parameters} that both constructions share. For full
products and for explicit lists, we then check the source-matrix
deletion bounds and the final interlace.

Recall $n=|V|$, $\lambda=255/256$, $\mu=1/256$, and $\zeta=\gamma/24$.
The source promise gives, with $\delta_v$ as in \cref{src:item-witness},
\[
 n/2\le |E|\le\Delta n/2,\qquad
 \delta_v=\frac{\gamma|E|}{2\Delta n}\ge\frac{\gamma}{4\Delta}.
\]
From \cref{res:destinations} we use the bound $E_{\rm exc}$ on the sum of
$\kappa$ over stopping nodes with no assigned copy, and from the proof of
\cref{res:incoming} the weighted moment constant $C_{128}$:
\begin{equation}\label{eq:common-budgets}
E_{\rm exc}=2e(\theta,\eta/2),\qquad
C_{128}=\left(\frac{2^{14}}{(\lambda-1/2)^2}\right)^{128}.
\end{equation}
The definition
of $e(v,\tau)$ before \cref{det:conditioning} gives the estimate
\begin{equation}\label{eq:conditioning-size-bound}
\begin{aligned}
e(v,\tau)
&\le8+2^{23}\tau^{-2}v^{-1/512}+2^{83}v^{-5/512}\\
&\le2^{84}\tau^{-2}v^{-1/64}
\qquad(0<v,\tau\le1).
\end{aligned}
\end{equation}

\begin{lemma}[Common numerical bounds]\label{lem:common-schedule}
There is an effective dyadic constant $s_0$, depending only on the fixed
graph promise, such that the recipe in \cref{def:amplification-parameters}
with $n\le\rho^{-1}$, $\rho\le s^{-4}$, and $s\ge s_0$ satisfies all source
and internal-interlace bounds listed in \eqref{eq:width-margins} below. The choice of
$s_0$ is independent of the base matrix and the number of rounds.
\end{lemma}

\begin{proof}
The middle column of the table below lists upper bounds on the quantities
in the first column, obtained by substituting the recipe. Each of these bounds
tends to zero as $s$ increases. The last column
lists the bound that the proof requires for each quantity.
\begin{equation}\label{eq:width-margins}
\begin{array}{c|c|c}
\text{quantity}&\text{upper bound}&\text{required allowance}\\ \hline
s\eta&s^{-1}&2^{-16}\min\{1,\delta_v,\mu\zeta\}\\
4s|E|/q_2&2\Delta/s&1\\
s^2\rho&s^{-2}&1/4\\
sq_1/q_2&s^{-4}&\zeta/128\\
\max\{q_2x,sx,wx\}&s^{-24}&10^{-4}\\
w^2x/\eta&s^{-14}&1/(32\lambda)\\
w^2q_2x&s^{-10}&1\\
wq_2x/\eta&s^{-16}&1/(128\mu)\\
e(\theta/4)/(sq_2)&2^{85}s^{-53/8}&1\\
C_{128}s^{-128}/\theta&512C_{128}s^{-104}&1/4\\
2sE_{\rm exc}/(\eta w)&2^{89}s^{-5/8}&1/16
\end{array}
\end{equation}
For the two rows involving $e$, where $e(v)=e(v,1)$, \eqref{eq:conditioning-size-bound} with
$\theta=2^{-9}s^{-24}$ and $\eta=s^{-2}$ gives
$e(\theta/4)\le2^{85}s^{3/8}$ and
$E_{\rm exc}\le2^{88}s^{35/8}$, while $sq_2=s^3/\rho\ge s^7$.
The other rows follow by substituting the widths of
\cref{def:amplification-parameters} and using $|E|\le\Delta n/2$, $n\le\rho^{-1}$,
and $\rho\le s^{-4}$.
We choose a dyadic $s_0\ge256$ so that every middle entry is below its allowance,
using $\gamma/(4\Delta)$ in place of $\delta_v$. Clearing the fixed rational
exponents makes this a finite set of integer-power comparisons. Each
middle entry decreases with $s$, so the same choice works for every $s\ge s_0$.
None of these comparisons involves the base matrix or the number of rounds.

The table's second row gives $q_2\ge4s|E|$. Since $q_2$ is a power of two,
$q_2\ge q_0=\operatorname{pow}_2(4s|E|)$, as \cref{def:source-matrix} requires. In particular, $|E|\ge n/2$ gives
$q_2\ge2sn\ge2(n+s-3)$, so all item labels fit: $|\mathcal I|=4s(n+s-3)\le2sq_2$, as
\cref{lem:amplifier-rank} requires. The table's last bound
also implies the two bounds used to count assigned nodes:
\[
E_{\rm exc}\le\frac{\eta w}{32s}<\frac w4,\qquad
e(\theta)\le e(\theta,\eta/2)=\frac{E_{\rm exc}}2
\le\frac{\eta w}{64s}<\frac w{128}.
\]
Here $e(\theta)=e(\theta,1)\le e(\theta,\eta/2)$ because $e(v,\tau)$ is
nonincreasing in the error bound $\tau$.

The remaining checks of fixed fractions appear in the proofs that use them.
For each such proof, we list the table rows it uses and derive here only the
bounds that it does not display.
\begin{itemize}
\item \emph{Block counts} (proof of \cref{src:profiles}). Since $s\ge256$
and $s\rho<10^{-3}$, the three lower bounds for $m/u$ satisfy
\[
\frac35(1-2/s)-q_1\rho>\frac9{16},\qquad
\frac35-q_2x>\frac9{16},\qquad
\frac{44}{75}-sx>\frac9{16},
\]
the last since $2^j(\beta_j-(1-\lambda))/\lambda\ge((3/5)-4(1-\lambda))/\lambda=44/75$
for $j\in\{0,1\}$. The bound on $m/\kappa$ in the
proof of \cref{res:alignment} uses $wx\le1/256$ and $4e(\theta)/w<1/32$
from the display above.
\item \emph{Item deletion and collisions} (proof of \cref{src:near-full}).
The row $s\eta$ gives $4096\eta sn\le\delta_v n/16$ and
$4096\eta s(s-3)\le(s-3)/16<s/2$. The rows $s\eta$ and $sq_1/q_2$ give
$8sq_1+4s\eta q_2/\mu\le(1/16+2^{-14})\zeta q_2<\zeta q_2/4$.
\item \emph{Row and template thresholds} (proofs of
\cref{src:near-full,res:destinations,cl:two-stream}). The rows $s^2\rho$, $s\eta$
and $\max\{q_2x,sx,wx\}$ give $2s\rho\le1/8$, $2wx+\eta/(16\lambda)<1$,
$2sx\le10^{-3}$, and hence $999/1000-2(s-1)x>998/1000$ in the proof of
\cref{src:near-full} and
$(99/100+\lambda^2-1)/\lambda^2-10^{-3}>49/64$ in \cref{cl:two-stream}.
\item \emph{Conservation} (proof of \cref{res:alignment}). That proof uses
the rows $w^2x/\eta$, $wq_2x/\eta$ and $2sE_{\rm exc}/(\eta w)$ and displays
the resulting bounds.
\item \emph{Moment bound} (proof of \cref{res:incoming}). That proof
bounds the tail estimate $C_{128}D_0^{-128}$ obtained from \cref{lem:conditioning-moment}
by $\theta/4$, using $D_0\ge512s\kappa\ge512s$ and the row $C_{128}s^{-128}/\theta$.
\end{itemize}
\end{proof}

\subsection{Full products}

In the randomized reduction, the graph-encoding width $q_2=s^22^L$ is the only
width that contributes a term linear in $L$ to the per-round overhead $h$. The widths
$s$, $w=s^8$, and $Q=s^{24}$ are polynomial in $\log n$. A smaller overhead allows more
amplification rounds before the final sample becomes too large.

\begin{lemma}[Parameters for full products]\label{lem:full-schedule}
Put $L=\ceil{\log n}$, $\rho=2^{-L}$, and
$s=\operatorname{pow}_2(2^{16}(L+1))$, and use the recipe of \cref{def:amplification-parameters}.
For every sufficiently large source graph, these choices satisfy the hypotheses of
\cref{lem:common-schedule}, the numerical conditions for the full-product deletion bounds for the source matrices, and
the numerical hypotheses of \cref{cl:bridge} with $b=L$ and $\delta=\delta_U$. The per-round overhead is
\[
h=2\log s+\log q_2+\log w+\log Q=L+36\log s=L+O(\log L).
\]
There is an effective fixed cutoff beyond which also
$L\ge16$, $d_0\le L$, and $h\le7L$, where $d_0$ is the cost of the initial matrix in
\cref{cl:seed}. Computing the widths has polynomial bit complexity and
uses neither the base matrix nor the number of rounds.
\end{lemma}

\begin{proof}
Since $s=\operatorname{pow}_2(2^{16}(L+1))$, we have $2^{16}(L+1)\le s<2^{17}(L+1)$. Hence $s\ge s_0$ and
$2^{-L}\le s^{-4}$ beyond an effective fixed cutoff. Since also
$n\le2^L=\rho^{-1}$, the common lemma applies. In each application of the deletion bounds for the source matrices, the block parameter
satisfies $u\ge s/2$. Since $\log(1/x)=8L$,
\[
u\ge2^{15}(L+1)>4096\log(1/x),
\]
which gives $x>2^{-u/4096}$ and hence the block threshold
$x\ge2^{-u/1024}$ of \cref{lem:many-product-blocks}. \Cref{lem:common-schedule}
gives the block counts and row-weight bounds required in the proof of \cref{src:profiles}.
The remaining hypotheses of \cref{cl:bridge} for the final interlace are
\[
0<\delta_U=\frac1{16s^{10}}<\frac1{64},\qquad
Q\theta=\frac1{512}<\frac1{20},\qquad
Q\delta_U=\frac{s^{14}}{16}\ge80L.
\]
Thus \cref{cl:bridge} applies with $b=L$.

Substituting the widths gives the displayed value of $h$. Writing
$\ell=\ceil{\log(L+1)}$, the initial matrix of \cref{cl:seed} has $d_0=\log p_0+1=\ell+7$ and
$\log s=16+\ell$, so $d_0\le L$ and $h\le7L$ eventually.
Since $s=\Theta(L+1)$ with explicit constants, the ratios $s^42^{-L}$,
$d_0/L$ and $(h-L)/L$ are at most fixed multiples of $(L+1)^42^{-L}$ or
$(1+\log(L+1))/L$. These bounds decrease to zero beyond a computable point.
Hence the cutoff is effective and valid at every larger size. All widths except $q_1=s-2$ are powers of two
with exponents $O(L)$; forming those powers and subtracting two from $s$ gives the stated bit complexity.
\end{proof}

\subsection{Explicit lists}

For the deterministic reduction, we choose the least dyadic $s\ge s_0$ with
$s^4\ge n$ and set $\rho=s^{-4}$. Then every width is polynomial in $s$,
including the final-interlace width $Q=s^{24}$ required by \cref{det:restore}.

\begin{lemma}[Parameters for explicit lists]\label{lem:generated-schedule}
After increasing the effective dyadic constant $s_0$ if necessary, every
dyadic $s\ge\max\{n^{1/4},s_0\}$ satisfies the bounds below. Choose the least such $s$ and put
$\mathsf N=s^4$, $L_{\rm g}=\log\mathsf N$, and $\rho=\mathsf N^{-1}$.
With these choices, the common recipe satisfies all numerical hypotheses of the explicit-list
construction and of the final interlace in \cref{det:restore}. Its overhead is
\[
h_{\rm g}=2\log s+\log q_2+\log w+\log Q
=40\log s=10L_{\rm g}.
\]
With $\mathsf N_0=s_0^4$, the padding satisfies
$n\le\mathsf N<16\max\{n,\mathsf N_0\}$. Every width is polynomial in
$\mathsf N$. One round has a fixed polynomial cost in $\mathsf N$ and the
complete encoding of the base matrix, with exponent independent of the round number.
\end{lemma}

\begin{proof}
Here $\rho=s^{-4}$, so the common lemma applies. Only the deletion bounds for the source matrices
at tuple weight $x$ and the final interlace need additional checks.
By \eqref{eq:conditioning-size-bound},
\[
e(x)\le2^{84}s^{1/2}\le s/128,\qquad
e(\rho,\delta_U/2)\le2^{94}s^{20+1/16}.
\]
We increase $s_0$ so that $2^{84}s^{1/2}\le s/128$ and
$128(2^{94}s^{20+1/16}+1)\le s^{24}=Q$. Dividing each condition by its right
side leaves fixed multiples of negative powers of $s$, so this enlargement is
effective and valid for every larger $s$.

For the deletion bounds for the source matrices, $u\ge s/2$, and by the three
block-count bounds in the proof of \cref{src:profiles}, more than $9u/16$
blocks meet the required row-weight threshold. Discarding the blocks at the at most $e(x)\le s/128\le u/32$ exceptional
coordinates leaves more than $9u/16-u/32>u/2$ blocks. The row thresholds obey
$u\rho\le s\rho\le1$ for the first matrix and
$ux\le\max\{q_2x,sx\}\le1$ in the other two.
For the final interlace, $0<\rho\le1$, $0<\delta_U<1/64$,
$Q\theta=1/512<1/20$, and the bound $Q\ge128(e(\rho,\delta_U/2)+1)$ above are
the numerical hypotheses of \cref{det:restore} with $\phi=\rho$ and
$\delta=\delta_U$.

The least dyadic choice satisfies
$s<2\max\{n^{1/4},s_0\}$, proving the factor-$16$ size bound. This $s$ can be
found by comparing integer fourth powers, without approximating a root.
The widths are $q_2=s^6$, $w=s^8$, and $Q=s^{24}$, giving the stated
overhead. The independence order stays fixed at $8192$, so by
\cref{det:code} each of the fixed number of tuple-list constructions has cost
polynomial in these widths and in the complete rational encoding of the base matrix.
Neither the parameter choices nor the exponent of the per-round polynomial
cost depends on the number of rounds.
\end{proof}

\section{Coloring promises from SAT}
\label{sec:sources}

The interlacing reduction uses bounded-degree graph coloring
(\cref{src:gap-colouring}), and the strict-alternation reduction uses balanced
hypergraph coloring (\cref{yao:source}). This appendix derives both hardness results from the literature and checks
the input conventions that the communication constructions need.

\subsection{Bounded-degree graph coloring of almost-linear size}
\label{app:graph-source}

The graph source of \cref{src:gap-colouring} needs a constant coloring gap
and must keep the almost-linear size of sparse SAT. Composing the reductions of
Dinur and of Bogdanov, Obata and Trevisan gives both. We record this composition
and our graph conventions because the size bound determines the final additive-error bound.

\begin{proof}[Proof of \cref{src:gap-colouring}]
\phantomsection\label{proof:src:gap-colouring}
For a SAT instance of encoding length $L$, Dinur's almost-linear PCP
construction gives a constant-alphabet binary constraint graph of size
$L(\log L)^{O(1)}$, with perfect completeness and a constant
unsatisfied-constraint fraction in the NO case
\cite[Lemma~8.3 and proof of Theorem~8.1]{dinur2007pcp}.
Dinur's degree reduction makes the degree constant, preserves perfect
completeness, and changes the size and the NO-case unsatisfied fraction by
only constant factors~\cite[Lemma~1.7]{dinur2007pcp}.
A constant-size Boolean encoding of the labels and constraints, followed by
the reduction of \citet*[Lemma~11]{bogdanov2002lower}, gives bounded-occurrence $3$-SAT
with perfect completeness, a constant NO-case gap and almost-linear size. The gap-preserving reduction of
\citet*[Theorem~14 and its proof, pp.~98--99]{bogdanov2002lower}
then gives a bounded-degree graph of linear size.

It remains to match the conventions of \cref{src:gap-colouring}. We remove isolated vertices and add a
disjoint edge, so that the graph is nonempty. This preserves completeness and, when the graph already has
an edge, decreases the NO-case gap by at most a factor of two. For trivial inputs we use fixed promise graphs.
For a sparse formula on $N$ variables, $L=O(N\log(N+2))$, so the resulting
graph $H$ has $N(\log(N+2))^{O(1)}$ vertices and edges. Taking
$\max\{1,\ceil{N/|E(H)|}\}$ disjoint copies gives
$N\le |E(G)|\le N+|E(H)|$ without changing the degree or gap.
The remaining bounds $n/2\le|E(G)|\le\Delta n/2$ follow from the absence of
isolated vertices and the maximum degree $\Delta$. Parallel edges are retained, and edge
occurrences are counted with multiplicity. We choose a rational $\gamma>0$
below the resulting gap and enlarge the fixed exponent $p$ to cover the
constant factors. The cited reductions and these adjustments are
deterministic and polynomial-time.
\end{proof}

\subsection{Balanced hypergraph coloring with distinct vertices}
\label{app:hypergraph-source}

The hypergraph promise of \cref{yao:source} comes from the hardness result of
Guruswami and Lee~\cite{guruswami2018balanced}, whose edges may contain repeated
vertices. We replace repeated occurrences by distinct labeled clones and
check that the YES and NO conditions of \cref{yao:source} are preserved.

\begin{proof}[Proof of \cref{yao:source}]
\phantomsection\label{proof:yao:source}
Recall the fixed parameters of \cref{yao:source}
\[
s_{\rm a}=2^t,\qquad K_{\rm a}=s_{\rm a}^2,\qquad
S_{\rm a}=2^{s_{\rm a}}.
\]
By Theorem~1.1 of \citet{guruswami2018balanced}, it is NP-hard to distinguish
$K_{\rm a}$-uniform occurrence hypergraphs admitting an $s_{\rm a}$-coloring
with at least $s_{\rm a}-1$ occurrences of each color in every edge from those
with no independent set containing at least a $1/(2S_{\rm a})$ fraction of the
vertices. Here, as in their reduction, an edge of an occurrence hypergraph is a
family of $K_{\rm a}$ vertex occurrences indexed by $[s_{\rm a}]\times[s_{\rm a}]$,
listed in a fixed order. A vertex may occur more than once in an edge, and
colors in an edge are counted with multiplicity, as in their completeness
proof. Hypergraphs of the second kind admit no proper $S_{\rm a}$-coloring,
since some color class would contain at least a $1/S_{\rm a}$ fraction of the
vertices and so would not be independent.

We replace each vertex $u$ by $K_{\rm a}S_{\rm a}$ labeled clones. For each
ordered occurrence edge $(u_1,\ldots,u_{K_{\rm a}})$, we include every tuple
\[
(u_1^{(j_1)},\ldots,u_{K_{\rm a}}^{(j_{K_{\rm a}})})
\]
of pairwise distinct clones, in the displayed order, as an edge. If an
occurrence edge appears several times, we treat its copies as separately labeled edges. Each original edge produces at least
one such tuple, since each vertex has $K_{\rm a}S_{\rm a}\ge K_{\rm a}$ clones. Giving clones
their source vertex's color preserves the color multiset of every edge,
and hence the YES promise.

Conversely, suppose the cloned hypergraph has a proper $S_{\rm a}$-coloring.
We give each source vertex a color appearing on at least $K_{\rm a}$ of its
clones. Suppose some occurrence edge were monochromatic in this induced coloring.
Each source vertex occurs at most $K_{\rm a}$ times in that edge, so we can choose
different clones of that color for its positions. This gives a monochromatic
edge of distinct clones, contradicting properness. Hence the induced coloring
is proper, contradicting the NO promise.

Each vertex produces $K_{\rm a}S_{\rm a}$ clones and each edge produces at
most $(K_{\rm a}S_{\rm a})^{K_{\rm a}}$ cloned edges. These are constants for fixed
$t$, so enumeration is deterministic and polynomial-time, although the
polynomial may depend on $t$. If the vertex set is empty, we add one isolated labeled
vertex, since \cref{yao:source} requires a nonempty vertex set. Because each
ordered edge has distinct vertices, the position $\iota_\alpha(v)$ in
\cref{def:alternating-table} is well defined.
\end{proof}

\section{Finite-field arrays}\label{sec:arrays}
\label{sec:alternating_implementation}

The alternating lower bound needs a short, explicitly generated list of
arrays on which no prescribed bit pattern occurs much more often than it
would for independent fair bits (\cref{eq:alternating-pattern}). We construct that list here, completing
the algorithmic step promised in \cref{yao:array-list}.

We use the polynomial-root construction of Alon, Goldreich,
H{\aa}stad and Peralta~\cite[Construction~3 and Proposition~3]{alon1992smallbias}.
The conversion from parity bounds to pattern frequencies is also recorded
in that paper, with credit to Vazirani~\cite[Lemma~1 and Corollary~1]{alon1992smallbias}.
We use the following direct consequence and keep its seed labels explicit
so that the reduction can print the full array list.

\begin{lemma}[Construction~3, Proposition~3 and Corollary~1 in \cite{alon1992smallbias}]\label{yao:finite-field-list}
For integers $n,r\ge1$, a deterministic algorithm prints a nonempty labeled
list $(w^\ell)_{\ell\in\mathcal L}$ of $n$-bit arrays satisfying
\[
\frac{|\{\ell\in\mathcal L:w^\ell|_I=z\}|}{|\mathcal L|}
\le2^{-|I|}+2^{-r}
\qquad(I\subseteq\{0,\ldots,n-1\},\ z\in\{0,1\}^I).
\]
Its running time and complete output length are polynomial in $n2^r$.
\end{lemma}

To instantiate the construction, let $R$ be the least power of two with
$R\ge n2^r$. The construction lists $R^2<4n^24^r$ arrays, indexed by
$\mathbb F_2^{\log R}\times\mathbb F_R$, whose parities over nonempty coordinate sets have bias at most
$(n-1)/R\le2^{-r}$, and the bias-to-pattern conversion gives the displayed bound.
A field representation of $\mathbb F_R$ can be found by exhaustive search in time
polynomial in $R$, and we keep repeated arrays as distinct labels.

Finally, we apply \cref{yao:finite-field-list} on the positions
$\Omega_{\rm a}$ of \cref{def:alternating-table} with error $\epsilon_{\rm a}$,
and check that the list has size polynomial in $H_{\rm a}$ for each fixed $t$.

\begin{proof}[Proof of \cref{yao:array-list}]
\phantomsection\label{proof:yao:array-list}
We number $\Omega_{\rm a}$ by $0,\ldots,n-1$ and apply
\cref{yao:finite-field-list} with
\[
n=K_{\rm a}H_{\rm a},\qquad r=4(b+3t).
\]
Here $2^{-r}=\epsilon_{\rm a}$ and
$n2^r=K_{\rm a}2^{12t}H_{\rm a}^5$, which is polynomial in $H_{\rm a}$
for fixed $t$. The integers $n$ and $r$ have $O_t(b+1)$ bits, and the
polynomial bound also covers $H_{\rm a}=1$.
\end{proof}

\section*{AI methodology}
\label{sec:ai-methodology}

The original NP-hardness result and the subsequent ETH-based
inapproximability result were obtained manually by the authors, who
developed the underlying constructions and mathematical insights. The
authors also initiated the effort to sparsify interlacing while retaining
its communication lower bounds. AI assistance contributed to the later
development: completing proof details, finding and checking further
arguments, extending the results to strict alternation, simplifying the
proofs, and formalizing them in Lean.

We used OpenAI GPT models through ChatGPT Pro and Codex, together with
Anthropic's Claude models, to develop and check arguments, search for
counterexamples, and compare proposed proofs with the literature. GPT and
Claude also helped refine the writing and simplify the proofs by removing
redundant cases, combining shared arguments, and reducing parameter
bookkeeping. The inputs to this work included the authors' manuscripts,
specific mathematical questions, and proposed arguments.

The final sparsification approach was developed with AI assistance,
including the connection to rank-sensitive counting: bounding the number
of different column sets on which a short protocol can be correct in
terms of the matrix rank. The counting ingredient used here builds on
the row-sum argument of Oum, S{\ae}ther and
Vatshelle~\cite[Theorem~4.2]{oum2014faster}, in its protocol-specific form
given by Mackenzie~\cite[Lemma~4.3]{mackenzie2026condensation}; related
counting and covering methods for hard submatrices appear in
Hrube{\v s}~\cite{hrubes2025hard}. AI helped connect these existing methods
to the authors' sparsification objective and develop their application to
the reduction. The contribution here is their use to retain the required
communication lower bound in a much smaller table.

AI assistance also contributed to amplification followed by a final
sampling step, to the bounded-independence and conditioning arguments
that keep the deterministic construction small, and to the separate
construction for strictly alternating protocols. The latter uses the
classical small-bias construction of Alon, Goldreich, H{\aa}stad and
Peralta~\cite{alon1992smallbias}. ChatGPT Pro supplied substantive
mathematical proposals; Codex helped complete their proofs and check the
parameter choices and running-time bounds; Claude contributed
mathematical and editorial review.

The mathematical and editorial work was followed by AI-assisted
formalization in Lean. The companion repository~\cite{interlacing2026lean}
provides the machine-checked statements and proofs, together with build
instructions. The interactive proof inspector~\cite{interlacing2026inspector}
connects the manuscript to those proofs and lets readers examine their
definitions, hypotheses, and dependencies. Cited source reductions remain
explicit hypotheses of the formal statements. The authors retain
responsibility for the mathematical claims, the use and attribution of
prior work, and the final text.


\bibliographystyle{plainnat}
\bibliography{references}

@article{alon1992smallbias,
  author = {Noga Alon and Oded Goldreich and Johan H{\aa}stad and Ren{\'e} Peralta},
  title = {Simple Constructions of Almost {$k$}-wise Independent Random Variables},
  journal = {Random Structures \& Algorithms},
  volume = {3},
  number = {3},
  pages = {289--304},
  year = {1992},
  doi = {10.1002/rsa.3240030308},
  url = {https://web.math.princeton.edu/~nalon/PDFS/aghp4.pdf},
  note = {Our theorem numbering refers to the author-hosted version dated 22 February 2002: Lemma 1, Corollary 1, Construction 3 and Proposition 3}
}

@inproceedings{yao1979distributed,
  author = {Andrew Chi-Chih Yao},
  title = {Some Complexity Questions Related to Distributive Computing (Preliminary Report)},
  booktitle = {Proceedings of the Eleventh Annual ACM Symposium on Theory of Computing},
  year = {1979},
  pages = {209--213},
  publisher = {Association for Computing Machinery},
  doi = {10.1145/800135.804414},
  url = {https://doi.org/10.1145/800135.804414}
}

@article{guruswami2018balanced,
  author = {Venkatesan Guruswami and Euiwoong Lee},
  title = {Strong Inapproximability Results on Balanced Rainbow-Colorable Hypergraphs},
  journal = {Combinatorica},
  volume = {38},
  number = {3},
  pages = {547--599},
  year = {2018},
  doi = {10.1007/s00493-016-3383-0},
  url = {https://eccc.weizmann.ac.il/report/2014/043/},
  note = {Theorem numbering in our source discussion refers to ECCC TR14-043}
}

@article{nisan1993rounds,
  author = {Noam Nisan and Avi Wigderson},
  title = {Rounds in Communication Complexity Revisited},
  journal = {SIAM Journal on Computing},
  volume = {22},
  number = {1},
  pages = {211--219},
  year = {1993},
  doi = {10.1137/0222016},
  url = {https://doi.org/10.1137/0222016}
}

@misc{gaspers-he-mackenzie-npcc,
  author        = {Serge Gaspers and Tao Zixu He and Simon Mackenzie},
  title         = {{NP}-Completeness of Deterministic Communication Complexity via Relaxed Interlacing},
  year          = {2026},
  eprint        = {2508.05597},
  archivePrefix = {arXiv},
  primaryClass  = {cs.CC},
  doi           = {10.48550/arXiv.2508.05597},
  url           = {https://arxiv.org/abs/2508.05597v4},
  note          = {Earlier version of the present paper, version 4, 13 April 2026}
}

@misc{hirahara2025communication,
  author        = {Shuichi Hirahara and Rahul Ilango and Bruno Loff},
  title         = {Communication Complexity is {NP}-hard},
  year          = {2025},
  eprint        = {2507.10426},
  archivePrefix = {arXiv},
  primaryClass  = {cs.CC},
  doi           = {10.48550/arXiv.2507.10426},
  url           = {https://arxiv.org/abs/2507.10426v1},
  note          = {Preprint, version 1, 14 July 2025}
}

@misc{mackenzie2025refuting,
  author        = {Simon Mackenzie and Abdallah Saffidine},
  title         = {Refuting the Direct Sum Conjecture for Total Functions in Deterministic Communication Complexity},
  year          = {2026},
  eprint        = {2411.19003},
  archivePrefix = {arXiv},
  primaryClass  = {cs.CC},
  doi           = {10.48550/arXiv.2411.19003},
  url           = {https://arxiv.org/abs/2411.19003v2},
  note          = {Corrected full version, 14 April 2026; conference version appeared in STOC 2025}
}

@article{dinur2007pcp,
  author    = {Irit Dinur},
  title     = {The {PCP} Theorem by Gap Amplification},
  journal   = {Journal of the ACM},
  volume    = {54},
  number    = {3},
  articleno = {12},
  numpages  = {44},
  year      = {2007},
  month     = jun,
  doi       = {10.1145/1236457.1236459},
  url       = {https://www.wisdom.weizmann.ac.il/~dinuri/mypapers/combpcp.pdf},
  note      = {Article 12. Our theorem numbering refers to the author manuscript dated 13 February 2007}
}

@inproceedings{bogdanov2002lower,
  author    = {Andrej Bogdanov and Kenji Obata and Luca Trevisan},
  title     = {A Lower Bound for Testing 3-Colorability in Bounded-Degree Graphs},
  booktitle = {Proceedings of the 43rd Annual IEEE Symposium on Foundations of Computer Science},
  publisher = {IEEE Computer Society},
  pages     = {93--102},
  year      = {2002},
  doi       = {10.1109/SFCS.2002.1181886},
  url       = {https://lucatrevisan.github.io/pubs/bot.pdf}
}

@inproceedings{kushilevitz2009complexity,
  author    = {Eyal Kushilevitz and Enav Weinreb},
  title     = {On the Complexity of Communication Complexity},
  booktitle = {Proceedings of the 41st Annual ACM Symposium on Theory of Computing},
  publisher = {Association for Computing Machinery},
  pages     = {465--474},
  year      = {2009},
  doi       = {10.1145/1536414.1536479},
  url       = {https://doi.org/10.1145/1536414.1536479}
}

@article{bui2010rankwidth,
  author = {Binh-Minh Bui-Xuan and Jan Arne Telle and Martin Vatshelle},
  title = {{$H$}-join Decomposable Graphs and Algorithms with Runtime Single Exponential in Rankwidth},
  journal = {Discrete Applied Mathematics},
  volume = {158},
  number = {7},
  pages = {809--819},
  year = {2010},
  doi = {10.1016/j.dam.2009.09.009},
  url = {https://perso.lip6.fr/binh-minh.bui-xuan/files/BTV.pdf},
  note = {Our Proposition 3.6 locator follows the author-hosted manuscript}
}

@article{oum2014faster,
  author = {Oum, Sang-il and S{\ae}ther, Sigve Hortemo and Vatshelle, Martin},
  title = {Faster Algorithms for Vertex Partitioning Problems Parameterized by Clique-width},
  journal = {Theoretical Computer Science},
  volume = {535},
  pages = {16--24},
  year = {2014},
  doi = {10.1016/j.tcs.2014.03.024},
  url = {https://arxiv.org/abs/1311.0224v2},
  note = {Our theorem numbering refers to arXiv version 2}
}

@article{hrubes2025hard,
  author = {Pavel Hrube{\v s}},
  title = {Hard Submatrices for Non-Negative Rank and Communication Complexity},
  journal = {Computational Complexity},
  volume = {34},
  year = {2025},
  note = {Article 9},
  doi = {10.1007/s00037-025-00269-4},
  url = {https://doi.org/10.1007/s00037-025-00269-4}
}

@inproceedings{ilango2020multioutput,
  author = {Rahul Ilango and Bruno Loff and Igor Carboni Oliveira},
  title = {{NP}-Hardness of Circuit Minimization for Multi-Output Functions},
  booktitle = {35th Computational Complexity Conference (CCC 2020)},
  series = {Leibniz International Proceedings in Informatics (LIPIcs)},
  volume = {169},
  pages = {22:1--22:36},
  year = {2020},
  doi = {10.4230/LIPIcs.CCC.2020.22},
  url = {https://drops.dagstuhl.de/entities/document/10.4230/LIPIcs.CCC.2020.22}
}

@inproceedings{hirahara2021constant,
  author = {Shuichi Hirahara and Rahul Ilango and Bruno Loff},
  title = {Hardness of Constant-Round Communication Complexity},
  booktitle = {36th Computational Complexity Conference (CCC 2021)},
  series = {Leibniz International Proceedings in Informatics (LIPIcs)},
  volume = {200},
  pages = {31:1--31:30},
  year = {2021},
  doi = {10.4230/LIPIcs.CCC.2021.31},
  url = {https://drops.dagstuhl.de/entities/document/10.4230/LIPIcs.CCC.2021.31}
}

@misc{mackenzie2026condensation,
  author = {Simon Mackenzie},
  title = {Lossless Hardness Condensation in Deterministic Communication Complexity},
  year = {2026},
  eprint = {2609.28691},
  archivePrefix = {arXiv},
  primaryClass = {cs.CC},
  doi = {10.48550/arXiv.2609.28691},
  url = {https://arxiv.org/abs/2609.28691v1},
  note = {Preprint, version 1, 23 September 2026}
}

@article{koller1994sample,
  author = {Daphne Koller and Nimrod Megiddo},
  title = {Constructing Small Sample Spaces Satisfying Given Constraints},
  journal = {SIAM Journal on Discrete Mathematics},
  volume = {7},
  number = {2},
  pages = {260--274},
  year = {1994},
  doi = {10.1137/S0895480192228140},
  url = {https://theory.stanford.edu/~megiddo/pdf/stoc93.pdf},
  note = {First appeared in STOC 1993. Algorithm and theorem numbering here refers to the author-hosted conference version}
}

@misc{moakhar2026beyond,
  author = {Soltani Moakhar, Arshia and Gholami, Iman and Springer, Max and
            JafariRaviz, Mahdi and Hajiaghayi, MohammadTaghi},
  title = {{Beyond the Library}: An Agentic Framework for Autoformalizing Research Mathematics},
  year = {2026},
  eprint = {2606.31134},
  archivePrefix = {arXiv},
  primaryClass = {cs.AI},
  doi = {10.48550/arXiv.2606.31134},
  url = {https://arxiv.org/abs/2606.31134v3},
  note = {Version 3, 5 August 2026; direct-sum formalization in Section 3.2.2 and Appendix C.3}
}

@misc{gaspers2026interlacingeth,
  author = {Serge Gaspers and Tao Zixu He and Simon Mackenzie},
  title = {{NP}-Completeness and {ETH}-Based Additive Inapproximability of Deterministic Communication Complexity via Relaxed Interlacing},
  year = {2026},
  eprint = {2508.05597},
  archivePrefix = {arXiv},
  primaryClass = {cs.CC},
  doi = {10.48550/arXiv.2508.05597},
  url = {https://arxiv.org/abs/2508.05597v5},
  note = {Earlier ETH stage of the present paper, version 5, 22 October 2026}
}

@misc{beyondlibrary2026inspector,
  author = {{Beyond the Library}},
  title = {Formal Proof Inspector for {Refuting the Direct Sum Conjecture}},
  howpublished = {FormaliaXiv},
  year = {2026},
  url = {https://beyondthelibrary.github.io/formal_archive/p/Refuting/},
  note = {Accessed 25 October 2026}
}

@misc{interlacing2026lean,
  key = {Interlacing additive Lean},
  title = {Interlacing and Additive Inapproximability: {Lean} Companion},
  howpublished = {GitHub repository},
  year = {2026},
  url = {https://github.com/SimonWMackenzie/interlacing-additive-lean},
  note = {Companion formalization and proof resources for this paper}
}

@misc{interlacing2026inspector,
  key = {Interlacing additive proof inspector},
  title = {Interlacing and Additive Inapproximability: Paper and {Lean} Proof Inspector},
  howpublished = {Interactive proof website},
  year = {2026},
  url = {https://simonwmackenzie.github.io/interlacing-additive-lean/inspector/},
  note = {Companion formalization and proof resources for this paper}
}
\end{document}